\documentclass[mnsc,nonblindrev]{informs3_nojournal}

\usepackage{tikz,mathrsfs}
\usepackage{pgf}
\usetikzlibrary{arrows,automata}
\usepackage{amsfonts,amssymb} 
\usepackage{placeins}
\usepackage{makecell} 
\tikzstyle{legend}=[rectangle, draw=black,fill=white, rounded corners, minimum width=1cm, minimum height=0.75cm]
\OneAndAHalfSpacedXI
\usepackage{endnotes}
\usepackage{ mathrsfs }
\usepackage{subcaption}
\usepackage{bbm}
\usepackage{bm}
\usepackage{enumerate}
\usepackage{algorithm}
\usepackage{algorithmic}
\usepackage[colorlinks,citecolor=black,linkcolor=black,bookmarks=false,hypertexnames=true]{hyperref} 
\usepackage{threeparttable}
\usepackage{booktabs,caption}
\usepackage[normalem]{ulem}
\usepackage{eurosym}

\newcommand{\R} { \mathcal{R} }

\newif\ifrevision
\revisiontrue   % 
\revisionfalse % 

\ifrevision
  \newcommand{\cor}[1]{\textcolor{blue}{#1}}
\else
  \newcommand{\cor}[1]{#1}
\fi

\usepackage{natbib}
 \bibpunct[, ]{(}{)}{,}{a}{}{,}%
 \def\bibfont{\small}%
 \def\bibsep{\smallskipamount}%
 \def\bibhang{24pt}%
\DeclareMathAlphabet{\mathpzc}{OT1}{pzc}{m}{it}
\TheoremsNumberedThrough     % Preferred (Theorem 1, Lemma 1, Theorem 2)
\ECRepeatTheorems

\EquationsNumberedThrough    % Default: (1), (2), ...
\MANUSCRIPTNO{}
\begin{document}
%%%%%%%%%%%%%%%%

\RUNAUTHOR{Elmachtoub and Shi}
\RUNTITLE{The Power of Static Pricing for Reusable Resources} 
\TITLE{The Power of Static Pricing for Reusable Resources}

\ARTICLEAUTHORS{%

\AUTHOR{Adam N. Elmachtoub}
\AFF{Department of Industrial Engineering and Operations Research \& Data Science Institute, Columbia University, New York, NY, 10027, \EMAIL{adam@ieor.columbia.edu}}
\AUTHOR{Jiaqi Shi}
\AFF{Department of Industrial Engineering and Operations Research, Columbia University, New York, NY, 10027, \EMAIL{js5778@columbia.edu}}
} 

\ABSTRACT{We consider the problem of pricing a reusable resource service system. Potential customers arrive according to a Poisson process and purchase the service if their valuation exceeds the current price.  If no units are available, customers immediately leave without service.  Serving a customer corresponds to using one unit of the reusable resource, where the service time has a general distribution. The objective is to maximize the steady-state revenue rate. This system is equivalent to the classical Erlang loss model with price-sensitive customers, which has applications in vehicle sharing, cloud computing, and spare parts management.

With general service times, the optimal pricing policy depends not only on the number of customers currently in the system but also on how long each unavailable unit has been in use. We prove several results that show a simple static policy is universally near-optimal for any service time distribution, arrival rate, and number of units in the system. When there are multiple classes of customers with regular valuation distributions, we prove that static pricing guarantees 78.9\% of the revenue of the optimal dynamic policy, achieving the same guarantee known for a single class of customers with exponential service times. When there is one class of customers who have a monotone hazard rate valuation distribution, we prove that a static pricing policy guarantees \cor{97.5\%} of the revenue from the optimal inventory-based policy. In addition, we prove that the optimal static policy can be easily computed, resulting in the first polynomial-time approximation algorithm for the multi-class problem. \cor{Finally, we consider the extensions to irregular valuation distributions and general multi-resource networks.} }

\maketitle

\section{Introduction}\label{Intro}
Resources are \textit{reusable} if they can be used to serve future customers upon completion of the previous service. These resources are common in many real-world applications such as car rentals (Hertz, Zipcar, Budget), cloud computing (AWS, Google Cloud), and sharing economy (Uber, Citi Bike, Airbnb). In these applications, customers arrive sequentially and use a unit of the reusable resources for some period of time. Once the service is completed, the unit is free to be used by another customer. Due to fierce competition and on-demand needs in these markets, customers who find no units available will immediately leave without service. For instance, in the car rental industry, each customer will use one car for a few days and usually return it at the same location, after which the same car can be rented by future customers. Those who find that all cars are occupied will typically rent from other nearby companies or not rent at all. Another example of a reusable resource is rotable spare parts in the aircraft industry, %of which the number of units is often small due to their size and cost. 
where customers exchange their broken part for a functional one with the firm. In turn, the firm repairs the broken part and it eventually becomes available again to serve future customers \citep{besbes2020pricing}. In some applications of reusable resources, customers are distinguished into different classes that differ in their service time and price sensitivities. For instance, cloud computing customers may be categorized into different classes based on the types of jobs they run. A car rental company may divide customers by a loyalty program or divide them into different groups according to the number of days they reserve since the rental time is specified upon arrival. A crucial component of the reusable resources economy is how to price the resources to maximize revenue.  While dynamic pricing strategies are generally optimal for reusable resource systems, in this paper we show that simple static pricing policies are universally near-optimal with few assumptions and no asymptotic notions required. 

Intuitively, an optimal pricing policy dynamically adjusts prices in response to the number of customers currently in the system. For example, the firm may charge higher prices when more customers are in the system (few units available) and lower prices when few customers are in the system (more units available). This class of pricing policies, which depends solely on the number of customers in the system, is referred to as an \textit{inventory-based} pricing policy. When the service time is exponential, the memoryless property ensures that an inventory-based policy is optimal. For general service times (as considered in this paper), however, the optimal pricing policy must also depend on the service times consumed so far by busy units.
We refer to this broader class of pricing policies as \textit{fully dynamic} pricing. With an inventory-based or fully dynamic pricing strategy, the firm can balance the supply (available units) and demand to mitigate inefficiencies arising from the randomness in the demand and service durations.

In practice, however, dynamic pricing may suffer from various drawbacks. First, dynamic pricing is hard to implement since not all companies have the infrastructure to track their inventory in real time or adjust their decisions on the fly \citep{ma2021dynamic}. For rotable spare parts, companies also typically publish prices in a catalog upfront, resulting in considerable costs if prices change frequently. Moreover, finding an optimal dynamic pricing policy becomes intractable especially when the state space of the system is large due to the curse of dimensionality and having general service time distributions. Going beyond the seller's perspective, customers may feel it is unfair if someone else's price is lower than their own for the same service.  Dynamic pricing can thus result in customers strategically waiting for lower prices, which can result in lower than expected revenue.  To avoid these potential issues, we consider a static pricing policy that offers a fixed price for all customers from the same class.
\footnote{\cor{Allowing the fixed price to depend on the customer class is essential, as Section~\ref{sec:uniform-static} shows that requiring a uniform price across all classes can lead to arbitrarily poor performance.} }  
We prove that static pricing guarantees a large fraction of the revenue of the optimal dynamic pricing or inventory-based policy for any arrival rate, service time distribution, and number of units.

In our framework, we consider a service provider that is endowed with a fixed number of units, $C$, of a single type of reusable resource. There are (possibly) multiple classes of customers, differing in market size, service distribution, and valuation distribution. The firm can dynamically change the price of service as the state changes (including the class types of customers and their service times consumed so far).  Customers arrive according to a Poisson process. Each customer has a valuation for the service drawn from a known class-dependent distribution. Upon arrival, customers purchase the service if their valuation exceeds the current price. If no units are available, customers immediately leave without service. Serving a customer corresponds to using one unit of the reusable resource, where the class-dependent service time has a general distribution. After service completion, the resource unit is free to serve other customers. The system (without pricing) is equivalent to the classical Erlang loss system. The goal is to find an optimal pricing policy that maximizes the long-run average revenue rate. 

In this paper, we first consider two widely used classes of valuation distributions: the class of regular distributions and the class of monotone hazard rate (MHR) distributions. MHR distributions include uniform, exponential, logistic, and truncated normal \citep{bagnoli2005log}. Regular distributions are a superset of MHR and further incorporate heavy-tailed distributions (e.g. log-normal, log-logistic, equal-revenue, subsets of Pareto,...). Note that regular assumption is equivalent to the revenue function being concave in demand \citep{ewerhart2013regular}, which is standard in the revenue management and pricing literature. MHR distributions are also common and capture the family of log-concave distributions. \cor{Beyond these standard classes, we also examine irregular valuation distributions using the ironing method \citep{myerson1981optimal}.}

\begin{figure}[!]
\FIGURE{\centering
\includegraphics[width=5.7 in]{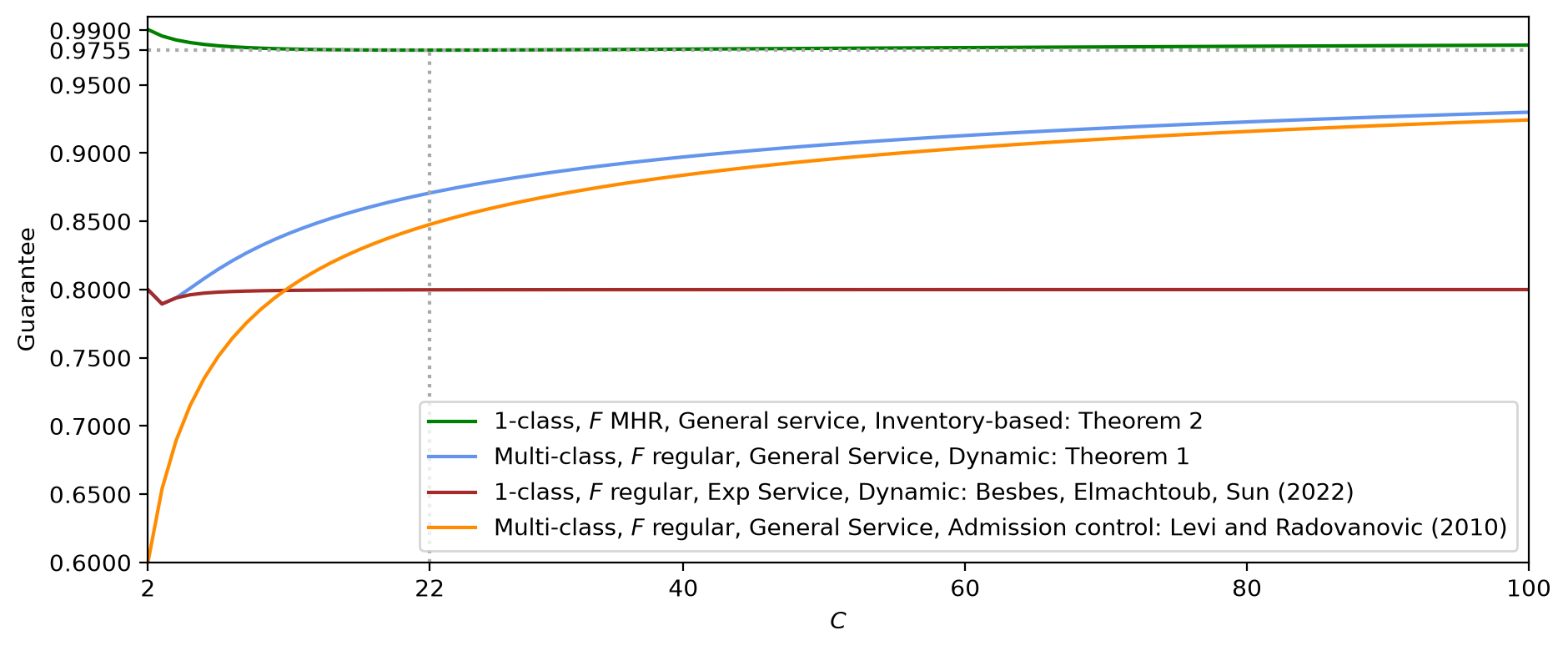}\label{Fig:Guarantee}
}{Our regular and MHR guarantees in comparison to guarantees from prior work. }
{The red line is the bound derived in \cite{besbes2022static} for one class with exponential service times. The orange line is the bound established in \cite{levi2010provably} and \cite{benjaafar2023pricing} under less generality. The former studies admission control policies, while the latter uses the inventory-based policy in the 1-class system. 
% \cor{It is also the guarantee in section \ref{sec:fluid-bound} that uses a fluid-based static policy.}
The blue line is our bound for regular distributions and multiple classes. The green line is our bound for MHR distributions and one class.
% Combined, they result in our 0.904 guarantee for MHR distributions and one class. 
} 
\end{figure}

We summarize our contributions regarding the power of static pricing below and in Table~\ref{table:main}. 
\begin{enumerate}
    \item We first consider the general case where there are multiple classes of customers, differing in market size, service time distribution, and regular valuation distribution. We prove that offering a single static price for each class can guarantee at least $1 - \frac{(C-1)^C/C!}{ \sum_{i=0}^C (C-1)^i/i! } $ of the revenue from the optimal fully dynamic policy. This result is universal and holds for \textit{any} instance of the problem. %In this setting, the optimal dynamic policy requires continuous real-time price adjustments, while the static policy requires only one price each class. 
    %We also provide an instance showing that even with exponential service time distributions and uniform valuation distributions, the best static policy obtains at most 78.99\% of the optimal dynamic policy.
    Moreover, our guarantee has its minimum when $C=3$, which results in the $\frac{15}{19} \approx 0.789$ lower bound as provided in \cite{besbes2022static} (and we provide an instance showing the guarantee is within 0.01\% of the best possible). However, \cite{besbes2022static} considers only one class of customer with exponential service time, while our bound applies to general service times and multiple classes which each require new proof ideas. Moreover, our new bound has a much improved dependency on $C$, which is critical for our next result. % In addition, our proof is much simpler and shorter.
    
    \item For a single-class system with MHR valuation distributions, \cor{we introduce a new flow-weighted geometric mean static policy that does not appear in previous literature, using the additional structure implied by MHR.} We prove that it achieves at least \cor{97.55\%} of the revenue of the optimal inventory-based policy. This guarantee holds across any $C$, market size, service time distribution, and MHR valuation distribution (see   Figure~\ref{Fig:Guarantee}).% shows the guarantees we can obtain for different values of $C$, the worst of which is \cor{0.9755 when $C =22$}.  %The set of MHR distributions includes a variety of distributions (e.g., uniform, exponential, logistic, and normal) that are commonly adopted in practice, indicating the excellent performance of static pricing policies in reality.
    
    \item We narrow in on the special case of a single-class system with $C =2$, which is the most common setting found in the rotable spare parts application described in \cite{besbes2020pricing}. We provide a guarantee of \cor{99.06\%} for MHR valuation distributions. For uniform valuations, we improve the 95.5\% guarantee provided in \cite{besbes2022static} to 99.53\%.
    
    % \item We extend the 1-class setting to incorporate multiple classes of customers, differing in market size, service time, and valuation distribution. We show that for regular valuation distributions, offering a single static price for each class also guarantees $1 - \frac{(C-1)^C/C!}{ \sum_{i=0}^C (C-1)^i/i! } > 0.789$ of the optimal policy. In this setting, the optimal policy grows exponentially large in the number of classes while the static policy requires only one price each class. We also provide an instance showing that the best static policy obtains at most 78.99\% of the optimal policy.
    
    \item We prove that the optimization problem corresponding to finding the best static pricing policy \cor{is a concave fractional program, and thus the optimal static policy can be efficiently computed.} Our combined results imply the first polynomial time approximation algorithm for pricing reusable resources with multiple customer classes and general service times distributions. 

    \item  \cor{ We also investigate irregular valuation distributions by applying the standard ironing approach. We show that a randomized static policy retains the same guarantee as under regular valuations. If pricing is restricted to a single deterministic price for each class, static pricing guarantees $50\%$ of the concave-envelope benchmark, and this guarantee is tight. }

     \item \cor{  We consider a general multi-resource network in which each customer class may require distinct resources. We show that static pricing still admits a universal guarantee in the special unit-bundle networks, where each class requires at most one unit of each resource.}
    % Finally,  we explore the robustness of our results in a general multi-resource network where each customer class may require multiple units or a bundle of distinct resources. We show that a fluid-based static policy admits a positive universal guarantee in the special unit-bundle networks (each class requires at most one unit of each resource), but may have arbitrarily poor performance beyond this special case. 
\end{enumerate}

Although we focus on maximizing revenue in this paper, our analysis for regular distributions applies to any objective that is concave in demand, which includes social welfare (\citealp{banerjee2022pricing}), throughput, and service level. \cor{The intuition for why static pricing performs extremely well is that it can match the optimal dynamic policy’s average arrival rate whenever units are available, so its revenue loss is driven entirely by its higher blocking probability. This advantage of dynamic pricing is inherently limited because avoiding future blocking requires sacrificing current revenue, while occupied units eventually return to the system (see Section \ref{sec: intuition} for further intuition).} The general proof technique we use is to choose a static policy that is explicitly constructed from the optimal policy. From the regularity or MHR assumption, this allows us to express the worst-case instances in terms of the steady-state probabilities of the optimal policy. For the multi-class system, we apply Little's law to introduce an elegant change of variables. We also identify the optimal static policy to obtain a constant-factor approximation algorithm for the multi-class setting. \cor{For the 1-class system with MHR valuations, we leverage the balance equations to formulate a two-layer optimization problem whose inner problem is convex.}
% For the 1-class system, our proofs leverage various properties of the optimal policy to formulate an optimization problem that generates a worst-case scenario. Since the number of variables depends on $C$, we also provide a novel reduction to a worst-case scenario with only a constant number of variables by leveraging properties of the MHR distribution. 

%Figure~\ref{Fig:Guarantee} also plots our guarantees and compares them to that of \cite{besbes2022static}. %Overall, this paper provides universal guarantees of static pricing for reusable resources under various settings, demonstrating the power of static pricing. 

\begin{table}[!]
\centering
\begin{tabular}{c|c|c|c|c}
\textbf{System}  & \textbf{Valuation Dist.} & \textbf{Benchmark Policy} & \textbf{Guarantee} & \textbf{Theorem} \\ \hline \hline
Multi-class   & regular  & fully dynamic  & $1 - \frac{(C-1)^C/C!}{\sum_{i=0}^C (C-1)^i/i! } \geq 78.9\%$ & \ref{Thm: Mregular}   \\ \hline
1-class   & MHR    & inventory-based  & \cor{97.55\%}  & \ref{Thm_975}   \\ \hline
1-class, $C= 2$ & MHR   & inventory-based  & \cor{99.06\%}  & \ref{Thm:linear}   \\ \hline
1-class, $C= 2$ & uniform    & inventory-based  & 99.53\%  & \ref{Thm:linear}   \\ \hline
\cor{Multi-class}   & \cor{irregular}    & \cor{fully dynamic }  & \cor{ $1 - \frac{(C-1)^C/C!}{\sum_{i=0}^C (C-1)^i/i! } $ or $ \frac{C}{2C -1}  $} & \ref{thm:irregular}    \\  
\end{tabular}
\caption{Our guarantees under different settings}
\end{table}\label{table:main}

% \begin{table}[!]
% \centering
% \begin{tabular}{c|c|c|c|c}
% \textbf{System}        & \textbf{Valuation Dist.} & \textbf{Service Time Dist.} & \textbf{Guarantee} & \textbf{Theorem} \\ \hline \hline
% Multi-class   & regular                & exponential  & $1 - \frac{(C-1)^C/C!}{\sum_{i=0}^C (C-1)^i/i! } \geq 78.9\%$ & \ref{Thm: Mregular}   \\ \hline
% % 1-class       & regular                & exponential  & $1 - \frac{(C-1)^C/C!}{\sum_{i=0}^C (C-1)^i/i! } \geq 78.9\% $  & \ref{Thm: regular}  \\ \hline
% 1-class       & MHR                    & exponential  & 90.4\%  & \ref{Thm_91}   \\ \hline
% 1-class, $C= 2$ & MHR                & exponential  & 98.0\%  & \ref{Thm 98}   \\ \hline
% 1-class, $C= 2$ & uniform                & exponential  & 99.5\%  & \ref{Thm:linear}   \\ 
% % Multi-class   & regular                & general      & $1 - \frac{C^C/C!}{\sum_{i=0}^C C^i/i! } \geq 60\% $ & \ref{Thm: general service}      \\  
% \end{tabular}
% \caption{Our guarantees under different settings}
% \end{table}\label{table:main}

\subsection{Literature Review}
Broadly speaking, our work is related to the literature on the effectiveness of static pricing policies and revenue management with reusable resources.

The majority of the revenue management literature focuses on perishable resources, where a firm has limited inventory to sell over a finite time horizon (e.g., surveys by \citealp{talluri2004theory} and \citealp{den2015dynamic}). One of the main results here is that static pricing is asymptotically optimal and achieves a universal $1-1/e$ guarantee (see \citealp{gallego1994optimal}, \citealp{chen2019efficacy}, and \citealp{ma2021dynamic} for results under various assumptions).  In contrast, our paper considers pricing reusable resources over an infinite horizon.

Reusable resources are related to the well-studied Erlang loss system \citep{erlang1917solution,sevast1957ergodic,takacs1969erlang,brumelle1978generalization,burman1981insensitivity,kaufman1981blocking,reiman1991critically,davis1995sensitivity}, in which customers arriving to the system are lost whenever no servers are available. For revenue management with reusable resources, there has been a stream of literature considering static and dynamic pricing for both single resource and multi-resource settings. 

\textit{Single resource:} 
\cite{reiman1999pricing} provide structural properties of a revenue maximization problem of a generalization of the classical Erlang loss model to multiple classes of customers, where each class may require multiple servers. \cite{Paschalidis2000} study a service provider managing a single resource with multiple classes of customers. They demonstrate that a static pricing policy derived from a nonlinear program is asymptotically optimal for the case of many small users. \cite{ziya2006optimal} provide conditions that the optimal static price should satisfy for the Erlang loss system. \cite{gans2007pricing} consider dynamic pricing to maximize discounted revenue for two classes of customers and analytically establish that static policies perform optimally when both the offered load and capacity are large and nearly balanced. \cite{shakkottai2008price} look at the performance of simple pricing rules (flat entry fees) and numerically show that the loss of revenue from using simple pricing rules is small (10\%-15\%) for some typical utility functions. \cite{xu2013dynamic} study dynamic pricing for cloud computing service providers and propose several properties of the optimal pricing policy and the optimal revenue. \cite{besbes2022static} prove that a static pricing policy can provide a universal 78.9\% guarantee of the profit, market share, and service level from the optimal dynamic policy under regular valuations, single class, and exponential service times. \cite{ks2022optimal} consider a loss system where customers arrive according to a renewal process and show that static pricing is asymptotically optimal as the system capacity grows large. \cite{park2024pricing} numerically computed the optimal two-step (initially flat, then increasing) pricing policy in a case study of airport customer services, which can be modeled as Erlang loss systems. \cor{\cite{balseiro2026dynamic} study a complementary question, showing that inventory-based policies can converge faster than static pricing to a fluid upper bound and that this rate can be achieved by a simple two-price policy. In contrast, we establish non-asymptotic multiplicative guarantees for static pricing relative to the optimal fully dynamic or inventory-based policy for every fixed capacity and market size.}
% \cite{balseiro2026dynamic} shows that inventory-based policies exhibit faster convergence rates in terms of the initial stock compared to static pricing, which can be achieved by a simple two-price policy.

\textit{Multi-resource:} \cite{paschalidis2002pricing} extend the asymptotic optimality of static pricing in \cite{Paschalidis2000} to the case of multiple resources and incorporate potential demand substitution effects. \cite{lei2020real} consider the problem of pricing reusable resources with deterministic service times and derive heuristic policies that are asymptotically optimal. \cite{doan2020pricing} consider a case with ambiguous demand and service time distributions and efficiently construct fixed-price policies. \cite{owen2018price} and \cite{rusmevichientong2020dynamic} both consider pricing from a discrete set under a general choice model and provide constant-factor guarantees for their proposed policies.

Previous work has also considered admission control policies, which is a particular case of dynamic pricing where a resource is either priced at a nominal price or infinity \citep{iyengar2004exponential}. \cite{levi2010provably} show that static LP-based policies can guarantee at least $1 - \frac{ C^C/C! }{ \sum_{i=0}^{C}  C^{i}/i! }$  of the revenue achieved by the optimal dynamic admission policy. \cite{chen2017revenue} consider a generalization of the model to allow advanced reservations.  Assortment optimization problems of reusable resources have also been considered \citep{owen2018price, rusmevichientong2020dynamic,goyal2025asymptotically,gong2022online, zhang2022online, huo2022online,feng2022near}.

There is also extensive literature on static and dynamic pricing in queues where customers can wait for service, differing from the loss system of reusable resources 
\citep{maglaras2005pricing, maglaras2006revenue, kim2018value, bergquist2026static, di2026pricing}. \cor{Most closely related, \cite{bergquist2026static} study an $M/M/C$ queue with regular valuations and an objective that balances revenue and congestion. They show that a threshold-static price achieves a universal $1/2$ guarantee, relative to the optimal dynamic policy.} All of the above literature on reusable resources/queues assume that the distribution information of the underlying model is known to a service provider a \textit{priori}. Recent work considers online learning (for arrival rates and service rates) with incomplete information (\citealt{chen2024online}, \citealt{jia2024online}, and \citealt{jia2022online2}).

Recently, a sequence of papers consider pricing for a ride-hailing platform that can be modeled as a single type of reusable resource moving over a spatial network. \cite{waserhole2016pricing} show that a static pricing policy based on solving a maximum flow relaxation of the problem can provide a guarantee of $C/(C + n-1)$ when service times are assumed to be zero, where $C$ is the number of units and $n$ is the number of nodes. \cite{banerjee2022pricing} extend these results in several directions including multiple objectives. For non-zero service times, \cite{banerjee2022pricing} show that if the number of units is sufficiently large ($C\geq 100$) and the total flow under the static pricing policy is at most $C - 2\sqrt{C\log(C)}$, then a static policy guarantees at least $\left(1 -2\sqrt{\frac{\log(C)}{C}}\right) \left(\frac{\sqrt{C\log(C)}}{\sqrt{C\log(C)} +n-1} - \frac{3}{\sqrt{C\log(C)}} \right)$ of the optimal dynamic policy. \cite{benjaafar2023pricing} propose a simple alternative approach to bounding the performance of static pricing policies and provide a lower bound of  $1 - \sqrt{\left(\frac{n-1}{2C} +\frac{1}{2(C+n)} +1 \right)^2-1} + \frac{n-1}{2C} +\frac{1}{2(C+n)} $ for static policies. In contrast, our paper provides stronger guarantees for any number of units but does not consider a spatial network.

% \subsection{Outline}
% The remainder of the paper is organized as follows. In section~\ref{Model}, we present the model and prove a fundamental property of the ratio of profit rates. In section~\ref{Sec:regular}, we establish our $15/19$ guarantee for the 1-class system with exponential service time and regular distributions. Moving to MHR distributions, in Section~\ref{sec:MHR}, we improve our guarantee to 90.41\%.  In section~\ref{sec:multi-class}, we extend our results to settings where multiple classes of customers are incorporated, and the service time may follow general distributions. Conclusions and future research directions are in Section~\ref{Sec: conclusion}.

\section{Model and Preliminaries}\label{Model}

We consider a service provider that has $C$ units of a single reusable resource. There are $M$ classes of price-sensitive customers and all these units are shared across the $M$ classes. Customers of class $j$ arrive according to a Poisson process with rate $\Lambda_j >0$. Each class $j$ customer has an i.i.d. valuation for the service drawn from a known distribution whose cumulative distribution and probability density functions are denoted by $F_j$ and $f_j$, respectively. Upon arrival, customers of class $j$ purchase the service if their valuation exceeds the current price $p^j$. If no units are available, customers immediately leave without service. Serving a class $j$ customer corresponds to using one unit of the reusable resource, which has a general service time distribution $G_j$ with mean $1/\mu^j$. Naturally, $G_j$ belongs to the class of non-negative distributions with finite means, denoted by $\mathcal{D}_{service}$. We assume that the service times are independent of one another and of customers valuations. This model directly corresponds to the classical Erlang loss system with general service times. The firm may incur a cost to serve customers, but we can assume this cost is zero without loss of generality.

We denote by $\lambda^j(p^j) := \Lambda_j (1-F_j(p^j))$ the effective arrival rate for class $ j $ customers at price $ p^j $ (when all units are unavailable, implicitly the price is $\infty$ and $\lambda^j(\infty)=0$). \cor{We assume that $F_j$ has interval support $[\underline{v}_j, \bar{v}_j ] \subseteq  \mathbb{R}_+$, where $\bar{v}_j$ may be infinite, and is continuously differentiable with $f_j(p) >0 $ on $(\underline{v}_j, \bar{v}_j)$. Thus, for every $\lambda^j \in (0, \Lambda_j) $, there is a unique price in $ (\underline{v}_j, \bar{v}_j) $ that induces effective arrival rate $\lambda^j$, denoted by $p^j(\lambda^j)$. We also set $p^j(\Lambda_j) = \underline{v}_j  $ and $p^j(0) = \bar{v}_j $. }
% We assume that there is a one-to-one mapping between prices and effective arrival rates. For each class $j$, $\lambda^j(p^j)$ has a unique inverse, denoted by $p^j(\lambda^j)$.  
As is common in the literature, the effective arrival rate is more convenient to work with. Therefore, one can view the effective arrival rates $\lambda^j$ as the decision variables. We shall focus on two standard classes of valuation distributions: the set of regular distributions and the set of monotone hazard rate (MHR) distributions, denoted by $\mathcal{D}_{reg}$ and $\mathcal{D}_{mhr}$, respectively. \cor{A valuation distribution $F_j$ with density $f_j$ is regular if its virtual valuation function $ x-\frac{1-F_j(x)}{f_j(x)} $ is non-decreasing in $x$}. Note that assuming a regular distribution is equivalent to the assumption that the revenue rate function $\lambda^jp^j(\lambda^j)$ is concave in $\lambda^j$.\footnote{When $\bar{v}_j = \infty $, we interpret the revenue rate at $\lambda^j =0$ by 
the limit: $\lim_{\lambda \to 0} \lambda p^j(\lambda) $.} The MHR assumption requires that the hazard rate $f_j(x)/(1-F_j(x))$ is non-decreasing. Moreover, $\mathcal{D}_{mhr}$ is a subset of $\mathcal{D}_{reg}$.

Let $x_j$ denote the number of class $j$ customers in the system. Given the Poisson assumption on arrivals, it is worthwhile to note that the system state can be described as $\bm{x} = (x_1,\ldots,x_M)$ if and only if the service time of each class is exponentially distributed. For general service times, however, the system state requires further tracking the service time consumed so far by every busy unit. Specifically, the state can be characterized by  $(\bm{x}, \bm{y}_1,\ldots,\bm{y}_M)$, where $\bm{x} = (x_1,\ldots,x_M)$ and $\bm{y}_j = (y_{j1},y_{j2},\ldots,y_{jx_j}) \in \mathbb{R}^{x_j}_+  $  with $y_{jk}$ denoting the service time spent so far of the class $j$ customer with label $k$. We clarify that when $x_j = 0$, $y_j$ does not contribute to the state representation and we interpret $\mathbb{R}^0_+$ as the vector space only containing the empty vector. Then, the state space of the system is defined as 
\begin{align*}
 \mathcal{S} := \{ (\bm{x} = (x_1,\ldots,x_M),\bm{y}_1,\ldots,\bm{y}_M): \sum_{j=1}^M x_j \leq C,  x_j \in \mathbb{N}\cup \{0\}, \ \bm{y}_j = (y_{j1} ,\ldots, y_{jx_j} ) \in \mathbb{R}_+^{x_j}, \ 
 \forall j\in [M]  \}.
\end{align*}

In this paper, we consider three classes of pricing policies: fully dynamic pricing, inventory-based pricing, and static pricing. Fully dynamic pricing allows the price to depend on the entire system state. Specifically, when a class $j$ customer arrives to the system in state $\bm{s}=(\bm{x}, \bm{y}_1,\ldots,\bm{y}_M)\in\mathcal{S}$, the service provider offers a price $p^j_{\bm{s}} $. For each price $ p_{\bm{s}}^j $, the corresponding effective arrival rate is $ \lambda_{\bm{s}}^j=\lambda^j(p_{\bm{s}}^j)$. Inventory-based only depends on the number of customers in the system, without accounting for the service times already consumed. Specifically, when a class $j$ customer arrives to the system in state $\bm{s}=(\bm{x}, \bm{y}_1, \ldots,\bm{y}_M)\in\mathcal{S}$, the chosen effective arrival rate is $\lambda^j_{\bm{x}}$ (corresponding to price $p^j(\lambda_{\bm x}^j)$), which is independent of $\bm{y}_1,\ldots,\bm{y}_M.$ Static pricing simply offers a single static price $\lambda^j$ for class $j$ customers at all times, regardless of the system state. We remark that an inventory-based policy still adjusts prices dynamically but is more practical and reasonable in real-world applications, as it avoids the added complexity of monitoring the service times consumed so far. Moreover, when service times are all exponentially distributed, the fully dynamic policy based on $\cal S$ simplifies to the inventory-based policy. This is because the memoryless property of exponential distributions reduces the system state to $\bm{x} = (x_1,\ldots,x_M) $. %When $C=1$, static pricing and dynamic pricing are equivalent so without loss of generality we restrict attention to $C\geq 2$.

Let $ \bm{\lambda} := \{\lambda^j_{\bm s}\}_{\bm s \in \mathcal{S}, j \in [M]}$ represent the arrival rate decisions (i.e., pricing policy) across all states for some arbitrary policy. \cor{We denote by $ \Pi^{\bm \lambda} $ the stationary probability measure on $\mathcal{S}$ under policy $ \bm{\lambda} $. For $\bm{x} \in \{\bm{x} |\sum_{j=1}^M x_j \leq C , x_j \in \mathbb{N} \cup \{0\}\}$, we let $\mathcal{S}_{\bm x}:= \{ (\bm{x},\bm{y}_1, \ldots,\bm{y}_M) \in \mathcal{S}  \} $ denote the corresponding subset of $\mathcal{S}$ and  denote by $ \mathbb{P}_{\bm{x}}(\bm{\lambda}):= \Pi^{\bm \lambda}(\mathcal{S}_{\bm x}) $ the steady-state probability of being in state with $\bm{x}$ customers under policy $ \bm{\lambda} $.} For $i \in [C] \cup \{0\} $, we denote $ \mathbb{P}_i (\bm{\lambda})  := \sum_{ x_1+\cdots+x_M =i}  \mathbb{P}_{\bm{x}} (\bm{\lambda})  $ to be the steady-state probability of $i$ units being occupied under policy $\bm{\lambda}$. 
% We denote by $ \mathbb{P}_{(\bm{x}, \bm{y}_1,\ldots, \bm{y}_M)}(\bm{\lambda}) $ the steady-state probability density of being in state $ (\bm{x}, \bm{y}_1,\ldots, \bm{y}_M) \in \mathcal{S}$ under policy $ \bm{\lambda} $. For $\bm{x} \in \{\bm{x} |\sum_{j=1}^M x_j \leq C , x_j \in \mathbb{N} \cup \{0\}\}$, we denote by $ \mathbb{P}_{\bm{x}}(\bm{\lambda}) $ the steady-state probability of being in state with $\bm{x}$ customers under policy $ \bm{\lambda} $. Note that by definition, $ \mathbb{P}_{\bm{x}}(\bm{\lambda}) :=  \int_{0}^{\infty} \cdots \int_{0}^{\infty} \mathbb{P}_{(\bm{x},\bm{y}_1,\ldots,\bm{y}_M)} (\bm{\lambda}) d\bm{y}_1 \cdots d\bm{y}_M $. For $i \in [C] \cup \{0\} $, we denote $ \mathbb{P}_i (\bm{\lambda})  := \sum_{x_1+\cdots+x_M =i} \int_{0}^{\infty} \cdots \int_{0}^{\infty} \mathbb{P}_{(\bm{x},\bm{y}_1,\ldots,\bm{y}_M)} (\bm{\lambda})  \,d\bm{y}_1 \cdots d\bm{y}_M $ to be the steady-state probability of $i$ units being occupied under policy $\bm{\lambda}$. 
Our metric when selling reusable resources is the long-run average revenue rate (although our analysis for regular valuations applies to any concave objective such as social welfare, throughput, and service level). For any pricing policy that corresponds to price decisions $ \bm \lambda $, the long-run average revenue rate $\cal R$ is given by
\begin{align}\label{eq:revenue-expression}
\mathcal{R}: =  \sum_{i=0}^{C-1} \sum_{x_1+\cdots+x_M =i} \int_{\mathcal S_{\bm x} }\sum_{j=1}^M \lambda^j_{\bm s} p^j(\lambda^j_{\bm{s}}) \  \Pi^{\bm \lambda}(d \bm{s}) .
\end{align}
% \begin{align}\label{eq:revenue-expression}
% \mathcal{R}: =  \sum_{i=0}^{C-1} \sum_{x_1+\cdots+x_M =i} \int_{0}^{\infty} \cdots \int_{0}^{\infty} \sum_{j=1}^M \left(\lambda^j_{(\bm{x},\bm{y}_1,\ldots,\bm{y}_M)} p^j(\lambda^j_{(\bm{x},\bm{y}_1,\ldots,\bm{y}_M)}) \right) \mathbb{P}_{(\bm{x},\bm{y}_1,\ldots,\bm{y}_M)} (\bm{\lambda})  \,d\bm{y}_1 \cdots d\bm{y}_M.
% \end{align}
% We clarify that when $i=0$, the revenue rate only has one term,  $\sum_{j=1}^M \lambda^j_{\bm{0}} p^j(\lambda^j_{\bm{0}}) \mathbb{P}_{\bm{0}}(\bm{\lambda})$, and $\mathbb{P}_{\bm{0}}(\bm{\lambda})$ is a point mass.
We let $\cal R^*$, $\cal R^{inv*}$, and $\cal R^{sta*}$ denote the long-run average revenue rates under the optimal fully dynamic, inventory-based, and static policies, respectively, and let $\bm{\lambda}^*$, $\bm{\lambda}^{inv*}$, $\bm{\lambda}^{sta*}$ be the corresponding optimal policies. In this paper, we focus on universal performance guarantees for static pricing in comparison to the optimal fully dynamic and inventory-based policies across all possible parameters of our model. Formally, we let $\Omega_{reg}^M(C)$ denote the family of multi-class instances with regular valuations,  i.e.,
\begin{align*}
\Omega_{reg}^M(C) := \{(C, \Lambda_1, \ldots,\Lambda_M,F_1,\ldots,F_M,  G_1,\ldots,G_M): \Lambda_j >0,  F_j \in \mathcal{D}_{reg}, G_j \in \mathcal{D}_{service}, \ j\in [M]  \}.
\end{align*}
We let $\Omega_{mhr}^1(C)$ denote the family of single class instances with MHR valuations, i.e., 
 \begin{align*} 
\Omega_{mhr}^1(C) := \{(C, \Lambda, F,G): \Lambda>0, F \in \mathcal{D}_{mhr}, G \in \mathcal{D}_{service} \}. 
\end{align*}
In Section \ref{sec:mul-exp}, we provide a lower bound on the ratio between the best static policy and best fully dynamic policy over all $C \in \mathbb{N}$ and the instances in $\Omega_{reg}^M(C)$, i.e., 
\begin{equation*}
 \inf_{C,M \in \mathbb{N}} \inf_{\Omega^M_{reg}(C)} \frac{\cal R^{sta*}}{\cal R^*}.
\end{equation*}
In Section \ref{sec:MHR}, we provide a lower bound on the ratio between the best static policy and best inventory-based policy over all $C \in \mathbb{N}$ and the instances in  $\Omega_{mhr}^1(C)$, i.e., 
\begin{equation*}
 \inf_{C \in \mathbb{N}}  \inf_{\Omega_{mhr}^1 (C) } \frac{\cal R^{sta*}}{\cal R^{inv*} } .
\end{equation*}
We note that since dynamic pricing, inventory-based pricing, and static pricing are equivalent when $C=1$ (as arrivals only occur when there are no customers in the system), then the ratios are always  1 in this case. In Sections \ref{sec:mul-exp} and \ref{sec:MHR}, we describe static price constructions that are convenient for our analysis, although we can find the best static price efficiently which we describe in Section \ref{sec:opt_sta}.%Next, we focus on the first ratio $\cal R^{sta*}/\cal R^*$ and defer the discussion on the second ratio to Section \ref{sec:MHR}.

\section{Multiple Classes and Regular Valuations }\label{sec:mul-exp}

We start with the general multi-class setting of our model and focus on the set of regular valuation distributions $\mathcal{D}_{reg}$. This is motivated by the fact that in some applications, customers are categorized into different classes based on their service time distributions $G_j \in \mathcal{D}_{service}, $ %average service time $1/\mu^j$,
valuation distribution $F_j \in \mathcal{D}_{reg}  $, and market size $\Lambda_j$.  

Under the multi-class setting, finding an optimal dynamic pricing policy is challenging since there is a price for each class and each state. Recall that a state in $\cal S$ not only tracks the number of customers in each class currently in the system but also includes information regarding how long each unavailable unit has been in use. As a result, the state space is infinite. However, even if we focus on the discrete components of the state space (i.e., $\bm{x}$), the number of possible states is $O(C^M)$, which makes finding, describing, and executing an optimal policy very challenging. 
This provides even more motivation for considering the class of static policies that always offers the same price for customers in the same class. We next describe a specific static policy we use for our analysis,  although we show how to find the best static price efficiently in Section \ref{sec:opt_sta}.

\textbf{Static Policy Construction.}  %\label{sec:ratio_analysis} 
We choose a static pricing policy $\tilde{\bm{\lambda}}$ that is good enough to generate strong guarantees, while having enough structural properties to be analyzed. Under the optimal fully dynamic policy $\bm{\lambda}^*$, recall that $ \lambda_{\bm s}^{j*} $ is the corresponding effective arrival rate of class $ j $ customers in state $ \bm{s} \in \cal S $, and $ \Pi^{\bm \lambda^*} $ is the corresponding stationary probability measure.
% $ \mathbb{P}_{\bm{s}} (\bm{\lambda}^*)  $ is the corresponding steady-state probability density.  
The single price for each class $ j $ is constructed using the optimal fully dynamic policy such that the corresponding arrival rate, $ \tilde{\lambda}^j $, is equal to the expected arrival rate of class $ j $ customers under the optimal fully dynamic policy when at least one unit is available. Recall that $ \mathbb{P}_i (\bm{\lambda}^*)$ denotes the steady-state probability under policy $\bm{\lambda}^*$ when $i$ units are occupied.  Formally, we construct the static arrival rate of class $ j $ customers $\tilde{\lambda}^j$ as
\begin{align}\label{eq:constructed-policy}
 \tilde{\lambda}^j := \frac{ \sum_{i=0}^{C-1} \sum_{x_1+\cdots+x_M =i} \int_{\mathcal{S}_{\bm x}}\lambda^{j*}_{\bm s} \ \Pi^{\bm \lambda^*}(d \bm{s})
}{1 -  \mathbb{P}_C(\bm{\lambda}^*)}, \qquad j=1,\ldots,M.
\end{align}
% \begin{align}\label{eq:constructed-policy}
%  \tilde{\lambda}^j := \frac{ \sum_{i=0}^{C-1} \sum_{x_1+\cdots+x_M =i} \int_{0}^{\infty} \cdots \int_{0}^{\infty} \lambda^{j*}_{(\bm{x},\bm{y}_1,\ldots,\bm{y}_M)} \mathbb{P}_{(\bm{x},\bm{y}_1,\ldots,\bm{y}_M)} (\bm{\lambda}^*)  \,d\bm{y}_1 \cdots d\bm{y}_M 
% }{1 -  \mathbb{P}_C(\bm{\lambda}^*)}, \qquad j=1,\ldots,M.
% \end{align}
Let $\tilde{\bm{\lambda}} :=(\tilde{\lambda}^1, \ldots,\tilde{\lambda}^M)$ denote this static pricing policy and let $ \mathcal{R}^{\tilde{\bm{\lambda}}} $ denote the long-run average revenue rate under the constructed static policy $\tilde{\bm{\lambda}} $. The static policy  $\tilde{\bm{\lambda}} $ is constructed by mimicking the optimal fully dynamic policy in expectation when it is able to sell. \cor{In Section \ref{sec:fluid-bound}, we provide further intuition for this static price construction by contrasting it with the standard fluid policy. }
One may observe that the revenue rate of a policy, \eqref{eq:revenue-expression}, involves the demand function $p^j(\cdot)$ and the stationary probability measure induced by the policy, which makes it difficult to derive a universal guarantee for static pricing. 
% Throughout the paper, we will focus on this constructed static pricing policy $\tilde{\bm{\lambda}}$ and establish a series of guarantees. Let $ \mathbb{P}_i(\tilde{\bm{\lambda}})$ denote the steady-state probabilities under the constructed static policy $\tilde{\bm{\lambda}}$ when $ i $ units are occupied. 
Lemma \ref{lem:prof_serv} proves that we can use $\tilde{\bm{\lambda}}$ to lower bound the ratio of the revenues between the optimal static and fully dynamic policies by the ratio of the service levels of policy $\tilde{\bm{\lambda}}$ and policy $\bm{\lambda}^*$.  Lemma \ref{lem:prof_serv} generalizes an analogous lemma of \cite{besbes2022static}  shown for the single class case with exponential service times.

\begin{lemma}\label{lem:prof_serv}
Fix an instance where the valuation distribution is regular. Then,
\begin{equation*}
 \frac{\cal R^{sta*}}{\cal R^*} \geq  \frac{\mathcal{R}^{\tilde{\bm{\lambda}}}}{\mathcal{R}^*} \geq  \frac{ 1 -\mathbb{P}_C(\tilde{\bm{\lambda}}) }{1- \mathbb{P}_C(\bm{\lambda}^*)} .
\end{equation*}
    
\end{lemma}

 \begin{proof}{Proof.}  
According to the definition of  $\cal R^{\tilde{\bm{\lambda}}}$ and $\cal R^*$, 
 \begin{align*}
    \frac{\mathcal{R}^{\tilde{\bm{\lambda}}}}{\mathcal{R}^*} =& \frac{ \left(\sum_{j=1}^M  \tilde{\lambda}^{j}p^j(\tilde{\lambda}^{j})\right)  \left(  \sum_{i=0}^{C-1} \mathbb{P}_i(\tilde{\bm{\lambda}}) \right)}{   \sum_{i=0}^{C-1} \sum_{x_1+\cdots+x_M =i} \int_{\mathcal{S}_{\bm x}} \sum_{j=1}^M  \lambda^{j*}_{\bm s} p^j(\lambda^{j*}_{\bm s}) \ \Pi^{\bm \lambda^*} (d \bm{s}) }  \\
    = &\frac{\sum_{j=1}^M  \tilde{\lambda}^{j}p^j(\tilde{\lambda}^{j}) \cdot  \sum_{i=0}^{C-1} \mathbb{P}_i(\tilde{\bm{\lambda}}) / (1-\mathbb{P}_C(\bm{\lambda}^*) ) }{  \sum_{i=0}^{C-1} \sum_{x_1+\cdots+x_M =i} \int_{\mathcal{S}_{\bm x}} \sum_{j=1}^M  \lambda^{j*}_{\bm s} p^j(\lambda^{j*}_{\bm s})\frac{\Pi^{\bm \lambda^*} (d \bm{s}) }{1-\mathbb{P}_C(\bm{\lambda}^*)} }    \\
    \geq & \frac{\sum_{j=1}^M  \tilde{\lambda}^{j}p^j(\tilde{\lambda}^{j})}{\sum_{j=1}^M    \tilde{\lambda}^{j} p^j(\tilde{\lambda}^{j})}\cdot \frac{ \sum_{i=0}^{C-1}\mathbb{P}_i(\tilde{\bm{\lambda}}) }{1- \mathbb{P}_C(\bm{\lambda}^*)}  \\
    =&  \frac{ 1 -\mathbb{P}_C(\tilde{\bm{\lambda}}) }{1- \mathbb{P}_C(\bm{\lambda}^*)}  .
\end{align*}
%  \begin{align*}
%     \frac{\mathcal{R}^{\tilde{\bm{\lambda}}}}{\mathcal{R}^*} =& \frac{ \left(\sum_{j=1}^M  \tilde{\lambda}^{j}p^j(\tilde{\lambda}^{j})\right)  \left(  \sum_{i=0}^{C-1}\mathbb{P}_i(\tilde{\bm{\lambda}}) \right)}{   \sum_{i=0}^{C-1} \sum_{x_1+\cdots+x_M =i} \int_{0}^{\infty} \cdots \int_{0}^{\infty} \sum_{j=1}^M  \lambda^{j*}_{(\bm{x},\bm{y}_1,\ldots,\bm{y}_M)} p^j(\lambda^{j*}_{(\bm{x},\bm{y}_1,\ldots,\bm{y}_M)}) \mathbb{P}_{(\bm{x},\bm{y}_1,\ldots,\bm{y}_M)}(\bm{\lambda}^*)   \,d\bm{y}_1 \cdots d\bm{y}_M} \\
%     = &\frac{\sum_{j=1}^M  \tilde{\lambda}^{j}p^j(\tilde{\lambda}^{j}) \cdot  \sum_{i=0}^{C-1}\mathbb{P}_i(\tilde{\bm{\lambda}}) / (1-\mathbb{P}_C(\bm{\lambda}^*) ) }{  \sum_{i=0}^{C-1} \sum_{x_1+\cdots+x_M =i} \int_{0}^{\infty} \cdots \int_{0}^{\infty} \sum_{j=1}^M \lambda^{j*}_{(\bm{x},\bm{y}_1,\ldots,\bm{y}_M)} p^j(\lambda^{j*}_{(\bm{x},\bm{y}_1,\ldots,\bm{y}_M)}) \frac{\mathbb{P}_{(\bm{x},\bm{y}_1,\ldots,\bm{y}_M)}(\bm{\lambda}^*) }{1-\mathbb{P}_C(\bm{\lambda}^*)}  \,d\bm{y}_1 \cdots d\bm{y}_M }    \\
%     \geq & \frac{\sum_{j=1}^M  \tilde{\lambda}^{j}p^j(\tilde{\lambda}^{j})}{\sum_{j=1}^M    \tilde{\lambda}^{j} p^j(\tilde{\lambda}^{j})}\cdot \frac{ \sum_{i=0}^{C-1}\mathbb{P}_i(\tilde{\bm{\lambda}}) }{1- \mathbb{P}_C(\bm{\lambda}^*)}  \\
%   =&  \frac{ 1 -\mathbb{P}_C(\tilde{\bm{\lambda}}) }{1- \mathbb{P}_C(\bm{\lambda}^*)} .
% \end{align*}
The inequality holds since (i) for each class $j$,  $ \lambda^jp^j(\lambda^j) $ is concave in $ \lambda^j $ and (ii) applying Jensen's inequality to a random variable that takes value $ \lambda_{\bm s }^{j*} $ with probability measure $ \Pi^{\bm \lambda^*}(d \bm{s}) /(1-\mathbb{P}_C(\bm{\lambda}^*))  $, for $ \bm s \in \mathcal{S}$ with $x_1 + \cdots+ x_M=i\leq C-1$. 
% $ \lambda_{(\bm{x},\bm{y}_1,\ldots,\bm{y}_M)}^{j*} $ with probability density $ \mathbb{P}_{(\bm{x},\bm{y}_1,\ldots,\bm{y}_M)}(\bm{\lambda}^*)/(1-\mathbb{P}_C(\bm{\lambda}^*))  $, for $ (\bm{x},\bm{y}_1,\ldots,\bm{y}_M)\in \mathcal{S}$ with $x_1 + \cdots+ x_M=i\leq C-1$. 
\hfill \Halmos
 \end{proof}

 From Lemma \ref{lem:prof_serv}, our problem is transformed into minimizing the ratio of service levels between our constructed static policy and the optimal dynamic policy. We shall leverage the well-known insensitivity property of the Erlang loss system to derive explicit expressions for the steady-state probabilities of our static policy $\tilde {\bm{\lambda}}$. A summary of the insensitivity properties used in this paper is provided in Appendix \ref{sec:insensitivity}.  %In Section \ref{sec:mul-exp}, we sho, we lower bound the ratio of service levels . %For completeness, we provide upper bounds for different settings considered in this paper in Appendix~\ref{sec:ub}.

\textbf{Main Result.} % Recall that our constructed static policy in \eqref{eq:constructed-policy} uses the optimal dynamic policy. We remark that this constructed policy is beneficial for the theoretical analysis, but in practice one can simply identify a good static policy using standard search or optimization methods. We defer the discussion on finding the optimal static policy to Section~\ref{sec:opt_sta}. 
We present our first main result in Theorem \ref{Thm: Mregular}.

\begin{theorem}	\label{Thm: Mregular} 
 Consider the multi-class system with regular valuation distributions. Let  $ G(C):= 1 - \frac{\frac{1}{C!} (C-1)^C }{ \sum_{i=0}^{C} \frac{1!}{i!} (C-1)^{i} }  $. For our constructed static policy $\tilde{\bm{\lambda}}$,
\begin{align*}
0.78951 \geq \inf_{C, M \in \mathbb{N} } \inf\limits_{\Omega^{M}_{reg}(C)}  \frac{\mathcal{R}^{sta*} }{\mathcal{R}^{*}}	\geq  \inf_{C, M \in \mathbb{N} }  \inf\limits_{\Omega^{M}_{reg}(C)}  \frac{\mathcal{R}^{\tilde{ \bm \lambda}}}{\mathcal{R}^{*}}  = \inf_{C \in \mathbb{N} }  G(C) \geq \frac{15}{19} > 0.78947. 
\end{align*}
\end{theorem}

Theorem \ref{Thm: Mregular} establishes that for regular distributions, static pricing can guarantee at least
 $ G(C) >0.789 $ of the revenue achieved by the optimal fully dynamic policy. It is worthwhile to note that our result is non-asymptotic, makes no assumption on demand rates and service times distributions, and holds for any number of classes $M$. Moreover, our policy can be described with only $M$ prices, rather than the infinite prices that the fully dynamic policy requires.  Finally, our $ G(C) $ guarantee serves as an important step in deriving our \cor{0.9755} guarantee in Section \ref{sec:MHR}.

The key proof idea is that one can leverage Little's law to establish a relationship between the steady-state probabilities of the number of occupied units under the optimal dynamic policy and those under the constructed static policy defined in \eqref{eq:constructed-policy}. We also apply the insensitivity result of the Erlang loss system to develop explicit expressions of the steady-state probability under the constructed static policy. Then, we use a change of variables to provide a lower bound on the ratio of service levels.

In Table \ref{table:upper_example} in Appendix \ref{sec:ub}, we present an example ($C=3, M=1$) in which the optimal static policy $\bm{\lambda}^{sta*}$ achieves about 0.78951 of the optimal dynamic policy. This value is remarkably close to $15/19>0.78947$, showing our guarantee is almost tight. 
% In Appendix \ref{sec:ub}, we present Example \ref{ex:multi} (where $C =3$ and $ M=2$), which shows that our static policy $\tilde{\bm{\lambda}}$ and the optimal static policy  $\bm{\lambda}^{sta*}$ achieve about 0.7899 of the optimal dynamic policy. This is true even under exponential service time distributions and uniform valuation distributions. This value is remarkably close to $15/19>0.7894$, showing our guarantee is almost tight. 
 
\textbf{Comparison to Prior Results.} Our bound achieves its minimum when $C=3$, recovering the 15/19 bound from \cite{besbes2022static}, which exclusively studies the 1-class setting of our model with exponential service times. Since our guarantee holds for any number of classes, for $C\geq 4$, Theorem \ref{Thm: Mregular} improves the $ 1 - \frac{(C-1)^4/C!}{\sum_{i=C-4}^C  (C-1)^{4+i-C}/i!} $ guarantee provided in \cite{besbes2022static} to $  G(C)= 1 - \frac{ (C-1)^C/C! }{ \sum_{i=0}^{C}  (C-1)^{i}/i! } $. Note that the previous bound increases slowly and is strictly upper bounded by $ 0.8 $. In contrast, our bound converges to $ 1 $ as $ C $ goes to $ \infty $, proving that our static policy is asymptotically optimal. Moreover, we note that the proof of Theorem \ref{Thm: Mregular} is much simpler and shorter than the corresponding proof in \cite{besbes2022static}, even though we consider general service times distributions and multiple classes. 
\cor{We clarify that the static policy construction in Equation \eqref{eq:constructed-policy} and the reduction to the ratio of service levels in Lemma \ref{lem:prof_serv} follow the high-level framework of \cite{besbes2022static}, while our subsequent analysis extends this framework to multiple classes and general service times and yields the sharper $C$-dependent guarantee in Theorem \ref{Thm: Mregular}. }

%Interestingly, the 0.904 guarantee for MHR distributions does not extend to the multi-class case. One possible explanation for why we cannot obtain a better bound is that the monotonicity property in Lemma~\ref{lemma Structural} cannot be generalized to the multi-class system (see \citealt{Paschalidis2000}).

\subsection{Proof of Theorem~\ref{Thm: Mregular}} \label{sec:proofmulticlass}
By leveraging Little's law on the optimal fully dynamic policy $\bm{\lambda}^*$, Lemma \ref{lem:1-sys-connection} relates the steady-state probabilities of $\bm{\lambda}^*$ to the constructed static policy $\tilde{\bm{\lambda}}$ defined in \eqref{eq:constructed-policy}.
% We describe the relationship in Lemma \ref{lem:1-sys-connection} below. 

% Describing and Analyzing this ratio is particularly challenging for two reasons: (i) The blocking probability $\mathbb{P}_C^*$ lacks a unified expression in the multi-class system with state-dependent arrival rates,  even under exponential service times. It is worth noting that the celebrated product-form expression for steady-state probabilities $\mathbb{P}^*_x$ does not apply in this context. In Appendix \ref{sec:no-product-form}, we provide an example with $C=M=2$ to illustrate this phenomenon. (ii) Although the blocking probability under static policy, $\mathbb{P}_C^{\tilde{\lambda}}$ admits a unified expression in terms of $(\tilde{\lambda}^1, \ldots,\tilde{\lambda}^M)$, each $\tilde{\lambda}^j$ depends on all $ \lambda_x^{j*} $. However, the exact dependency relationship is unknown, as specified in \eqref{eq:static-policy}.

\begin{lemma}\label{lem:1-sys-connection}
Consider the steady-state probabilities of the optimal dynamic policy when $i$ units are occupied, $\mathbb{P}_i(\bm{\lambda}^*)$, and the constructed static policy $\tilde{\lambda}^j$. We have that
        \begin{align*}%\label{eq: Little's law}
	\sum_{i=1}^C i \mathbb{P}_i(\bm{\lambda}^*) 
    % = \sum_{j=1}^M \tilde{\lambda}^j \cdot \left( 1 - \mathbb{P}_C^*\right) \cdot \frac{\sum_{j=1}^M (\tilde{\lambda}^j/\mu^j)  }{ \sum_{j=1}^M \tilde{\lambda}^j } 
    = \sum_{j=1}^M \frac{\tilde{\lambda}^j}{\mu^j}  \cdot \left( 1 - \mathbb{P}_C(\bm{\lambda}^*)\right).
	\end{align*}
\end{lemma}

From Lemma~\ref{lem:prof_serv}, the ratio of revenue rates is lower bounded by the corresponding ratio of service levels, i.e., $ \frac{\cal R^{\tilde{\bm{\lambda}}}}{ \cal R^* } \geq \frac{1 - \mathbb{P}_C(\tilde{\bm{\lambda}}) }{ 1 -\mathbb{P}_C(\bm{\lambda}^*)}$. We next analyze this ratio using a change of variables. Let $\alpha^*:= 1 - \mathbb{P}_C(\bm{\lambda}^*) $ denote the service level and $\beta^*: =  \frac{\sum_{i=1}^{C-1} i \mathbb{P}_i(\bm{\lambda}^*)  }{1 - \mathbb{P}_C(\bm{\lambda}^*) }  $  be the expected number of customers in the system when at least one unit is available. By definition, $\alpha^* \in [0,1]$ and $\beta^* \in [0,C-1]$.
Based on the definition of $\alpha^*$ and $\beta^*$, Lemma \ref{lem:1-sys-connection} establishes that
\begin{align}
\sum_{j=1}^M \frac{\tilde{\lambda}^j}{\mu^j}  =\frac{\sum_{i=1}^{C} i \mathbb{P}_i(\bm{\lambda}^*)}{1 -\mathbb{P}_C(\bm{\lambda}^*)}  = \beta^* + \frac{C(1-\alpha^*)}{\alpha^*} . \label{eq:CTS}
\end{align}
%In addition, for our constructed static pricing policy $\tilde{\lambda}^j$, $j =1,\ldots, M$, the existing insensitivity result shows that the blocking probability, $\mathbb{P}_C(\tilde{\bm{\lambda}})$, is independent of the service time distribution beyond its mean (see, e.g., \citealt{kaufman1981blocking}), and can be written as
Therefore, 
\begin{align*}
    1 - \mathbb{P}_C(\tilde{\bm{\lambda}}) =  \frac{ \sum_{i=0}^{C-1}\frac{1}{i!} \left(\sum_{j=1}^M \frac{\tilde{\lambda}^j}{\mu^j}  \right)^i }{ \sum_{i=0}^{C} \frac{1}{i!} \left(\sum_{j=1}^M \frac{\tilde{\lambda}^j}{\mu^j}  \right)^i } = \frac{ \sum_{i=0}^{C-1}\frac{1}{i!} \left(C(1-\alpha^*)/\alpha^* + \beta^* \right)^i }{ \sum_{i=0}^{C} \frac{1}{i!} \left(C(1-\alpha^*)/\alpha^* + \beta^* \right)^i }.
\end{align*}
The first equality above follows from the insensitivity property of the multi-class Erlang loss system with static arrival rates (see, e.g., \citealt{kaufman1981blocking}), which we review in Appendix \ref{sec:insensitivity}. The second equality follows from \eqref{eq:CTS}.
Then, the ratio of service levels is given by
\begin{align*}
	\frac{1 - \mathbb{P}_C(\tilde{\bm{\lambda}})}{1-\mathbb{P}_C(\bm{\lambda}^*)} =  \frac{1}{\alpha^*} \cdot \frac{\sum_{i=0}^{C-1} \frac{1}{i!} ( C(\frac{1}{\alpha^*}-1) + \beta^*)^{i}}{\sum_{i=0}^C \frac{1}{i!} {( C (\frac{1}{\alpha^*}-1) + \beta^*)^{i}}}  
	=: R(\alpha^*,\beta^*).
\end{align*}

We next establish monotonicity properties of $ R(\alpha,\beta) $ in Lemma \ref{lem:0_ineq} and Lemma \ref{lem:mono} below.
\begin{lemma}\label{lem:0_ineq}
 Consider any $ \beta \in [0,C-1]$ and $\alpha_1, \alpha_2 \in[0,1]$ such that $R(\alpha_1,\beta) \leq 1 $ and $R(\alpha_2,\beta) \leq 1 $. If $\alpha_1 \geq \alpha_2 $, then  $R(\alpha_1, \beta) \leq R(\alpha_2,\beta)$.
% there exists $\alpha^*\in [0,1) $ such that $R(\alpha,\beta)$ is increasing in $(0,\alpha^*)$ and decreasing in $(\alpha^*,1)$. Moreover, we have $\lim\limits_{ \alpha \to 0^+} R(\alpha,\beta) = 1 $ and $R(\alpha,\beta) \geq R(1,\beta)$ for all $\alpha \in [0,1] $ and $\beta \in [0,C-1]$.
\end{lemma}
%Furthermore, we observe that $ R(\alpha,\beta) $ is non-increasing in $ \beta $ in Lemma \ref{lem:mono}.
\begin{lemma}\label{lem:mono}
 For all $ \alpha \in [0,1]$ and $\beta \in [0,C-1] $, we have  $
		\frac{\partial R(\alpha,\beta)}{ \partial \beta} \leq 0 $.
\end{lemma}
%Since we are interested in $R(\alpha,\beta) \leq 1 $, Lemmas~\ref{lem:0_ineq} and \ref{lem:mono} together imply that $R(\alpha,\beta)$ are both non-increasing in $\alpha$ and $\beta$. 
Combining Lemma~\ref{lem:0_ineq}, Lemma~\ref{lem:mono}, and the fact that $\alpha^* \leq 1$ and $\beta^* \leq C-1$, we can conclude that
\begin{align}
R(\alpha^*,\beta^*) \geq R(1,\beta^*)  \geq R(1, C-1)
% = \frac{\sum_{i=0}^{C-1} \frac{1}{i!} (\beta^*)^{i}}{\sum_{i=0}^C \frac{1}{i!} (\beta^*)^{i}}
= \frac{\sum_{i=0}^{C-1} \frac{1}{i!} (C-1)^{i}}{\sum_{i=0}^C \frac{1}{i!} (C-1)^{i}} 
% = 1 - \frac{  \frac{1}{C!}(C-1)^{C}}{\sum_{i=0}^C \frac{1}{i!} (C-1)^{i}}
= G(C) \label{eq:C_bound}.
\end{align}
This proves that $ \inf_{\Omega^{M}_{reg}}  \frac{\mathcal{R}^{\tilde{ \bm \lambda}}}{\mathcal{R}^{*}} \geq  G(C) $. Next, we show that
the bound $G(C)$ is tight by presenting an instance with a particular regular valuation distribution where our static price achieves $G(C)$ guarantee as $\mu$ goes to 0. 
\begin{lemma}\label{lem:tight}
Consider $M =1 $ and \cor{any} $a,b, \Lambda>0$. Let $F_1(p) = 1 -\frac{a/\Lambda}{p-b} $. Then, $
\lim_{\mu \to 0}  \frac{ \mathcal{R}^{ \tilde{\bm \lambda}} }{ \mathcal{R}^{*} }  = G(C)$.
\end{lemma}

Observe that $ G(1) = 1, G(2)=\frac{4}{5}, G(3) = \frac{15}{19}$. To prove $ G(C)\geq \frac{15}{19} $ in our theorem, we show in Lemma \ref{lem:C_incr} that $G(C)$ defined in (\ref{eq:C_bound}) is increasing in $ C $ for $ C\geq3 $.
\begin{lemma}\label{lem:C_incr}
	For $ C\geq 3 $, we have $G(C+1)>G(C)$. 
\end{lemma}
This completes the proof of Theorem \ref{Thm: Mregular}. 

\subsection{Comparison with Fluid-Based Static Policy} \label{sec:fluid-bound}
\cor{It is useful to compare our constructed policy $\tilde{\bm\lambda}$ with the static policy obtained from the standard fluid relaxation \citep{levi2010provably,benjaafar2023pricing,balseiro2026dynamic}. This relaxation only requires the capacity constraint to be satisfied in expectation, i.e., the expected number of customers in the system cannot exceed $C$. Moreover, concavity of the objective makes the fluid relaxation a valid revenue upper bound. Let $\bm\lambda^F = (\lambda^{1,F}, \ldots, \lambda^{M,F} )$ denote the fluid policy obtained by solving the fluid relaxation:
\begin{align}\label{eq:fluid}
\max_{\lambda^j\ge 0} \quad \sum_{j=1}^M \lambda^j p^j(\lambda^j) \qquad \text{s.t.} \quad \sum_{j=1}^M \frac{\lambda^j}{\mu^j}\le C .
\end{align}
Then, using concavity and a similar argument in Lemma \ref{lem:prof_serv}, we show that the fluid policy $\bm\lambda^F$ establishes a 0.6 universal guarantee.
}

\cor{
\begin{proposition}\label{prop:fluid-bound}
 Consider the multi-class system with regular valuation distributions. For the static policy $\bm{\lambda}^F$ that solves \eqref{eq:fluid},
	\begin{align*}
	\inf\limits_{\Omega^{M}_{reg}(C)}  \frac{\mathcal{R}^{ \bm{\lambda}^F } }{\mathcal{R}^{*}} 
    % \geq  1 - \mathbb{P}_C(\bm{\lambda}^F)
=  1 - \frac{\frac{1}{C!} C^C }{ \sum_{i=0}^{C} \frac{1}{i!} C^{i} } \geq \frac{3}{5} . 
	\end{align*}
\end{proposition} 
}

\cor{
\begin{proof}{\textbf{Proof of Proposition \ref{prop:fluid-bound}}.}
We let $\hat{\lambda}^j $ denote the unconditional expected effective arrival rate of class $j$ under the optimal fully dynamic policy, i.e., 
$ \hat{\lambda}^j:= \tilde{\lambda}^j \left( 1 -  \mathbb{P}_C(\bm{\lambda}^*)\right) = \sum_{i=0}^{C-1} \sum_{x_1+\cdots+x_M =i}\int_{\mathcal{S}_{\bm x} } \lambda^{j*}_{\bm s} \ \Pi^{\bm{\lambda}^*}(d \bm{s})  $.
% $ \hat{\lambda}^j:= \tilde{\lambda}^j \left( 1 -  \mathbb{P}_C(\bm{\lambda}^*) \right)  = \sum_{i=0}^{C-1} \sum_{x_1+\cdots+x_M =i} \int_{0}^{\infty} \cdots \int_{0}^{\infty} \lambda^{j*}_{(\bm{x}, \bm{y}_1, \ldots, \bm{y}_M)} \mathbb{P}_{(\bm{x}, \bm{y}_1,\ldots,\bm{y}_M)} (\bm{\lambda}^*)  \,d\bm{y}_1 \cdots d\bm{y}_M. $
From Lemma \ref{lem:1-sys-connection}, it 
holds that $\sum_{j=1}^M \hat{\lambda}^j /\mu^j = \sum_{i=1}^{C} i \mathbb{P}_i (\bm{\lambda}^*) \leq C  $, which shows that $(\hat{\lambda}^1, \ldots,\hat{\lambda}^M ) $ is a feasible solution to \eqref{eq:fluid}. Moreover, using the concavity of $\lambda^j p^j(\lambda^j)$ and a similar argument in Lemma~\ref{lem:prof_serv}, we have that 
\begin{align*}
   \mathcal{R}^* \leq \sum_{j=1}^M \hat{\lambda}^j p^j (  \hat{\lambda}^j )  \leq  \sum_{j=1}^M \lambda^{j,F} p^j ( \lambda^{j,F} ) ,
\end{align*}
where the last inequality follows from the optimality of the fluid policy. Hence, we conclude that
\begin{align*}
    \frac{\mathcal{R}^{\bm\lambda^F}}{\mathcal{R}^*} = \frac{ \left(\sum_{j=1}^M \lambda^{j,F} p^j(\lambda^{j,F}) \right) \left( 1 -  \mathbb{P}_C(\bm{\lambda}^F ) \right)  } {\mathcal{R}^*} \ge  1 -  \mathbb{P}_C(\bm{\lambda}^F ) =  \frac{ \sum_{i=0}^{C-1}\frac{1}{i!} \left(\sum_{j=1}^M \frac{\lambda^{j,F} }{\mu^j}  \right)^i }{ \sum_{i=0}^{C} \frac{1}{i!} \left(\sum_{j=1}^M \frac{\lambda^{j,F}}{\mu^j}  \right)^i } \geq \frac{ \sum_{i=0}^{C-1}\frac{1}{i!} C^i }{ \sum_{i=0}^{C} \frac{1}{i!} C^i }.
\end{align*}
The second equality holds by the insensitivity property of the multi-class Erlang loss system with static arrival rates. The last inequality follows from $\sum_{j=1}^M \lambda^{j,F}/\mu^j  \leq C $ and the service level is decreasing in the total offered load. Finally, recall that when $C=1$, static and dynamic are equivalent. For $C\ge 2$, we prove the universal lower bound $3/5$, which is achieved when $C=2$. Indeed, we observe that $\frac{\sum_{i=0}^{C-1}C^i/i!}{\sum_{i=0}^{C-1}C^i/i! + C^{C}/C!} \ge \frac{C^{C-2}/(C-2)! + C^{C-1}/(C-1)! }{C^{C-2}/(C-2)! + C^{C-1}/(C-1)! + C^{C}/C!} = \frac{2C -1 }{3C - 1} \ge \frac{3}{5}$. The tightness of $  1 - \frac{\frac{1}{C!} C^C }{ \sum_{i=0}^{C} \frac{1}{i!} C^{i} }  $ is shown in Example \ref{ex:fluid-tight} in Appendix~\ref{sec: network-proof}. \hfill \Halmos
\end{proof}
}

\cor{
This $ 1 - \frac{ C^C/C! }{ \sum_{i=0}^{C}  C^{i} /i! }  $ guarantee is exactly the orange line in Figure \ref{Fig:Guarantee}, whereas our new guarantee $ 1 - \frac{ (C-1)^C/C! }{ \sum_{i=0}^{C} (C-1)^{i}/i! } $ strictly dominates it. The intuition is that, in a loss system, prices only generate revenue when at least one unit is available, so a better static price should mimic the dynamic policy's behavior in those selling states rather than its average behavior over the entire states. The fluid policy ignores this feature and chooses arrival rates so that the average workload ($\sum_{j=1}^M \lambda^{j,F} / \mu^j $) is at most $C$. Thus, its guarantee is based on the service level at load $C$. However, our constructed policy $\tilde{\bm\lambda}$ is calibrated over the states when units are available. Since every such state has at most $C-1$ occupied units, Theorem \ref{Thm: Mregular} obtains a sharper effective load $C-1$ rather than $C$, leading to a stronger guarantee.
}

\subsection{Limitation of Uniform Static Pricing}\label{sec:uniform-static}

\cor{Our static pricing policy is class-dependent, allowing prices to differ across customer classes. A natural question is whether a single uniform static price across all classes can achieve a similar universal performance guarantee. The following example shows that this is not possible, even relative to the optimal class-dependent static policy. The intuition is that one long-service class can occupy the system for a very long time, and a uniform price cannot exclude that class without also excluding the desirable short-service class. In contrast, class-dependent static pricing can price out the long-service class while continuing to serve the short-service class.}

% \cor{The intuition is that a low-revenue class with very long service times can occupy the resource for a long time and block a more profitable class with short service times. A uniform price cannot reduce demand from the former without also reducing demand from the latter. In contrast, class-dependent static pricing can price out the long-service, low-revenue class while continuing to serve the short-service, high-revenue class.}

\cor{
\begin{example}
 Consider a one-unit system ($C=1$) with two classes. Both classes have uniform valuation distribution on $[0,1]$ and revenue rate function $\lambda^j p^j(\lambda^j) = \lambda^j(1 - \lambda^j)  $ for $j=1,2$. The service times are exponential with rate $\mu^1$ and $\mu^2= 1$. Under a uniform price policy $\lambda$, the long-run average revenue rate is $ \mathcal{R}(\lambda)=\frac{2\lambda(1-\lambda)}{1+\lambda/\mu^1+\lambda}$ for $ \lambda \in [0,1]$. 
We observe that $\mathcal{R}(\lambda)\le \frac{2\lambda(1-\lambda)}{\lambda/\mu^1}\le 2\mu^1$. Thus the best uniform static revenue converges to $0$ as $\mu^1\to0$. In contrast, a class-specific static policy can set $\lambda^1=0$ and $\lambda^2=1/2$, thereby excluding the long-service class and serving only class $2$. This yields revenue $\frac{(1/2)(1-1/2)}{1+1/2}=\frac{1}{6}$. Therefore, the ratio between the best uniform static revenue and the best class-dependent static revenue converges to $0$ as $\mu^1 \to 0$.  \hfill \Halmos
\end{example} }

\section{Single Class and MHR Valuations }\label{sec:MHR}
 
In the previous section, we show that static pricing yields a universal 15/19 guarantee for any demand rates, general service time distributions, number of units, and number of customer classes. This guarantee is benchmarked against the optimal fully dynamic pricing policy that tracks the service time already consumed by each busy unit. Although the optimal dynamic pricing policy maximizes revenue, its practical implementation is highly challenging, as prices must be continuously adjusted in real time. A more practical and feasible alternative for real-world applications is the inventory-based pricing policy, which relies solely on the number of customers in the system, avoiding the additional complexity of monitoring service times consumed so far.
 
Recall that $\cal R^{inv*}$ represents the long-run average revenue rate under the optimal inventory-based policy. 
% Because $\cal R^{*} \geq \cal R^{inv*} $ by definition, it holds that
% \begin{equation*}
%  \inf_{\Omega_{reg}^M} \frac{\cal R^{sta*}}{\cal R^{inv*} } \geq \inf_{\Omega_{reg}^M} \frac{\cal R^{sta*}}{\cal R^*} \geq \frac{15}{19},
% \end{equation*}
% where the last inequality follows from Theorem \ref{Thm: Mregular}. 
 When all service times follow exponential distributions, the optimal dynamic policy is equivalent to the optimal inventory-based policy, i.e., $\cal R^{*} = \cal R^{inv*} $. In Appendix \ref{sec:ub}, we present Example \ref{ex:multi} (where $C =3$ and $ M=2$), which
 shows that the $15/19$ guarantee remains nearly tight for the multi-class system, even when the benchmark is the revenue from the optimal inventory-based policy and under uniform valuation distributions. To sharpen the guarantee and further demonstrate the effectiveness of static pricing, we next focus on the single-class system.

For simplicity, we omit the dependency on the class in the notation since there is only one class. In the 1-class system, an inventory-based pricing policy can be fully characterized by arrival rates $\lambda_0, \ldots, \lambda_{C-1}$ with $\lambda_i$ denoting the effective arrival rate when $i$ units are occupied. For the 1-class Erlang loss system with inventory-based arrival rates, \cite{brumelle1978generalization} shows that it has an insensitivity property, i.e., the steady-state probability is independent of the service-time distribution beyond its mean. Thus, a standard calculation for $\mathbb{P}_i(\bm{\lambda})$ yields that
\begin{align}\label{eq:steady}
\mathbb{P}_0(\bm{\lambda})
=\frac{ 1}{1 + \sum_{k=1}^C \frac{1}{k!} \Pi_{j=1}^k\frac{\lambda_{j-1}}{\mu}}, \qquad
\mathbb{P}_i(\bm{\lambda})
=\frac{ \frac{1}{i!}\Pi_{j=1}^i \frac{\lambda_{j-1}}{\mu}}{1 + \sum_{k=1}^C \frac{1}{k!}\Pi_{j=1}^k\frac{\lambda_{j-1}}{\mu}}, \qquad i=1,\ldots,C.
\end{align}
Then, the revenue expression in \eqref{eq:revenue-expression} can be simplified as $\cal R = \sum_{i=0}^{C-1} \lambda_ip(\lambda_i)\mathbb{P}_i(\bm{\lambda}) $. Let $\bm{\lambda}^{inv*} = (\lambda_0^{inv*}, \ldots, \lambda_{C-1}^{inv*}) $ denote the optimal inventory-based policy and recall that $\mathbb{P}_{i}(\bm{\lambda}^{inv*})$ and $ \mathcal{R}^{inv*}$ are the corresponding steady-state probability and long-run average revenue rate, respectively.

% It is natural to first consider a static policy similar to the one constructed in \eqref{eq:constructed-policy}:
% \begin{align}\label{eq:1-class-constructed}
%     \tilde{\lambda}: = \frac{ \sum_{i=0}^{C-1} \lambda_i^{inv*} \mathbb{P}_i(\bm{\lambda}^{inv*}) }{1  - \mathbb{P}_C (\bm{\lambda}^{inv*}) }.
% \end{align}
% Recall that $\cal{R}^{\tilde{\lambda}}$ represents the revenue corresponding to this policy. First, we observe that the analysis for $G(C)$ bound developed in Theorem \ref{Thm: Mregular} is tight by presenting an instance with a particular regular valuation distribution where our static price achieves $G(C)$ performance as $\mu$ goes to 0.
% \begin{lemma}\label{lem:tight}
% Consider $M=1$ and \cor{any} $a,b>0$. Let $F(p) = 1 -\frac{a/\Lambda}{p-b} $. Then, $
% \lim_{\mu \to 0}  \frac{ \mathcal{R}^{\tilde{\lambda}} }{ \mathcal{R}^{inv*} }  = G(C)$.
% \end{lemma}
% In the worst-case scenario, the optimal dynamic policy will never stock out $(\alpha =1)$ and the number of customers in the system approaches $C-1$ ($\beta = C-1$).

Recall that Lemma \ref{lem:tight} shows that the $G(C)$ guarantee is tight for regular valuations. This motivates us to further consider the class of MHR distributions $ \mathcal{D}_{mhr} $ for valuation distributions. The class of MHR distributions is widely used in the literature and incorporates a broad set of common distributions used in practice such as the uniform, exponential, logistic, and truncated normal distributions. Note that the distribution $ F(p) = 1 - \frac{a/\Lambda}{p- b}  $ that achieves the lower bound $G(C)$ in Lemma~\ref{lem:tight} does not satisfy the MHR condition since $f(p)/(1-F(p))= 1/(p-b)$ is decreasing in $p$. 

\textbf{Static Policy Construction.} 
% We aim to provide a universal lower bound on $\cal R^{sta*}/\cal R^{inv*} $ using a constructed static policy. 
\cor{The MHR result uses a different construction from the averaging policy in \eqref{eq:constructed-policy}. A useful implication of the MHR condition is that $p(e^x)$ is concave in $x$, which suggests averaging the logarithms of the effective arrival rates under the optimal inventory-based policy. Before defining this policy, we first establish that all effective arrival rates under the optimal inventory-based policy are strictly positive.
\begin{lemma}\label{lem:mhr-1-class-positive}
For $M=1$ and $F \in \mathcal{D}_{mhr} $, the optimal policy satisfies $ \lambda_i^{inv*} > 0 $, $i=0,\ldots,C-1$.
\end{lemma}
}

% A useful implication of the MHR condition is that $p(e^x)$ is concave in $x$, which motivates an alternative static policy that directly exploits this additional structure. 
\cor{Formally, we construct $\lambda^g$ as the flow-weighted geometric mean of the effective arrival rates under the optimal inventory-based policy:
\begin{align}\label{eq:lambda-geo}
\lambda^g:=\exp\left(\frac{\sum_{i=0}^{C-1}\mathbb{P}_i (\bm \lambda^{inv*}) \lambda_i^{inv*}\log\lambda_i^{inv*}}{\sum_{i=0}^{C-1}\mathbb{P}_i(\bm \lambda^{inv*})\lambda_i^{inv*}}\right).
% =\prod_{i=0}^{C-1}\omega_i^{\frac{\mathbb{P}_i \omega_i}{\sum_{l=0}^{C-1}\mathbb{P}_l\omega_l}},
\end{align}
We remark that $ \lambda^g $ is a new constructed static policy that does not appear in \cite{besbes2022static}. Let $\mathbb P_i(\lambda^g)$ denote the steady-state probability that $i$ units are occupied under $\lambda^g$, and let $\mathcal R^{\lambda^g}$ denote its long-run average revenue rate.
}

\textbf{Main Result.} \cor{Theorem~\ref{Thm_975} establishes that for MHR distributions, a simple static pricing policy $\lambda^g$ guarantees $ 97.55\% $ revenue rate of an optimal inventory-based pricing policy.} 
% While for the static policy $\tilde{\lambda}$, we show in Theorem~\ref{Thm_91} that the guarantee for MHR distributions is at least 0.9041. 
In comparison to the case of regular distributions, this guarantee is substantially higher. We note that providing simple policies with approximation ratios over 90\% is rare 
\citep{roundy198598, arnosti2016adverse}.
\cor{
\begin{theorem}\label{Thm_975}
    For the 1-class system with MHR valuation distributions, our static policy $\lambda^g$ guarantees at least 97.55\% of the revenue of the optimal inventory-based policy, i.e., 
	\begin{align*}
0.97554 \geq \inf_{C \in \mathbb{N} }  \inf_{\Omega_{mhr}^1(C)} \frac{\mathcal{R}^{sta*} }{\mathcal{R}^{inv*}}  \geq	\inf_{C \in \mathbb{N} } \inf_{\Omega_{mhr}^1(C)} \frac{\mathcal{R}^{\lambda^g} }{\mathcal{R}^{inv*}} \geq  0.97550.
	\end{align*}
\end{theorem}
}

% \begin{theorem}\label{Thm_91}
%     For the 1-class system with MHR valuation distributions, our static policy $\tilde{\lambda}$ guarantees at least 90.41\% of the revenue of the optimal inventory-based policy, i.e., 
% 	\begin{align*}
% 	\inf_{\Omega_{mhr}^1} \frac{\mathcal{R}^{\tilde{\lambda}}}{\mathcal{R}^{inv*}} \geq  0.9041.
% 	\end{align*}
% \end{theorem}

\cor{In the proof of Theorem~\ref{Thm_975}, the concavity of $p(e^x)$ reduces the revenue ratio to the throughput ratio $ \lambda^g\bigl(1-\mathbb P_C(\lambda^g)\bigr) / \sum_{i=0}^{C-1}\mathbb P_i(\bm\lambda^{inv*}) \lambda_i^{inv*}$. The balance equations then allow us to eliminate $\lambda_i^{inv*}$ and formulate the worst-case problem directly in terms of the steady-state probabilities $\mathbb P_i(\bm\lambda^{inv*})$. In Table~\ref{Table:upperbounds} in Appendix~\ref{sec:ub}, we provide a $97.554\%$ upper bound on the performance ratio of the optimal static policy, showing that our $97.55\%$ guarantee is nearly tight. }

\subsection{Proof of Theorem~\ref{Thm_975}}\label{sec:proof of 9775}
\cor{
We first use the MHR condition to lower bound the revenue ratio by the corresponding throughput ratio. All omitted proofs in this section are given in Appendix~\ref{proof:mhr}.
\begin{lemma}\label{lem:geo-policy}
Fix an instance where $M=1$ and the valuation distribution is MHR. Then,
\begin{align*}
 \frac{\mathcal{R}^{ \lambda^g }}{\mathcal{R}^{inv*} } \geq  \frac{\lambda^g\left(1-\mathbb{P}_C(\lambda^g)\right)}
{\sum_{i=0}^{C-1}\mathbb{P}_i  (\bm \lambda^{inv*}) \lambda_i^{inv*}}.
% \geq  \frac{1 -\mathbb{P}_C(\tilde{\lambda}) }{1- \mathbb{P}_C(\bm \lambda^{inv*} )} .
\end{align*}
\end{lemma}
}

\begin{proof}{Proof.}  
\cor{
We first show that the MHR condition implies that $ p(e^x)$ is concave in $x$. To see this, let $h(p):=f(p)/(1-F(p))$ denote the hazard rate. Since $\lambda=\Lambda(1-F(p(\lambda)))$, we have that $p'(\lambda)=-\frac{1}{\Lambda f(p(\lambda))}.$ Therefore, the first derivative is computed as $ \frac{d}{dx}p(e^x)=e^xp'(e^x)=-\frac{1-F(p(e^x))}{f(p(e^x))}=-\frac{1}{h(p(e^x))} $. Because $h(p)$ is non-decreasing in $p$ and $ p(e^x) $ is non-increasing in $x$, $-1/h(p(e^x))$ is non-increasing in $x$, which proves the concavity. 
% and $\frac{d^2}{dx^2}p(e^x)=-\frac{h'(p(e^x))}{h^3(p(e^x))}\leq 0$, 
% where the inequality follows from the MHR condition that $h(p)$ is non-decreasing.
According to the definitions of $ \mathcal{R}^{ \lambda^g }  $ and $\mathcal{R}^{inv*} $, 
\begin{align*}
    \frac{\mathcal{R}^{ \lambda^g }}{\mathcal{R}^{inv*}}  = &\frac{ \lambda^g p(\lambda^g) (1 - \mathbb{P}_C (\lambda^g) )  }{ \sum_{i = 0}^{C-1} \lambda_i^{inv^*} p(\lambda_i^{inv^*}) \mathbb{P}_i (\bm \lambda^{inv*})  } \\
    =& \frac{ \lambda^g p(\lambda^g) (1 - \mathbb{P}_C (\lambda^g) )  }{ \sum_{i = 0}^{C-1} \frac{\lambda_i^{inv^*}  \mathbb{P}_i (\bm \lambda^{inv*})}{\sum_{j=0}^{C-1}\mathbb{P}_j (\bm \lambda^{inv*})\lambda_j^{inv*}} p(\lambda_i^{inv^*})  } \cdot \frac{1}{\sum_{j=0}^{C-1}\mathbb{P}_j (\bm \lambda^{inv*})\lambda_j^{inv*}}  \\
    \geq&   \frac{ \lambda^g p(\lambda^g) (1 - \mathbb{P}_C (\lambda^g) )  }{  p(\lambda^g)  \sum_{i = 0}^{C-1} \lambda_i^{inv^*} \mathbb{P}_i (\bm \lambda^{inv*})  } =  \frac{ \lambda^g  (1 - \mathbb{P}_C (\lambda^g) ) }{ \sum_{i = 0}^{C-1} \lambda_i^{inv^*} \mathbb{P}_i (\bm \lambda^{inv*})  },
\end{align*}
where the inequality follows from applying Jensen's inequality to $p(e^{x})$:
\begin{align*}
 \sum_{i=0}^{C-1}  \frac{ \mathbb{P}_i (\bm \lambda^{inv*}) \lambda_i^{inv*} }{\sum_{j=0}^{C-1}\mathbb{P}_j (\bm \lambda^{inv*})\lambda_j^{inv*}} p(e^{\log \lambda_i^{inv*} } ) \leq p \left( \exp \left( \sum_{i=0}^{C-1} \frac{ \mathbb{P}_i (\bm \lambda^{inv*}) \lambda_i^{inv*} }{\sum_{j=0}^{C-1}\mathbb{P}_j (\bm \lambda^{inv*})\lambda_j^{inv*}}  \log \lambda_i^{inv*}  \right)  \right) = p(\lambda^g)  . \hfill \Halmos 
\end{align*}
}

\end{proof}

\cor{
Lemma \ref{lem:geo-policy} reduces the problem to finding a lower bound on the throughput ratio between our constructed static policy $\lambda^g$ and the optimal inventory-based policy. For convenience, we let $\mathbb{P}_i:=\mathbb{P}_i(\bm \lambda^{inv*})$ for $i=0,1,\ldots,C$, $\omega_i:=\lambda_i^{inv*}/\mu$ for $i=0,1,\ldots,C-1$, and $\omega^g = \lambda^g/\mu$. Note that the optimal inventory-based policy satisfies the balance equations $ \mathbb{P}_i \omega_i = (i+1) \mathbb{P}_{i+1} $ for $i=0,\ldots,C-1$, implying that we can replace $\omega_i   $ with $  (i+1)\mathbb{P}_{i+1} / \mathbb{P}_i $. Let $A_C(x): = x \cdot \frac{\sum_{i=0}^{C-1} x^i/i! }{\sum_{i=0}^{C} x^i/i! } $ denote the carried load of an Erlang loss system with offered load $x$. We also define $d: = \sum_{i=1}^C i \mathbb{P}_i $ as the expected number of customers in the system. Note that by Lemma~\ref{lem:mhr-1-class-positive} and \eqref{eq:steady}, $d \in (0,C)$. We observe that
\begin{align}\label{eq:mhr-d-lower}
\frac{\mathcal{R}^{ \lambda^g }}{\mathcal{R}^{inv*}} \geq  \frac{\omega^g\left(1-\mathbb{P}_C(\lambda^g)\right)}
{\sum_{i=0}^{C-1}\mathbb{P}_i \omega_i }  = \frac{A_C( \omega^g )}{d } 
= \frac{A_C \left( \exp \left( \frac{1}{d}\sum_{i=1}^{C} i \mathbb{P}_{i}   \log \frac{ i \mathbb{P}_i }{\mathbb{P}_{i-1} }   \right)  \right)}{d} 
\geq \min_{0 < d < C }  \frac{ A_C ( e^{\mathcal{I}_C (d) /d}  )}{d},
\end{align}
where $ \mathcal{I}_C(d) $ is defined as the optimal value of the following convex optimization problem over $ \mathbb{P}_i $:
\begin{equation}\label{min-convex-d}
\begin{aligned}
\mathcal{I}_C(d): = \ \min_{ \mathbb{P}_0, \ldots, \mathbb{P}_{C} \geq 0 }\quad  \sum_{i=1}^C i  \mathbb{P}_i \log \left(\frac{i \mathbb{P}_i}{\mathbb{P}_{i-1}} \right) 
\quad \textrm{s.t.} \quad 
 \sum_{i=0}^C \mathbb{P}_i = 1, \quad
 \sum_{i=1}^C i \mathbb{P}_i =d .
\end{aligned}
\end{equation}
Here, the first inequality follows from Lemma \ref{lem:geo-policy}. The first equality follows from the definition of $A_C(\cdot)$ and $ \sum_{i=0}^{C-1} \omega_i \mathbb{P}_i  = \sum_{i=1}^C i \mathbb{P}_i := d  $ from balance equations. The second equality follows from $ \eqref{eq:lambda-geo} $ and the balance equations that $  \log \omega^g = \frac{\sum_{i=0}^{C-1}\mathbb{P}_i  \omega_i \log\omega_i}{\sum_{i=0}^{C-1}\mathbb{P}_i \omega_i } = \frac{\sum_{i=0}^{C-1}\mathbb{P}_{i+1} (i+1)  \log\omega_i}{\sum_{i=1}^{C} i \mathbb{P}_i  } = \frac{\sum_{i=1}^{C} i \mathbb{P}_{i}   \log  i \mathbb{P}_i /\mathbb{P}_{i-1}  }{d }  $. The last inequality follows because $(\mathbb{P}_0,\ldots,\mathbb{P}_C)$ induced by $\bm\lambda^{inv*}$ is feasible for \eqref{min-convex-d} and $A_C(\cdot)  $ is increasing (Theorem 1 in \citealp{oner2009monotonicity}). 
}

\cor{
Thus, it remains to lower bound the two-layer optimization problem in \eqref{eq:mhr-d-lower}. The following lemma provides the desired bound for $2\leq C\leq974$.
\begin{lemma}\label{lem:0.9755-(2,974)}
For $ 2 \leq C \leq 974 $, $ \min_{0<d<C}\frac{A_C\left(e^{\mathcal{I}_C(d)/d}\right)}{d}\geq0.975501 $.
\end{lemma}
   }

\cor{   
The proof of Lemma \ref{lem:0.9755-(2,974)} is given in Appendix~\ref{proof:mhr}, where we leverage the dual problem of \eqref{min-convex-d} and a modified brute-force search method (see Algorithm~\ref{alg:dual-certificate} in Appendix~\ref{proof:mhr}) to provide a provable lower bound. For $ 2 \leq C \leq 974 $, we implement our modified brute-force algorithm and the results are summarized in Table~\ref{table:bounds-2-100} and Figure~\ref{Fig:mhr>=100} in Appendix~\ref{app:9755 tables}, with the smallest value $0.975501$ attained at $C=22$. For $C\geq975$, we can rely on the $G(C)$ bound for regular valuation distributions due to the Lemma below.
\begin{lemma}\label{lem:geo>G(C)}
Fix $M=1$ and an MHR valuation distribution. Then, $ \frac{\mathcal R^{\lambda^g}}{\mathcal R^{inv*}} \geq G(C) $.
\end{lemma}
}

\cor{
Combining Lemmas \ref{lem:geo>G(C)} and \ref{lem:C_incr}, we conclude that for $C\geq975$, $ \frac{\mathcal R^{\lambda^g}}{\mathcal R^{inv*}} \geq G(C) \geq G(975) \geq0.9755$.  Together with Lemma \ref{lem:0.9755-(2,974)}, this proves Theorem~\ref{Thm_975}. }

\subsection{Improved results for two units}
In particular, for $C=2$, our approach establishes the $0.9906$ guarantee under MHR valuation distributions. We are also interested in the case where $C=2$ and the valuations are uniformly distributed (linear demand), which is the most common setting observed (over 30\%) in the rotable spare parts application described in \cite{besbes2020pricing}. Surprisingly, we are able to improve our bound to 99.53\% in this case, as described in Theorem \ref{Thm:linear}.
\begin{theorem}[$C=2$]\label{Thm:linear}
For $C=2$, our constructed static pricing policy $\lambda^g$ guarantees $99.06\%$ of the revenue of the optimal inventory-based pricing policy under MHR valuation distributions and $99.53\%$ under uniform valuation distributions.
% Furthermore, the analysis is tight.
\end{theorem}

We remark that the technique for uniform valuation distribution is slightly different. For this proof, we directly analyze the ratio of revenue rates instead of the throughput ratio, and the analysis is tight.

\subsection{Why Does Static Pricing Perform Well?}\label{sec: intuition}
\cor{
We next provide intuition for why static pricing performs well and how its performance depends on the number of units, valuation distributions, and the number of customer classes. }

\cor{
First, we point out that the main advantage of dynamic pricing over static pricing is its ability to adjust price based on the system state. It can induce more demand when the system is relatively empty and suppress demand when the system is close to full, thereby reducing blocking probability. Under regular valuation distributions, however, this is essentially the only advantage. When units are available, although the optimal fully dynamic policy can leverage state information to post infinitely many prices, the concavity of $r^j(\lambda^j)$ ensures that the instantaneous revenue rate does not decrease when these prices are replaced by a static price constructed by averaging the fully dynamic pricing policy. Recall that Lemma~\ref{lem:prof_serv} gives 
\begin{align*}
    \mathcal{R}^* \leq  \sum_{j=1}^M    \tilde{\lambda}^{j} p^j(\tilde{\lambda}^{j}) \cdot \left( 1- \mathbb{P}_C(\bm{\lambda}^*) \right), \quad \mathcal{R}^{\tilde{\bm \lambda}} =\sum_{j=1}^M  \tilde{\lambda}^{j} p^j(\tilde{\lambda}^{j}) \cdot \left( 1 - \mathbb{P}_C(\tilde{\bm{\lambda}})  \right) ,
\end{align*}
Thus, the revenue loss from using the static policy is driven entirely by its higher blocking probability. However, the loss is small because reducing blocking is not free for dynamic pricing.
% To avoid becoming full in the future, the fully dynamic policy must sometimes raise prices even when units are currently available, sacrificing current revenue and potentially leaving units idle. Moreover, because the resources are reusable, an occupied unit eventually returns, and note that future customers come from the same markets. Thus, the value of reserving a unit for future customers is limited here. 
To see this, suppose that only one unit remains available. To protect this last unit, a dynamic policy must raise the price, which sacrifices current revenue and risks leaving the unit idle. Because the resource is reusable, stockouts create only a temporary bottleneck, as one of the occupied units will eventually return. Thus, the value of protecting the last unit is the expected revenue saved during a temporary stockout, not the full value of the unit. Moreover, because future customers are drawn from the same markets, reserving the unit for them provides only limited additional value. }
% To see this, suppose that only one unit remains available. Accepting a customer makes the system full, but the resulting cost comes only from customers who arrive before one of the occupied units returns. Thus, the value of protecting the last unit is the expected revenue saved during a temporary stockout, not the full value of the unit. The fully dynamic policy can protect this unit only by raising the current price, sacrificing current revenue and potentially leaving units idle. Because occupied units eventually return and future customers are drawn from the same markets, the value of reserving a unit for future customers is limited here. }

\cor{
Our proofs of Theorem~\ref{Thm: Mregular} and Lemma \ref{lem:tight} make this intuition precise by identifying the worst-case scenario. The service level comparison is weakest when the optimal dynamic policy avoids blocking ($\mathbb{P}_C(\bm{\lambda}^* ) \approx 0  $) while keeping the system highly utilized, with almost $C-1$ units occupied. Our constructed static policy then matches its average arrival rate but cannot prevent occasional stockouts. Its offered load approaches $C-1$, and its service level approaches $ G(C)=1-\frac{(C-1)^C/C!}{\sum_{i=0}^C (C-1)^i/i!}$. This service level remains high because congestion cannot accumulate in a loss system. In particular, consider a 1-class system with exponential service times. Under an offered load of $C-1$, the system enters the full state from state $C-1$ at rate $(C-1)\mu$, whereas it leaves the full state at rate $C\mu$. Thus, even when the system becomes full, it tends to leave that state quickly, which keeps the service level uniformly high, with $G(C)\geq 15/19$ for every $C$. The insensitivity property extends the guarantee to general service times. Moreover, $G(C)=1-\sqrt{\frac{2}{\pi C}}+o(C^{-1/2})$, so the blocking probability is of order $C^{-1/2}$ and the service level converges to one as $C$ grows. In larger systems, random fluctuations become small relative to the system size, so the benefit of dynamically responding to these fluctuations vanishes. 
% Our constructed static policy then matches its average arrival rate, but it cannot prevent occasional stockouts. Its offered load approaches $C-1$, and its service level approaches $G(C)=1-\frac{(C-1)^C/C!}{\sum_{i=0}^C (C-1)^i/i! } = 1- O(\sqrt{\frac{2}{\pi C }}) + o(C^{-1/2}) \geq 15/19 $. 
% Thus, the blocking probability is of order $C^{-1/2}$, and $G(C)$ converges to one as $C$ grows. This is because, in larger systems, random fluctuations become small relative to the system size, so the benefit of dynamically responding to these fluctuations vanishes. 
}

\cor{In the 1-class system, the MHR condition further limits the advantage of inventory-based pricing. Under regular valuation distributions, the optimal inventory-based policy may induce substantially different effective arrival rates across states. For example, a worst-case construction similar to the one in Lemma~\ref{lem:tight} yields $\lambda_0^{inv*}= \cdots=\lambda_{C-2}^{inv*}=\Lambda$ and $\lambda_{C-1}^{inv*}=0$. The policy therefore admits customers aggressively in most states but completely rejects all customers when only one unit remains available. Under MHR valuations, however, further proportional reductions in the effective arrival rate generate progressively smaller increases in price, making such extreme variation in effective arrival rates across states less valuable.
% Under MHR valuations, however, the upper tail becomes thinner as the price increases, so raising the price quickly reduces the effective arrival rate. Consequently, the optimal effective arrival rates cannot differ as sharply across states, ruling out the extreme behavior in the worst case of regular valuation. 
Our flow-weighted geometric mean static price $\lambda^g$ exploits this additional structure through the concavity of $p(e^x)$.  As a result, a single static effective arrival rate more closely captures the decisions of the optimal inventory-based policy, leading to the stronger 97.55\% guarantee.
}

% \cor{
% In the 1-class system, the MHR condition further limits the advantage of inventory-based pricing. Under regular valuation distributions, the optimal inventory-based policy may induce very different effective arrival rates across states. For example, the worst-case construction in Lemma~\ref{lem:tight} has $\lambda_0^{inv*}= \cdots=\lambda_{C-2}^{inv*}=\Lambda$ and $\lambda_{C-1}^{inv*}=0$. The policy therefore admits customers aggressively in most states but completely rejects all customers when only one unit remains available. However, the MHR condition rules out such extreme behavior. In fact, Lemmas~\ref{lemma Structural} and~\ref{lem:cons} imply that $\lambda_j^{inv*}\leq \lambda_0^{inv*}\leq \frac{C}{C-j}\lambda_j^{inv*},\ j=0,\ldots,C-1$. Thus, if the optimal inventory-based policy induces a high arrival rate when the system is empty ($\lambda_0^{inv*}$), then it must also maintain a non-negligible arrival rate when the system is more occupied ($\lambda_j^{inv*}$). This additional restriction weakens the main advantage of dynamic pricing. As a result, a single static arrival rate more closely represents the decisions of the optimal inventory-based policy across states, leading to a stronger performance guarantee.
% }

\cor{
Finally, we note that the MHR improvement is specific to the 1-class system. As mentioned before, Example \ref{ex:multi} shows that in the multi-class system ($M \geq 2$), even with uniform valuation distributions, the guarantee of static policy can remain close to 15/19. This is because in a multi-class system, dynamic pricing can adjust the relative mix of admitted classes across states.
% dynamic pricing can dynamically reallocate units across heterogeneous customer classes. 
Since classes may differ substantially in their valuation and service time distributions, the dynamic policy may aggressively admit one class while strongly restricting another class in the same state. However, the MHR condition places no restriction on this cross-class heterogeneity and thus cannot improve the 15/19 guarantee.
}

\section{Optimizing Static Policies}\label{sec:opt_sta}

In the previous sections, we examined two constructed static pricing policies and their effectiveness in maximizing revenue. While these pricing policies have shown promising results, 
\cor{they are primarily analytical constructions and are not directly implementable as they require access to the optimal dynamic policy. For implementation, it is therefore important to explore the possibility of finding optimal static prices that can further improve revenue outcomes. }

Let $\bar{\lambda}^j \in \argmax_{\lambda^j \in [0, \Lambda_j] }  \lambda^jp^j(\lambda^j)  $ be a maximizer of the concave function $\lambda^j p^j(\lambda^j)$. Since $\lambda^jp^j(\lambda^j)$ is concave and the service level is decreasing in $\lambda^j$, it suffices to solve the optimal static policy restricted to the box  constraint $[0,\bar{\lambda}^1]\times \cdots \times[0,\bar{\lambda}^M]$, i.e.,
\begin{align}\label{op_sta}
\max_{\lambda^j \in [0,\bar{\lambda}^j ]} \cal R^{sta}\left(\lambda^1,\ldots,\lambda^M \right) : = \left(\sum_{j=1}^M \lambda^jp^j(\lambda^j) \right) \cdot \frac{\sum_{i=0}^{C-1} \frac{1}{i!}(\sum_{j=1}^M \frac{\lambda^j}{\mu^j}  )^i}{\sum_{i=0}^C \frac{1}{i!}(\sum_{j=1}^M \frac{\lambda^j}{\mu^j}  )^i } .
\end{align}
\cor{Again, the insensitivity property implies that this static revenue holds for general service time distributions. We note that computing the optimal static policy requires only the price functions ($p^j(\cdot)$) and the mean service times ($\frac{1}{\mu^j}$), and does not require knowledge of the optimal fully dynamic policy. In practice, we can solve \eqref{op_sta} and then implement the prices corresponding to the optimal static arrival rates. }
% \cor{
% We observe that $\cal R^{sta}\left(\lambda^1,\ldots,\lambda^M \right)$ is not necessarily concave because the service level $\frac{\sum_{i=0}^{C-1} ( \frac{\lambda}{\mu}  )^i/i!}{\sum_{i=0}^C (\frac{\lambda}{\mu}  )^i/i! } $ is first concave then convex w.r.t $\lambda$ in $[0,\infty)$ (see \citealp{harel1990convexity}). Nevertheless, the reciprocal of the service level $\frac{\sum_{i=0}^{C} ( \frac{\lambda}{\mu}  )^i/i!}{\sum_{i=0}^{C-1} (\frac{\lambda}{\mu}  )^i/i! } $ is convex in $\lambda$. 
% Since each revenue rate function $\lambda^j p^j(\lambda^j)$ is concave, \eqref{op_sta} is indeed a concave fractional program. We summarize this result in Theorem~\ref{thm:optimal-static} below. }
\cor{
Theorem \ref{thm:optimal-static} shows that \eqref{op_sta} can be solved efficiently by observing that the reciprocal of the service level $\frac{\sum_{i=0}^{C} ( \frac{\lambda}{\mu}  )^i/i!}{\sum_{i=0}^{C-1} (\frac{\lambda}{\mu}  )^i/i! } $ is convex in $\lambda$. }

\cor{
\begin{theorem}\label{thm:optimal-static}
Consider the multi-class system with regular valuation distributions. The static revenue maximization problem \eqref{op_sta} is a concave fractional program. 
\end{theorem} }

The proof, in Appendix \ref{Proofs_static}, relies on the fact that the carried load $A_C(x)= x \frac{\sum_{i=0}^{C-1} x^i/i!}{\sum_{i=0}^Cx^i/i! }   $  
% $ \frac{\lambda}{\mu} \cdot \frac{\sum_{i=0}^{C-1} ( \frac{\lambda}{\mu}  )^i/i!}{\sum_{i=0}^C (\frac{\lambda}{\mu}  )^i/i! } $ 
is concave in $\lambda$ (\citealp{harel1990convexity}). \cor{Using a standard change of variables technique for the concave fractional program, we can reformulate \eqref{op_sta} as a concave optimization problem, and thus the optimal static pricing can be efficiently computed. }

\subsection{A Heuristic Based on the Fluid Relaxation}

Since we are able to find the optimal policy efficiently due to Theorem \ref{thm:optimal-static}, this allows us to measure the optimality gap of heuristics motivated by previous works. We consider the static pricing heuristic suggested by the following concave problem \eqref{opt_fluid} proposed in \cite{levi2010provably}. 
\begin{equation}\label{opt_fluid}
\begin{aligned}
\max_{\lambda^j \geq 0} \quad &  \sum_{j=1}^M  \lambda^j p^j(\lambda^j) \\
\textrm{s.t.} \quad & \sum_{j=1}^M  \frac{\lambda^j}{\mu^j} \leq \Delta .
\end{aligned}
\end{equation}
We clarify that in \cite{levi2010provably}, $\Delta$ is equal to $C$. Inspired by the methodology provided in \cite{benjaafar2023pricing} for a spatial network of reusable resources, we also vary the value of $\Delta$ between $[0,3C  ]$ to identify the best static policy indicated by this fluid formulation. We conduct a set of numerical experiments to test the performance of this heuristic compared to the optimal static policy, computed by solving the concave fractional program \eqref{op_sta}. 
% computed using BFGS to solve \eqref{op_sta}.

We consider two types of demand functions, i.e., linear: $p^j = b^j - a^j\lambda^j$ and exponential: $p^j = a^j \log \frac{b^j}{a^j\lambda^j}$. Note that in both cases, the maximum demand rate is set to be $\frac{b^j}{a^j} $ corresponding to a price of 0. For each combination of $M $ and $C$, the parameters $a^j$ and $b^j$ are randomly generated uniformly in $[0.1,5]$ and $[0.5,10]$, respectively. The service rate $\mu^j$ is randomly generated uniformly in $[0.02,20]$. We divide the range of $\Delta$, i.e., $[0,3C]$ into 100 points and select the one that yields the highest revenue. We generate 1000 different instances of inputs and compute the revenue rate under the static policy obtained from \eqref{op_sta}, the static policy solved from \eqref{opt_fluid} with $\Delta = C$, and the best static policy with $\Delta \in [0,3C]$. We report the worst and average case performance of the fluid-based static policies compared to the optimal static policy in Table~\ref{table:worst/avg}.

\begin{table}[ht]
\centering
\caption{Worst/Average case revenue ratio}\label{table:worst/avg}
\begin{threeparttable}
\begin{tabular}{|ll|llll|llll|}
\hline
                          &     & \multicolumn{4}{c|}{Linear}                                                                                 & \multicolumn{4}{c|}{Exponential}                                                                            \\ \cline{3-10} 
                          &     & \multicolumn{2}{c|}{$\Delta = C$}                           & \multicolumn{2}{c|}{Best $\Delta \in [0,3C]$} & \multicolumn{2}{c|}{$\Delta=C$}                             & \multicolumn{2}{c|}{Best $\Delta \in [0,3C]$} \\ \hline
\multicolumn{1}{|c|}{$M$} & $C$ & \multicolumn{1}{l|}{Worst}   & \multicolumn{1}{l|}{Average} & \multicolumn{1}{l|}{Worst}       & Average    & \multicolumn{1}{l|}{Worst}   & \multicolumn{1}{l|}{Average} & \multicolumn{1}{l|}{Worst}       & Average    \\ \hline
\multicolumn{1}{|c|}{5}   & 5   & \multicolumn{1}{l|}{73.19\%} & \multicolumn{1}{l|}{97.60\%} & \multicolumn{1}{l|}{99.92\%}     & 99.99\%    & \multicolumn{1}{l|}{74.67\%} & \multicolumn{1}{l|}{97.58\%} & \multicolumn{1}{l|}{99.97\%}     & 99.99\%    \\
\multicolumn{1}{|c|}{}    & 10  & \multicolumn{1}{l|}{79.42\%} & \multicolumn{1}{l|}{99.27\%} & \multicolumn{1}{l|}{99.98\%}     & 99.99\%    & \multicolumn{1}{l|}{79.50\%} & \multicolumn{1}{l|}{98.62\%} & \multicolumn{1}{l|}{99.97\%}     & 99.99\%    \\
\multicolumn{1}{|c|}{}    & 15  & \multicolumn{1}{l|}{83.10\%} & \multicolumn{1}{l|}{99.55\%} & \multicolumn{1}{l|}{99.97\%}     & 99.99\%    & \multicolumn{1}{l|}{83.31\%} & \multicolumn{1}{l|}{98.55\%} & \multicolumn{1}{l|}{99.98\%}     & 99.99\%    \\
\multicolumn{1}{|c|}{}    & 20  & \multicolumn{1}{l|}{85.66\%} & \multicolumn{1}{l|}{99.79\%} & \multicolumn{1}{l|}{99.99\%}     & 99.99\%    & \multicolumn{1}{l|}{86.62\%} & \multicolumn{1}{l|}{98.70\%} & \multicolumn{1}{l|}{99.97\%}     & 99.99\%    \\ \hline
\multicolumn{1}{|l|}{10}  & 5   & \multicolumn{1}{l|}{72.12\%} & \multicolumn{1}{l|}{93.49\%} & \multicolumn{1}{l|}{99.85\%}     & 99.99\%    & \multicolumn{1}{l|}{72.91\%} & \multicolumn{1}{l|}{93.19\%} & \multicolumn{1}{l|}{99.96\%}     & 99.99\%    \\
\multicolumn{1}{|l|}{}    & 10  & \multicolumn{1}{l|}{79.16\%} & \multicolumn{1}{l|}{97.68\%} & \multicolumn{1}{l|}{99.93\%}     & 99.99\%    & \multicolumn{1}{l|}{79.27\%} & \multicolumn{1}{l|}{96.80\%} & \multicolumn{1}{l|}{99.71\%}     & 99.99\%    \\
\multicolumn{1}{|l|}{}    & 15  & \multicolumn{1}{l|}{82.24\%} & \multicolumn{1}{l|}{98.99\%} & \multicolumn{1}{l|}{99.94\%}     & 99.99\%    & \multicolumn{1}{l|}{82.94\%} & \multicolumn{1}{l|}{96.96\%} & \multicolumn{1}{l|}{99.97\%}     & 99.99\%    \\
\multicolumn{1}{|l|}{}    & 20  & \multicolumn{1}{l|}{85.10\%} & \multicolumn{1}{l|}{99.21\%} & \multicolumn{1}{l|}{99.98\%}     & 99.99\%    & \multicolumn{1}{l|}{85.47\%} & \multicolumn{1}{l|}{97.29\%} & \multicolumn{1}{l|}{99.98\%}     & 99.99\%    \\ \hline
\multicolumn{1}{|l|}{15}  & 5   & \multicolumn{1}{l|}{72.67\%} & \multicolumn{1}{l|}{91.05\%} & \multicolumn{1}{l|}{99.92\%}     & 99.99\%    & \multicolumn{1}{l|}{73.24\%} & \multicolumn{1}{l|}{89.05\%} & \multicolumn{1}{l|}{95.69\%}     & 99.99\%    \\
\multicolumn{1}{|l|}{}    & 10  & \multicolumn{1}{l|}{78.83\%} & \multicolumn{1}{l|}{95.40\%} & \multicolumn{1}{l|}{99.92\%}     & 99.99\%    & \multicolumn{1}{l|}{79.53\%} & \multicolumn{1}{l|}{95.63\%} & \multicolumn{1}{l|}{99.81\%}     & 99.99\%    \\
\multicolumn{1}{|l|}{}    & 15  & \multicolumn{1}{l|}{82.34\%} & \multicolumn{1}{l|}{97.84\%} & \multicolumn{1}{l|}{99.96\%}     & 99.99\%    & \multicolumn{1}{l|}{82.87\%} & \multicolumn{1}{l|}{97.13\%} & \multicolumn{1}{l|}{99.89\%}     & 99.99\%    \\
\multicolumn{1}{|l|}{}    & 20  & \multicolumn{1}{l|}{84.81\%} & \multicolumn{1}{l|}{98.58\%} & \multicolumn{1}{l|}{99.97\%}     & 99.99\%    & \multicolumn{1}{l|}{84.65\%} & \multicolumn{1}{l|}{96.43\%} & \multicolumn{1}{l|}{98.97\%}     & 99.99\%    \\ \hline
\multicolumn{1}{|l|}{20}  & 5   & \multicolumn{1}{l|}{75.59\%} & \multicolumn{1}{l|}{90.92\%} & \multicolumn{1}{l|}{99.93\%}     & 99.99\%    & \multicolumn{1}{l|}{74.78\%} & \multicolumn{1}{l|}{85.89\%} & \multicolumn{1}{l|}{94.51\%}     & 99.97\%    \\
\multicolumn{1}{|l|}{}    & 10  & \multicolumn{1}{l|}{79.31\%} & \multicolumn{1}{l|}{93.66\%} & \multicolumn{1}{l|}{99.94\%}     & 99.99\%    & \multicolumn{1}{l|}{78.95\%} & \multicolumn{1}{l|}{93.81\%} & \multicolumn{1}{l|}{99.94\%}     & 99.99\%    \\
\multicolumn{1}{|l|}{}    & 15  & \multicolumn{1}{l|}{82.17\%} & \multicolumn{1}{l|}{96.48\%} & \multicolumn{1}{l|}{99.96\%}     & 99.99\%    & \multicolumn{1}{l|}{82.57\%} & \multicolumn{1}{l|}{96.11\%} & \multicolumn{1}{l|}{98.79\%}     & 99.99\%    \\
\multicolumn{1}{|l|}{}    & 20  & \multicolumn{1}{l|}{84.66\%} & \multicolumn{1}{l|}{97.87\%} & \multicolumn{1}{l|}{99.96\%}     & 99.99\%    & \multicolumn{1}{l|}{85.11\%} & \multicolumn{1}{l|}{96.67\%} & \multicolumn{1}{l|}{99.41\%}     & 99.99\%    \\ \hline
\end{tabular}
\begin{tablenotes}[para,flushleft]
     \textit{Note.}  The revenue rate ratio is compared between static policies derived from \eqref{opt_fluid} and the optimal static policy corresponding to \eqref{op_sta}.
\end{tablenotes}
\end{threeparttable}
\end{table}
 
In the worst case, the fluid static policy derived from \eqref{opt_fluid} by setting $\Delta =C $ only gains 72\% of the revenue collected by the optimal static policy. \cor{For fixed $C$ and $M$, our numerical search suggests that the poorest performance occurs when the fluid policy allocates a large share of the total offered load to a class with a long mean service time but only a small contribution to the total revenue rate. Because the fluid relaxation ignores blocking, it does not capture the congestion imposed by this class in the objective, while the optimal static policy substantially reduces arrival rate from this class.  We also observe that the fluid policy performs better as $C$ grows because random fluctuations become smaller relative to the system size, and tends to incur larger losses when $C$ is small relative to the number of customer classes because the blocking probability is then more sensitive to changes in offered load.} In contrast, applying line search over $\Delta \in [0,3C]$ is very close to the optimal static policy. However, computing the optimal static policy only requires solving one concave optimization problem with $M+1$ variables rather than 100 concave optimization problems with $M$ variables for the fluid-based heuristic with line search.

\section{Multiple Classes, Fully Dynamic Policies, Irregular Valuations}\label{sec: irregular}
\cor{In this section, we extend our analysis to irregular valuation distributions while maintaining the other modeling assumptions in Section \ref{Model}. Recall that regularity is equivalent to the concavity of the revenue rate function $\lambda^j p^j(\lambda^j)$ in the effective arrival rate $\lambda^j$. When this revenue rate function is not concave, the standard ironing approach \citep{myerson1981optimal} replaces it with its concave envelope.
% We maintain the assumption that there is a one-to-one mapping between prices and effective arrival rates.\footnote{Our results in this section can be extended to discrete valuation distributions, where different prices may induce the same effective arrival rate. In such case, we may define $p(\lambda) = F^{-1} \left( 1 - \frac{\lambda}{\Lambda} \right)$, where $F^{-1}(x) = \inf \{ v : x \leq F(v) \} $ is the inverse CDF of the valuation distribution $F$.}
To simplify notation, for each class $j$, let $r^j(\lambda^j):=\lambda^j p^j(\lambda^j)$ denote the revenue rate function, and let $\bar r^j(\cdot)$ denote the concave envelope of $r^j(\cdot)$ on $[0,\Lambda_j]$. 
% The following example illustrates the distinction between $r^j$ and $\bar r^j$.
}

\cor{
For any long-run average revenue rate notation $\mathcal{R}$ defined under the original revenue rate function $r^j$, including $\mathcal{R}^*$, $\mathcal{R}^{sta*}$, and $\mathcal{R}^{\tilde{\bm{\lambda}}}$, we use $\bar{\mathcal{R}}$ to denote the corresponding quantity obtained by replacing $r^j$ with $\bar r^j$. For example, under a fully dynamic policy $\bm{\lambda}$,
\begin{align*}
\bar{\mathcal{R}}^{\bm{\lambda}}:=\sum_{i=0}^{C-1}\sum_{x_1 + \cdots + x_M=i}\int_{\mathcal{S}_{\bm x}} \sum_{j=1}^M \bar r^j\left(\lambda^j_{\bm s}\right) \ \Pi^{\bm{\lambda}} (d \bm{s}).
\end{align*}
% \begin{align*}
% \bar{\mathcal{R}}^{\bm{\lambda}}:=\sum_{i=0}^{C-1}\sum_{x_1+\cdots+x_M=i}\int_0^\infty\cdots\int_0^\infty\sum_{j=1}^M \bar r^j\left(\lambda^j_{(\bm{x},\bm{y}_1,\ldots,\bm{y}_M)}\right)\mathbb{P}_{(\bm{x},\bm{y}_1,\ldots,\bm{y}_M)}(\bm{\lambda})\,d\bm{y}_1\cdots d\bm{y}_M.
% \end{align*}
Under a randomized static policy $(\lambda^1,\ldots,\lambda^M)$, the expression becomes
\begin{align*}
\bar{\mathcal{R}}(\lambda^1,\ldots,\lambda^M):=\left(\sum_{j=1}^M \bar r^j(\lambda^j)\right)\frac{\sum_{i=0}^{C-1}\frac{1}{i!} \left(\sum_{j=1}^M \frac{\lambda^j}{\mu^j}\right)^i}{\sum_{i=0}^{C}\frac{1}{i!} \left(\sum_{j=1}^M \frac{\lambda^j}{\mu^j}\right)^i}.
\end{align*}
}
\cor{
We note that to implement a point $(\lambda^j,\bar r^j(\lambda^j))$ on the envelope, the service provider may need to randomize between two deterministic prices. Indeed, for any effective arrival rate $\lambda^j$, the point $(\lambda^j,\bar r^j(\lambda^j))$ can be represented by a convex combination of at most two points on the original revenue rate function $r^j(\cdot)$. We refer to this as a randomized static policy. The policy remains ``static'' in the sense that the induced effective arrival rate is fixed and does not depend on the system state. Under randomized static policies, the same 15/19-guarantee in Theorem~\ref{Thm: Mregular} continues to hold for irregular valuation distributions.  On the other hand, if pricing must be deterministic and static, meaning that the service provider can only post a single deterministic price for each class, then static pricing guarantees a fraction $C/(2C-1)$ of the concave-envelope benchmark. Let $\Omega_{irreg}^M(C)$ denote the family of multi-class instances with irregular valuations,  i.e., $ \Omega_{irreg}^M := \{(C,\Lambda_1, \ldots, \Lambda_M,F_1,\ldots,F_M,G_1,\ldots,G_M): \Lambda_j > 0, F_j \text{ is irregular}, G_j \in \mathcal{D}_{service}, \ j\in [M]  \}.$ The following theorem formalizes these two guarantees and shows that the deterministic static guarantee is tight. 
}

\cor{
\begin{theorem}\label{thm:irregular}
Consider the multi-class system with irregular valuation distributions. If a randomized static policy is allowed, then
\begin{align*}
\inf_{C, M \in \mathbb{N} }  \inf_{\Omega_{irreg}^M(C)} \frac{\bar{\mathcal{R}}^{sta*}}{\bar{\mathcal{R}}^*} = \inf_{C \in \mathbb{N} } G(C) \ge \frac{15}{19}.
\end{align*}
If pricing must be deterministic and static, then
\begin{align*}
\inf_{C,M \in \mathbb{N} }  \inf_{\Omega_{irreg}^M(C)} \frac{\mathcal{R}^{sta*}}{\bar{\mathcal{R}}^*} = \inf_{C \in \mathbb{N} }  \frac{C}{2C-1}\ge \frac{1}{2}.
\end{align*}
\end{theorem}
}

\cor{We note that the guarantee for deterministic static pricing under irregular valuations, $C/(2C-1)$, decreases with $C$ and converges to 1/2. This is because the main loss does not come from blocking and therefore does not disappear as $C$ grows. Instead, the best revenue may require randomizing between two prices, while deterministic static pricing must choose only one. As a result, a constant loss can remain even in a large system.}

\begin{proof}{\textbf{Proof of Theorem~\ref{thm:irregular}}.}
\cor{
We first prove the guarantee for the randomized static policy. Since $\bar r^j$ is concave for every class $j$, the same Jensen argument as in Lemma~\ref{lem:prof_serv} gives
\begin{align*}
\frac{\bar{\mathcal{R}}^{sta*}}{\bar{\mathcal{R}}^*}\ge \frac{\bar{\mathcal{R}}^{\tilde{\bm{\lambda}}}}{\bar{\mathcal{R}}^*}\ge \frac{1-\mathbb{P}_C(\tilde{\bm{\lambda}})}{1-\mathbb{P}_C(\bm{\lambda}^*)},
\end{align*}
where $\tilde{\bm{\lambda}}$ is constructed as in \eqref{eq:constructed-policy} from the optimal fully dynamic policy under the concave envelope. The remaining steps in the proof of Theorem~\ref{Thm: Mregular} use only Little's law and the insensitivity property of the Erlang loss system under static arrival rates. Therefore, the same argument yields that
\begin{align*}
\frac{\bar{\mathcal{R}}^{sta*}}{\bar{\mathcal{R}}^*}\ge \frac{\bar{\mathcal{R}}^{\tilde{\bm{\lambda}}}}{\bar{\mathcal{R}}^*}\ge 1-\frac{(C-1)^C/C!}{\sum_{i=0}^{C}(C-1)^i/i!} = G(C) \ge \frac{15}{19}.
\end{align*} 
Thus, the randomized static policy recovers the same guarantee for irregular valuation distributions. The tightness follows from an arbitrarily small irregular perturbation of the revenue rate function in Lemma 5 that leaves its concave envelope unchanged.}

\cor{
We next prove the guarantee for deterministic static pricing. Fix any $\lambda^1\in[0,\Lambda_1],\ldots,\lambda^M\in[0,\Lambda_M]$. By the definition of $\mathcal{R}^{sta*}$, it holds that
\begin{align*}
\sum_{j=1}^M r^j(\lambda^j) &\leq
\mathcal{R}^{sta*} \frac{\sum_{i=0}^{C}\frac{1}{i!}\left(\sum_{j=1}^M \frac{\lambda^j}{\mu^j} \right)^i}{\sum_{i=0}^{C-1}\frac{1}{i!}\left(\sum_{j=1}^M \frac{\lambda^j}{\mu^j} \right)^i} 
= \mathcal{R}^{sta*} \left[ 1+ \frac{ \frac{1}{C!}\left(\sum_{j=1}^M  \frac{\lambda^j}{\mu^j} \right)^C }{ \sum_{i=0}^{C-1}\frac{1}{i!}\left(\sum_{j=1}^M \frac{\lambda^j}{\mu^j} \right)^i } \right] 
\leq \mathcal{R}^{sta*} \left( 1+\frac{1}{C} \sum_{j=1}^M\frac{\lambda^j}{\mu^j} \right), 
\end{align*}
where the last inequality follows from $ \sum_{i=0}^{C-1}\frac{1}{i!}(\sum_{j=1}^M \lambda^j/\mu^j )^i \geq \frac{1}{(C-1)!} (\sum_{j=1}^M \lambda^j/\mu^j )^{C-1} $. 
We next extend this inequality from $r^j(\cdot)$ to its concave envelope $\bar{r}^j(\cdot)$. For each class $j$, the point $ (\lambda^j, \bar{r}^j(\lambda^j) ) $ can be represented as a convex combination of at most two points on the original revenue rate function. Combining these representations across classes gives a convex combination of deterministic effective arrival rate vectors. The above inequality holds for every deterministic effective arrival rate vector in this combination. Averaging these inequalities, the left-hand side becomes $\sum_{j=1}^M \bar{r}^j(\lambda^j) $. The right-hand side remains unchanged because it is linear in $(\lambda^1, \ldots,\lambda^M)$. Therefore, we have that
% Because the right-hand side is linear in $(\lambda^1,\ldots,\lambda^M)$ and  $\bar r^j$ is the concave envelope of $r^j(\cdot)$, the same upper bound holds with $r^j$ replaced by $\bar r^j (\cdot)$, i.e.,
\begin{align*}
\sum_{j=1}^M \bar r^j(\lambda^j)\le \mathcal{R}^{sta*}\left(1+\frac{1}{C}\sum_{j=1}^M\frac{\lambda^j}{\mu^j}\right).
\end{align*}
}

\cor{
Now consider any fully dynamic policy $\bm{\lambda}=\{\lambda^j_{\bm s}\}_{\bm s\in\mathcal{S},j\in[M]}$. Applying the above inequality for the expression of $\bar{\mathcal{R}}^{\bm{\lambda}}$ gives
\begin{align*}
\bar{\mathcal{R}}^{\bm{\lambda}} \le & \mathcal{R}^{sta*} \left[1-\mathbb{P}_C(\bm{\lambda}) + \frac{1}{C}\sum_{i=0}^{C-1} \sum_{x_1 + \cdots + x_M=i}\int_{\mathcal{S}_{\bm x}} \sum_{j=1}^M\frac{\lambda^j_{\bm s}}{\mu^j} \ \Pi^{\bm{\lambda}}(d \bm{s}) \right] \\
% \bar{\mathcal{R}}^{\bm{\lambda}} \le & \mathcal{R}^{sta*} \left[1-\mathbb{P}_C(\bm{\lambda}) + \frac{1}{C}\sum_{i=0}^{C-1} \sum_{x_1 + \cdots + x_M=i}\int_0^\infty\cdots\int_0^\infty\sum_{j=1}^M\frac{\lambda^j_{(x,y_1,\ldots,y_M)}}{\mu^j}\mathbb{P}_{(x,y_1,\ldots,y_M)}(\bm{\lambda})dy_1\cdots dy_M\right] \\
= &  \mathcal{R}^{sta*} \left( 1 - \mathbb{P}_C(\bm{\lambda}) + \frac{1}{C} \sum_{i=1}^C i\mathbb{P}_i(\bm{\lambda})  \right) \\
\leq & \mathcal{R}^{sta*}\left(1-\mathbb{P}_C(\bm{\lambda})+\frac{C-1+\mathbb{P}_C(\bm{\lambda})}{C}\right)\le \mathcal{R}^{sta*} \cdot \frac{2C-1}{C}.
\end{align*}
The first equality follows from Little's law that
\begin{align*}
\sum_{i=0}^{C -1} \sum_{x_1+\cdots+x_M=i} \int_{\mathcal{S}_{\bm x}}\sum_{j=1}^M\frac{\lambda^j_{\bm s}}{\mu^j} \ \Pi^{\bm{\lambda}}(d \bm{s}) =\sum_{i=1}^C i\mathbb{P}_i(\bm{\lambda}).
\end{align*}
% \begin{align*}
% \sum_{i=0}^{C -1} \sum_{x_1+\cdots+x_M=i} \int_0^\infty\cdots\int_0^\infty \sum_{j=1}^M\frac{\lambda^j_{(x,y_1,\ldots,y_M)}}{\mu^j} \mathbb{P}_{(x,y_1,\ldots,y_M)}(\bm{\lambda})dy_1\cdots dy_M=\sum_{i=1}^C i\mathbb{P}_i(\bm{\lambda}).
% \end{align*}
The second inequality follows from $\sum_{i=1}^C i\mathbb{P}_i(\bm{\lambda}) \leq (C-1)(1-\mathbb{P}_C(\bm{\lambda}))+C \mathbb{P}_C(\bm{\lambda}) = C-1 + \mathbb{P}_C(\bm{\lambda})$ and the third inequality follows from  $\mathbb{P}_C(\bm{\lambda}) \geq 0$. Taking the supremum over all fully dynamic policies shows that
\begin{align*}
\bar{ \mathcal{R}}^* \leq \mathcal{R}^{sta*} \cdot \frac{2C-1}{C} \Longleftrightarrow \frac{\mathcal{R}^{sta*}}{\bar{\mathcal{R}}^*}\ge \frac{C}{2C-1}.
\end{align*}
The tightness result is deferred to Appendix \ref{Proofs_static}. \hfill \Halmos
}

\end{proof}

% \subsection{A de-randomization guarantee for irregular revenue curves.}

\section{General Multi-Resource Network Setting}\label{sec:network-setting}
\cor{
In this section, we consider a more general multi-resource network setting where customers may require multiple units or bundles of resources. There are $K$ reusable resources in total with resource $k$ having $C_k$ units. Let $\bm C=(C_1,\ldots,C_K)$ denote the capacity vector. For $j \in [M] $, class $j$ requires $a_{kj} \in [C_k] \cup \{0\}  $ units of resource $k$, and we denote $\bm a_j=(a_{1j},\ldots,a_{Kj})$ as the requirement vector of class $j$. The original model in Section \ref{Model} corresponds to $K=1$ and $a_{1j}=1$ for all $j \in [M] $. Let $\bm A=(a_{kj})_{k\in[K], j \in [M] }$ denote the resource-requirement matrix. We still consider regular valuation distributions and general service time distributions for all customer classes.}

\cor{ The main message of this section is twofold. First, in the multi-resource network setting, different classes no longer share a common service level (blocking state). We show that the natural extension of the constructed policy from Section~\ref{sec:mul-exp} does not admit a positive universal performance guarantee. The key issue is that the constructed static policy cannot selectively reject a low-revenue class in states where accepting it would block a more profitable class.
% As a result, the construction from Section~\ref{sec:mul-exp} leads to a ratio of two revenue-weighted sums of class-dependent service levels. We show that this difficulty is fundamental: the constructed static policy does not admit a positive universal performance guarantee in the general multi-resource network setting. 
Second, a fluid-based static policy still admits a universal guarantee in the special unit-bundle setting, where each class requires at most one unit of each resource, i.e., $a_{kj}\in\{0,1\}$. However, this fluid-based approach fails to provide any positive universal guarantee beyond this special case. All omitted proofs in this section are presented in Appendix \ref{sec: network-proof}. }

 % For a system state $\bm s = (\bm x = (x_1, \ldots, x_M) , \bm y_1, \ldots, \bm y_M)$, let $I^j_{\bm s}$ indicate whether class $j$ is feasible in state $\bm s + \bm a_j$, i.e., whether adding one class-$j$ customer keeps the system feasible. Let $\mathcal B^j=\{k:a_{kj}>0\}$ be the set of resources used by class $j$, and let $\kappa=\max_j|\mathcal B^j|$. We continue to write $r^j(\lambda^j)=\lambda^jp^j(\lambda^j)$ for the revenue-rate function. Under regular valuation distributions, $r^j(\lambda^j)$ is concave in $\lambda^j$. We still consider regular valuation distributions and general service time distributions for all classes of customer. 

\subsection{Limitation of the Constructed Static Policy}\label{sec:network-constructed}
\cor{
Recall that $\bm{\lambda}^*$ denotes the optimal fully dynamic policy and $\Pi^{\bm \lambda^*} $ denote its stationary probability measure.
% $\mathbb{P}_{\bm s}(\bm{\lambda}^*)$ denotes its steady-state probability density. 
For $\bm{x} \in \{\bm{x} | \bm{A} \bm{x}  \leq \bm{C} , x_j \in \mathbb{N} \cup \{0\}\}$, $\mathbb{P}_{\bm{x}}(\bm{\lambda}^*) $ still represents the steady-state probability of being in state with $\bm{x}$ customers. Let $ \alpha^{j*} : = \sum_{ \bm{x}: \bm{A} \bm{x} \leq \bm{C} - \bm{a_j}  } \mathbb{P}_{ \bm{x} } (\bm{\lambda}^*)    $ be the service level of class $j$ under the optimal fully dynamic policy. Unlike the original model in Section~\ref{Model}, where all classes share the same service level, the multi-resource network setting naturally induces class-dependent service levels. This heterogeneity turns out to be the main obstacle to extending the previous analysis. }

\cor{
We now consider the natural extension of the static policy construction from Section~\ref{sec:mul-exp}. For each class $j$, we define the static arrival rate $\tilde{\lambda}^j$ as
\begin{align*}
 \tilde{\lambda}^j := \frac{ \sum_{\bm{x}: \bm{A} \bm{x} \leq \bm{C} - \bm{a_j}} \int_{\mathcal{S}_{\bm x}}\lambda^{j*}_{\bm{s} =(\bm{x}, \bm{y}_1,\ldots,\bm{y}_M)} \ \Pi^{\bm{\lambda}^*}(d \bm{s})}{ \alpha^{j*} }, \qquad j=1,\ldots,M.
\end{align*}
% \begin{align*}
%  \tilde{\lambda}^j := \frac{ \sum_{\bm{x}: \bm{A} \bm{x} \leq \bm{C} - \bm{a_j}} \int_{0}^{\infty} \cdots \int_{0}^{\infty} \lambda^{j*}_{(\bm{x},\bm{y}_1,\ldots,\bm{y}_M)} \mathbb{P}_{(\bm{x},\bm{y}_1,\ldots,\bm{y}_M)} (\bm{\lambda}^*)  \,d\bm{y}_1 \cdots d\bm{y}_M 
% }{ \alpha^{j*} }, \qquad j=1,\ldots,M.
% \end{align*}
% Let $\tilde{\bm{\lambda}}=(\tilde{\lambda}^1,\ldots,\tilde{\lambda}^M)$ denote this static policy. 
Unfortunately, Example \ref{ex:constructed-unit-bundle-no-bound} in Appendix \ref{sec: network-proof}
shows that the constructed static policy does not admit a positive universal performance guarantee in the general multi-resource network setting. In this example, there are 3 resources whose capacity vector is $(C_1,C_2,C_3)=(1,2,2)$ and 3 customer classes with resource requirements $\bm a_1=(1,1,1), \  \bm a_2=(0,1,0), \ \bm a_3=(0,0,1)$. Class 1 has a higher revenue rate than classes 2 and 3. The constructed static policy fails here because it cannot reject classes 2 and 3 selectively across states. The dynamic policy rejects class 2 (3) whenever one class 2 (3) customer is already in the system, thereby preserving one unit of resources 2 and 3 for the more profitable class 1. However, the static policy may continue accepting class 2 (3) until resource 2 (3) is fully occupied, at which point class 1 is blocked. This issue does not arise in the original model in Section \ref{Model} when all classes can be accepted whenever the system is not full.
% In this example, the failure occurs because the constructed static policy cannot reject class 1 selectively across states. The dynamic policy rejects class 1 in state $10$ to preserve two units for the more profitable class 2, whereas the static policy may continue accepting class 1 and move the system to state $20$, where class 2 is blocked even though one unit remains available. This issue does not arise when every class requires one unit, since all classes remain available whenever the system is not full.
}

\subsection{Fluid-Based Guarantee}\label{sec:fluid-network}
\cor{ In this section, we explore the possibility of deriving universal lower bounds in the multi-resource network setting using the fluid policy. The fluid relaxation is given by
\begin{align}\label{opt:network-fluid}
\Phi(\bm C):=\max_{\bm\lambda = (\lambda^1,\ldots,\lambda^M) \ge \bm{0} }\left\{\sum_{j=1}^M r^j(\lambda^j):\sum_{j=1}^M a_{kj}\frac{\lambda^j}{\mu^j}\le C_k,\ k\in[K]\right\}.
\end{align}
We denote by $\bm \lambda^F=(\lambda^{1,F}, \ldots,\lambda^{M,F})$ an optimal solution to \eqref{opt:network-fluid}, and refer to it as the fluid policy. To simplify notation, let $\mathcal B^j=\{k:a_{kj}>0\}$ denote the set of resources required by class $j$. The next proposition shows that the fluid policy admits a meaningful universal guarantee only in a special unit-bundle setting. }

\cor{
\begin{theorem}\label{prop: network}
For a general multi-resource network with regular valuation distributions, the following statements hold.
\begin{enumerate}
    % \item[(i)] The fluid relaxation is an upper bound on the optimal fully dynamic revenue: $\mathcal R^*\le \Phi(\bm C)$.
    \item[(i)] Suppose the multi-resource network is unit-bundle, i.e., $a_{kj}\in\{0,1\}$ for every resource $k$ and class $j$. Then, $\mathcal R^{sta*}/\mathcal R^*\ge (1/2)^{\max_j|\mathcal B^j|}$. Moreover, the analysis is tight for the fluid policy. 
%     Let $\bm\lambda^F$ be an optimal solution to the fluid problem, and define $\omega^{j,F}:=\lambda^{j,F}/\mu^j$. Then,
% \begin{align*}
% \frac{\mathcal R^{sta*}}{\mathcal R^*}\ge \frac{\mathcal R^{\bm\lambda^F}}{\mathcal R^*}\ge \sum_{j=1}^M\frac{r^j(\lambda^{j,F})}{\Phi(\bm C)}\prod_{k\in \mathcal B^j}\frac{\sum_{i=0}^{C_k-1}\left(\sum_{l:k\in\mathcal B^l} \frac{\lambda^{l,F}}{\mu^l} \right)^i/i!}{\sum_{i=0}^{C_k} \left(\sum_{l:k\in\mathcal B^l}\frac{\lambda^{l,F}}{\mu^l}\right)^i/i!} \geq  \sum_{j=1}^M \frac{r^j(\lambda^{j,F})}{\Phi(\bm C)} \prod_{k\in \mathcal B^j} \frac{\sum_{i=0}^{C_k-1}C_k^i/i!}{\sum_{i=0}^{C_k}C_k^i/i!} .
% \end{align*}
% In particular, if $C_k\ge2$ for every resource, then $\mathcal R^{sta*}/\mathcal R^*\ge (3/5)^{\max_j|\mathcal B^j|}$. Otherwise, $\mathcal R^{sta*}/\mathcal R^*\ge (1/2)^{\max_j|\mathcal B^j|}$. Moreover, the analysis is tight for the fluid policy. 
 \item[(ii)] The fluid policy $\bm\lambda^F$ does not admit a positive universal performance guarantee in general non-unit-bundle networks.
\end{enumerate}
\end{theorem}
}

\cor{
Note that in Example~\ref{ex:constructed-unit-bundle-no-bound} in Appendix \ref{sec: network-proof}, even in a unit-bundle network, our constructed static policy $\tilde{\lambda}^j$ does not provide a positive universal guarantee. The reason is that the service levels of classes 2 and 3 are very small, leading to large static arrival rates. These large rates allow the two low-revenue classes to occupy the second units of resources 2 and 3, so the resource bundle required by the high-revenue class 1 is rarely available. In contrast, the fluid policy directly limits the total offered load on each resource through the capacity constraints, which provides a positive lower bound on the service level of every class in a unit-bundle network.
}

\cor{
Recall that Proposition \ref{prop:fluid-bound} in Section \ref{sec:fluid-bound} demonstrates that the fluid-based static policy achieves a universal $3/5$ guarantee when there is only one resource ($K=1$). This is because static and dynamic pricing are equivalent when $C=1$ since customers can only be accepted at the empty state and $\frac{\sum_{i=0}^{C-1}C^i/i!}{\sum_{i=0}^{C}C^i/i!} \geq 3/5 $ for $C \geq 2$. In contrast, under the multi-resource setting, one class may be accepted in many states even when all resources have exactly one unit. This results in the $(1/2)^{\max_j|\mathcal B^j|}$ bound by setting $C=1$ in $\frac{\sum_{i=0}^{C-1}C^i/i!}{\sum_{i=0}^{C}C^i/i!} $. Beyond the unit-bundle setting, however, the fluid policy may not provide any positive universal guarantee. The intuition is that the fluid relaxation controls only average resource usage, whereas a class requiring multiple units of the same resource needs a large block of units to be available at the same time. A fluid-based static policy may allocate most of the average capacity to small one-unit classes, and those classes can occupy the resource so that the large class almost never finds enough free units together. }

\cor{
Building on the buffering idea of \cite{lei2020real}, we conjecture that the above issue can be addressed by introducing a buffered fluid-based static policy that leaves sufficient slack in every resource. We leave this as an open question.}

\section{Conclusion}\label{Sec: conclusion}
In this paper, we examine the problem of pricing a reusable resource, with a focus on the effectiveness of static pricing policies. We provide universally strong guarantees for the performance of static pricing under various settings. For regular valuation distributions and general service times, we show that a simple static pricing policy achieves at least 78.9\% of the revenue from the optimal dynamic policy. This result holds for any capacity, number of classes of customers, market size, and service time distribution. 
We also show that the optimal static pricing problem is a concave fractional program, 
% We also show that finding the best static policy reduces to finding the only stationary point of a Lipschitz continuous function, 
implying that we have a polynomial-time approximation algorithm. We sharpen the bound to \cor{97.55\%} for the 1-class system with MHR valuation distributions, benchmarked against the optimal inventory-based pricing policy, providing a very strong guarantee for commonly used distributions in practice. We also prove a 99.5\% guarantee in a special case where there are two units and demand is linear. \cor{Finally, we examine the robustness and limitations of static pricing under irregular valuations and in general multi-resource networks. }
For completeness, we provide upper bounds on these performance ratios in Appendix~\ref{sec:ub}.
%Under the most general setting (multi-class system with general service time), we establish the same 60.0\% guarantees as shown in \cite{levi2010provably} for admission control policy. 

Interesting directions for future research include deriving performance guarantees relative to the fully dynamic benchmark under MHR valuations and determining whether the optimal static pricing policy admits a constant universal guarantee in general multi-resource networks without the unit-bundle condition. One may also consider generalizing our results to the spatial network setting motivated by ride-sharing applications. Another potential direction is to consider the substitution effects in demand between different classes of customers.

% \vspace{-25pt} 
% \openup 2em    

\bibliographystyle{ormsv080} % outcomment this and next line in Case 1
%\bibliography{OpaqueBib} % if more than one, comma separated

%\clearpage
\bibliography{PRRBib} % if more than one, comma separated
%\ECSwitch 
%\ECDisclaimer
%\ECHead{Electronic Companion}

%\ECDisclaimer
%%%%%%%%%%%%%%%%%%%%%%%%%%%%%%%%%%%%%%%%%%%%%%%%%%%%%%%%%%
%%% Main head for the e-companion
% \ECHead{Online Appendix}
\newpage
\begin{APPENDICES}
%\section{Appendix}\label{Appendix}

\section{Existing Insensitivity Results}\label{sec:insensitivity}
An important property of the Erlang loss system is its insensitivity property: the steady-state probabilities of the Erlang loss system are independent of the service time distributions beyond their means. This property allows us to derive explicit expressions for steady-state probabilities, even when considering general service time distributions and multiple classes. In essence, the steady-state probability for any service time distribution reduces to that of the steady-state probability for exponential service distributions, which are simple to express and derive. Below, we summarize the established insensitivity properties utilized in this work:
\begin{enumerate}
    \item \textbf{Single-class Erlang loss system:} Consider a 1-class Erlang loss system with inventory-based arrival rates $\lambda_0, \ldots, \lambda_{C-1}$ and general service times with mean service time $1/\mu$. \cite{brumelle1978generalization} shows the insensitivity property. Thus, a standard calculation for $\mathbb{P}_i$ yields that
\begin{align*}
\mathbb{P}_0
=\frac{ 1}{1 + \sum_{k=1}^C \frac{1}{k!} \Pi_{j=1}^k\frac{\lambda_{j-1}}{\mu}}, \qquad
\mathbb{P}_i
=\frac{ \frac{1}{i!}\Pi_{j=1}^i \frac{\lambda_{j-1}}{\mu}}{1 + \sum_{k=1}^C \frac{1}{k!} \Pi_{j=1}^k\frac{\lambda_{j-1}}{\mu}}, \qquad i=1,\ldots,C.
\end{align*}
These expressions are used in Section \ref{sec:MHR}.
  \item \textbf{Multi-class Erlang loss system with static arrival rates:} Consider  a multi-class Erlang loss system with static arrival rates  $\lambda^j$, $j=1,\ldots,M$ and general service times with mean $1/\mu^j$. \citealt{kaufman1981blocking}  shows that the steady-state probability for $i$ units being occupied, $ \mathbb{P}_i $ is given by
 \begin{align*}%\label{eq:multi-probs}
    \mathbb{P}_i = \frac{ \frac{1}{i!}   (\sum_{j=1}^M \lambda^j/\mu^j)^i }{\sum_{i=0}^C \frac{1}{i!} (\sum_{j=1}^M \lambda^j/\mu^j )^i    }, \qquad i=0,\ldots,C.
 \end{align*}
This expression is used to analyze steady-state probabilities for static pricing policies used in Section \ref{sec:mul-exp}, \ref{sec:opt_sta}, and \ref{sec: irregular}. 
\end{enumerate}

\section{Upper Bounds on Approximation Guarantees} \label{sec:ub}
We provide upper bounds on the best possible guarantees for our constructed policies ($ \tilde{\lambda} $ and $\lambda^g $) and the optimal static policy in the following Table~\ref{Table:upperbounds}. Recall that the 78.94\% guarantee of $\tilde{\lambda}$ in Theorem~\ref{Thm: Mregular} when $M=1$, and the 99.53\% guarantee of $\lambda^g$ in Theorem \ref{Thm:linear} under uniform valuation distributions have a tight analysis, implying that we cannot prove a better guarantee for our static policies. The upper bounds in Table~\ref{Table:upperbounds} for the 1-class system are computed based on the concrete examples presented in Table~\ref{table:upper_example}. For the multi-class system $(M \geq 2)$, we obtain  $78.99\%$ upper bounds from Example \ref{ex:multi} ($C=3$, $M=2$, and linear demands) (the difference between the two upper bounds is less than $0.01\%$).

\begin{table}[ht]
\centering
\small
\begin{tabular}{c|c|c|c|c|c|c}
System    & \makecell{Valuation \\ Dist.} & Benchmark  & Guarantee  & Thm   & \makecell{Upper bound \\ $  \frac{\mathcal R^{\tilde{\lambda}}}{\mathcal R^*} $ }& \makecell{Upper bound \\ $ \frac{\mathcal R^{sta*}}{\mathcal R^*}$ }       \\ \hline \hline
Multi-class   & regular   & fully dynamic  & 15/19  & \ref{Thm: Mregular} & 15/19 (tight)    & 78.951\%              \\  \hline \hline
% System    & \makecell{Valuation \\ Dist.} & Benchmark  & Guarantee  & Thm   & \makecell{Upper bound \\ 
% $ \frac{\mathcal R^{\tilde{\lambda}}}{\mathcal R^{inv*}} $ } & \makecell{Upper bound \\ $ \frac{\mathcal R^{sta*}}{\mathcal R^{inv*}}$  }  \\ \hline \hline
% 1-class    & regular         & inventory-based  &  78.947\% & \ref{Thm: Mregular}  & 78.947\%         & 78.951\%              \\ \hline \hline
System    & \makecell{Valuation \\ Dist.} & Benchmark  & Guarantee  & Thm   & \makecell{Upper bound \\ 
$ \frac{\mathcal R^{\lambda^g} }{\mathcal R^{inv*}} $ } & \makecell{Upper bound \\ $ \frac{\mathcal R^{sta*}}{\mathcal R^{inv*}}$  }  \\ \hline \hline
1-class    & MHR     & inventory-based  & 97.55\%   &  \ref{Thm_975}      & 97.554\%       & 97.554\%             \\ \hline
1-class, $C= 2$ & MHR     & inventory-based  & 99.06\% & \ref{Thm:linear}   & 99.074\%     & 99.074\%   \\ \hline
1-class, $C= 2$ & uniform     & inventory-based  & 99.537\%    & \ref{Thm:linear}     & 99.537\% (tight)      & 99.539\%     \\      
\end{tabular}
\caption{Our guarantees and upper bounds under different settings}\label{Table:upperbounds}
\end{table}

\begin{table}[ht]
\centering
\begin{tabular}{c|c|c|c|c|c|c}
$C$       & $M$ & $\Lambda$ & Policy      & Demand    & Service Time     & Performance Ratio     \\ \hline \hline
3 & 1 & 10        & $\tilde{\lambda}$    & $p(\lambda)= 3 + 1/\lambda$            & exp, $\mu \to 0 $   & 15/19 \\ \hline
3 & 1 & 96        & optimal static    & $p(\lambda)= 30.5 + 1/\lambda$            & exp, $\mu=0.01$   & 78.951\% \\ \hline
22      & 1 & 25     & $\lambda^g$ & $p(\lambda)=\log(25/\lambda)$ & exp, $\mu =0.1115$ & 97.554\%         \\ \hline
22      & 1 & 25      & optimal static    & $p(\lambda)=\log(25/\lambda)$ & exp, $\mu = 0.1115$ & 97.554\% \\ \hline
2      & 1 & 10     & $\lambda^g$ & $p(\lambda)=\log(10/\lambda)$ & exp, $\mu =0.7117$ & 99.074\%         \\ \hline
2      & 1 & 10      & optimal static    & $p(\lambda)=\log(10/\lambda)$ & exp, $\mu = 0.7117$ & 99.074\% \\ \hline
2       & 1 & 6      & $\lambda^g$     & $p(\lambda)=6 -\lambda$    & exp, $\mu =1$   & 99.538\% \\ \hline
2       & 1 & 6       & optimal  static   & $p(\lambda)=6 -\lambda$    & exp, $\mu =1$   & 99.539\% \\ 
% 20      & 1 & 18     & $\tilde{\lambda}$ & $p(\lambda)=\log(18/\lambda)$ & exp, $\mu =0.05$ & 97.38\%         \\ \hline
% 20        & 1 & 25      & optimal static    & $p(\lambda)=\log(25/\lambda)$ & exp, $\mu = 0.14$ & 97.56\% \\ \hline
% 2        & 1 & 10        & $\tilde{\lambda}$  & $p(\lambda)=\log(10/\lambda)$   & exp, $\mu =0.73$  & 99.06\%         \\ \hline
% 2       & 1 & 10        & optimal static & $p(\lambda)=\log(10/\lambda)$   & exp, $\mu = 0.73$ &  99.07\% \\ \hline
% 2       & 1 & 6      & $\tilde{\lambda}$     & $p(\lambda)=5.7 -\lambda$    & exp, $\mu =1.00$   & 99.54\% \\ \hline
% 2       & 1 & 31       & optimal  static   & $p(\lambda)=9.3-0.3\lambda$    & exp, $\mu =5.30$   & 99.54\% \\ 
\end{tabular}
\caption{Examples of computing upper bounds}\label{table:upper_example}
\end{table}

\begin{example} \label{ex:multi}
% We provide an example that shows a 78.99\% guarantee even when valuations follow uniform distributions. 
Consider a system with two classes of customers ($M=2$) and three units of a single type of reusable resource ($C=3$). Both valuation distributions are uniformly distributed. The corresponding demand functions are $p^1(\lambda^1) =180 - 0.05 \lambda^1$ and $p^2(\lambda^2) =11 - 50 \lambda^2$. The service times are exponentially distributed with rates $\mu^1 = 0.001$ and $\mu^2 = 1000$. 
% The state space is given by $ \cal S_E^2 = \{ (0,0),(0,1),(0,2),(0,3),(1,0),(1,1),(1,2),(2,0),(2,1),(3,0)  \}.$
% \begin{align*}
%     \cal S_E^2 = \{ (0,0),(0,1),(0,2),(0,3),(1,0),(1,1),(1,2),(2,0),(2,1),(3,0)  \}.
% \end{align*}
The revenue maximization problem is a continuous time Markov Decision Process (MDP) and can be uniformized to a discrete time MDP. Using the standard relative value iteration method, the optimal pricing policy is computed as follows
\begin{align*}
    \lambda^1 =\begin{pmatrix}
2.68162 & 2.68161 & 2.67769\\
1.89643 & 1.89039 &  \\
0       &         &
\end{pmatrix},  \
\lambda^2=\begin{pmatrix}
0.11 & 0.10999 & 0.10999\\
0.10999 & 0.10999 &  \\
0.10999      &         & 
\end{pmatrix},
\end{align*} 
where the row and column indices $i,j$ represent the state $(i,j)$. The constructed static policy $\tilde{\lambda}$ is given by $\tilde{\lambda}^1 = 0.00199 $ and $\tilde{\lambda}^2 = 0.10999$.
The optimal revenue $\cal R^* = \cal R^{inv*} $ and the revenue under the constructed static policy $\cal R^{\tilde{\lambda}}$ are approximately equal to 0.96455 and 0.76183, respectively. Therefore, the ratio of revenue rates is computed by $  \cal R^{\tilde{\lambda}} / \cal R^*  \approx 0.78982 $. Solving the concave fractional program \eqref{op_sta}, the optimal static policy is $(0.00194,0.10999)$, which gives revenue $\mathcal{R}^{sta*} =0.76189 $. Hence, the revenue ratio $  \cal R^{sta} / \cal R^*  $ is approximately equal to $0.78989 < 0.7899$.
\Halmos
\end{example}

\section{Proofs in Section~\ref{sec:mul-exp}}\label{Proofs_regular}

\begin{proof}{\textbf{Proof of Lemma~\ref{lem:1-sys-connection}}.}
We shall apply Little's Law with respect to the optimal fully dynamic policy for the multi-class system. For each class $j$, the long-run average rate at which class $j$ customers enter service under the optimal fully dynamic policy is
\begin{align*}
\sum_{i=0}^{C-1}\sum_{x_1+\cdots+x_M=i}\int_{\mathcal S_{\bm x}}\lambda_{\bm s}^{j*} \ \Pi^{\bm\lambda^*}(d\bm s)=\tilde\lambda^j \left(1-\mathbb P_C (\bm\lambda^*)\right),
\end{align*}
where the equality follows from the definition of $\tilde\lambda^j$. Since an arriving customer enters service whenever at least one unit is available, the left-hand side is also the long-run throughput of class $j$. Applying Little's law separately to class $j$, the expected number of class $j$ customers in the system equals its throughput multiplied by its mean service time $1/\mu^j$. Therefore,
\begin{align*}
\sum_{\bm x:x_1+\cdots+x_M\leq C}x_j\mathbb P_{\bm x}(\bm\lambda^*)=\frac{\tilde\lambda^j}{\mu^j}\left(1-\mathbb P_C(\bm\lambda^*)\right).
\end{align*}
Summing over all classes yields
\begin{align*}
\sum_{i=1}^C i\mathbb P_i(\bm\lambda^*)&=\sum_{j=1}^M\sum_{\bm x:x_1+\cdots+x_M\leq C}x_j\mathbb P_{\bm x}(\bm\lambda^*)=\sum_{j=1}^M\frac{\tilde\lambda^j}{\mu^j}\left(1-\mathbb P_C(\bm\lambda^*)\right).
\end{align*}
This proves the result. \hfill \Halmos  \\

\end{proof}

\begin{proof}{\textbf{Proof of Lemma~\ref{lem:0_ineq}.}}
Given $\beta$, let $R^\beta(\alpha):= R(\alpha,\beta)$ denote the one variable function in terms of $\alpha$. To prove the monotonicity property of $R^{\beta}(\alpha)$ when $R^{\beta}(\alpha) \leq 1$, it suffices to show that $R^\beta(\alpha)$ has at most one stationary point in $(0,1)$ with  $R^\beta(\alpha)|_{\alpha= 0} =1$ and $(R^\beta(\alpha))'|_{\alpha= 1} < 0$. 

We first observe that the derivative of $ R^\beta(\alpha) $ can be expressed as a specific product form, which is essential in proving that there is at most one stationary point. To this end, the derivative of $R^\beta(\alpha)$ is given by
\begin{align}
    &\left(R^\beta(\alpha) \right)' \notag \\
   =& \frac{-1}{\alpha^2}  \frac{\sum_{i=0}^{C-1} \frac{1}{i!} (C(\frac{1}{\alpha}-1) + \beta)^i   }{\sum_{i=0}^{C} \frac{1}{i!} (C(\frac{1}{\alpha}-1) + \beta)^i } - \frac{C}{\alpha^3} \frac{(\sum_{i=0}^{C} \frac{1}{i!} (C(\frac{1}{\alpha}-1) + \beta)^i)(\sum_{i=0}^{C-2} \frac{1}{i!} (C(\frac{1}{\alpha}-1) + \beta)^i)- (\sum_{i=0}^{C-1} \frac{1}{i!} (C(\frac{1}{\alpha}-1) + \beta)^i)^2}{[\sum_{i=0}^{C} \frac{1}{i!} (C(\frac{1}{\alpha}-1) + \beta)^i]^2} \notag \\
    =& \left( \frac{1}{\alpha}\frac{\sum_{i=0}^{C-1} \frac{1}{i!} (C(\frac{1}{\alpha}-1) + \beta)^i   }{\sum_{i=0}^{C} \frac{1}{i!} (C(\frac{1}{\alpha}-1) + \beta)^i } - \frac{1}{C} \frac{[\sum_{i=0}^{C-1} \frac{1}{i!} (C(\frac{1}{\alpha}-1) + \beta)^i]^2}{[\sum_{i=0}^{C-1} \frac{1}{i!} (C(\frac{1}{\alpha}-1) + \beta)^i]^2 - [\sum_{i=0}^{C} \frac{1}{i!} (C(\frac{1}{\alpha}-1) + \beta)^i][\sum_{i=0}^{C-2} \frac{1}{i!} (C(\frac{1}{\alpha}-1) + \beta)^i]} \right) \notag \\
    & \cdot \frac{C}{\alpha^2} \frac{ [\sum_{i=0}^{C-1} \frac{1}{i!} (C(\frac{1}{\alpha}-1) + \beta)^i]^2 - [\sum_{i=0}^{C} \frac{1}{i!} (C(\frac{1}{\alpha}-1) + \beta)^i][\sum_{i=0}^{C-2} \frac{1}{i!} (C(\frac{1}{\alpha}-1) + \beta)^i]  }{[\sum_{i=0}^{C} \frac{1}{i!} (C(\frac{1}{\alpha}-1) + \beta)^i][\sum_{i=0}^{C-1} \frac{1}{i!} (C(\frac{1}{\alpha}-1) + \beta)^i]} \notag \\
    = & \left( R^\beta(\alpha) - \frac{1}{C} h\left(C(\frac{1}{\alpha}-1)+\beta\right) \right)  \cdot \frac{C}{\alpha^2} \frac{1 -B_C(C(\frac{1}{\alpha }-1)+\beta) }{h\left(C(\frac{1}{\alpha}-1)+\beta\right) }, \label{eq:derivate_t}
    % \frac{[\sum_{i=0}^{C} \frac{1}{i!} (Ct + s)^i][\sum_{i=0}^{C-2} \frac{1}{i!} (Ct + s)^i] - [\sum_{i=0}^{C-1} \frac{1}{i!} (Ct + s)^i]^2 }{[\sum_{i=0}^{C} \frac{1}{i!} (Ct + s)^i][\sum_{i=0}^{C-1} \frac{1}{i!} (Ct + s)^i]},
\end{align}
where $h(\omega) := \frac{\left(\sum_{i=0}^{C-1}  \omega^i /i!\right)^2}{\left(\sum_{i=0}^{C-1} \omega^i/i! \right)^2-\left(\sum_{i=0}^{C} \omega^i/i! \right)\left(\sum_{i=0}^{C-2}  \omega^i/i! \right)}$ and $B_C(\omega):= \frac{\omega^C/C!}{\sum_{i=0}^C \omega^i/i! }$ denotes the blocking probability when the total capacity is $C$. Here, $\omega$ can be seen as the traffic intensity of a standard Erlang loss system. Note that $h\left(C(\frac{1}{\alpha}-1)+\beta\right)$ is positive since $\left( \sum_{i=0}^{C-1} \frac{(C(\frac{1}{\alpha}-1)+\beta)^i}{i!} \right)^2-\left( \sum_{i=0}^{C} \frac{(C(\frac{1}{\alpha}-1)+\beta)^i}{i!}\right)\left( \sum_{i=0}^{C-2} \frac{(C(\frac{1}{\alpha}-1)+\beta)^i}{i!} \right) =\frac{(C(\frac{1}{\alpha}-1)+\beta)^{C-1}}{C!}\left(\sum_{i=0}^{C-1} (C-i)\frac{(C(\frac{1}{\alpha}-1) +\beta )^i}{i!}\right) >0$. Another important fact is that $h\left(C(\frac{1}{\alpha}-1)+\beta\right)$ is increasing w.r.t. $\alpha$ in $(0,1)$, which is a direct corollary of Lemma \ref{lemm:decrea} below. 
\begin{lemma}\label{lemm:decrea}
 $h(\omega)$ is decreasing in $(0,\infty)$, i.e., $ h'(\omega)  < 0.$
\end{lemma}
\begin{proof}{\textbf{Proof of Lemma~\ref{lemm:decrea}.}}
Since $h(\omega)>0$, it suffices to prove that $H(\omega) := 1 - \frac{1}{h(\omega)} = \frac{\left(\sum_{i=0}^{C}\omega^i/i!\right)\left(\sum_{i=0}^{C-2} \omega^i/i!\right)}{\left(\sum_{i=0}^{C-1} \omega^i/i!\right)^2}$ is decreasing in $(0,\infty)$. We observe that  
\begin{align*}
 H(\omega) =& 1 +  \frac{\frac{\omega^{C-1}}{(C-1)!} \left[ \left(\frac{\omega}{C}-1\right)\sum_{i=0}^{C-1} \frac{\omega^i}{i!}- \frac{\omega^C}{C!}  \right] }{\left(\sum_{i=0}^{C-1} \frac{\omega^i}{i!}\right)^2} = 1 + B_{C-1}(\omega) \left(\frac{\omega}{C} -1 - \frac{\omega}{C}B_{C-1}(\omega) \right) = 1- \frac{\omega  B'_{C-1}(\omega) + B_{C-1}(\omega)}{C},
\end{align*}
where the last equality follows that $B'_{C-1}(\omega) = B_{C-1}(\omega)\left(\frac{C-1}{\omega}-1 + B_{C-1}(\omega) \right) $. 
Note that the carried load $\omega\left(1 - B_{C-1}(\omega)\right)$
is concave in $\omega$ (\citealp{harel1990convexity}), implying that the first order derivative  $1 - B_{C-1}(\omega) -\omega B'_{C-1}(\omega)$ is decreasing in $\omega$. This immediately shows that $H(\omega)$ is decreasing in $\omega$. \hfill \Halmos \\
\end{proof}

Using the product form in \eqref{eq:derivate_t} and that $h\left(C(\frac{1}{\alpha}-1)+\beta \right)$ is increasing in $\alpha$, we now prove that $R^\beta(\alpha)$ has at most one stationary point in $(0,1)$.
Suppose by contradiction that there exists more than one stationary point and the two consecutive ones are denoted by $\alpha_a$ and $\alpha_b$ ($0 < \alpha_a < \alpha_b < 1$), with $R^\beta(\alpha_a) = \frac{1}{C}h\left(C(\frac{1}{\alpha_a}-1)+\beta\right) < \frac{1}{C}h\left(C(\frac{1}{\alpha_b}-1)+\beta\right) = R^\beta(\alpha_b)$. Since $\alpha_a$ and $\alpha_b$ are two consecutive stationary points, $R^\beta(\alpha)$ is increasing in $(\alpha_a,\alpha_b)$. According to the product form \eqref{eq:derivate_t}, we have that
 \begin{align*}
     R^\beta(\alpha)> \frac{1}{C} h\left(C(\frac{1}{\alpha}-1)+\beta\right), \qquad \forall \alpha \in (\alpha_a,\alpha_b).
 \end{align*}
On the other hand, because $\left(R^\beta(\alpha)\right)'|_{\alpha=\alpha_a} = 0$ and $\frac{1}{C} \frac{\partial}{\partial \alpha} [h  \left(C(\frac{1}{\alpha}-1)+\beta\right)] |_{\alpha=\alpha_a} >0$, we know that $R^\beta(\alpha) < \frac{1}{C} h\left(C(\frac{1}{\alpha}-1)+\beta\right) $ at the right neighbourhood of $\alpha_a$, which is a contradiction. Hence, $R^\beta(\alpha)$ has at most one stationary point within $(0,1)$.

In the remainder of proof, we show that $(R^\beta(\alpha))'|_{\alpha=1} <0$ and $\lim_{ \alpha \to 0^+} R^\beta(\alpha) = 1 $. After some algebra, $(R^\beta(\alpha))'$ evaluated at one is given by
\begin{align*}
     (R^\beta(\alpha))'|_{\alpha=1} = &  \frac{  \sum_{k=0}^{C-1} (C-k) \frac{1}{(C-1)!k!}  \beta^k \beta^{C-1} - (\sum_{i=0}^{C} \frac{1}{i!} \beta^i)(\sum_{j=0}^{C-1} \frac{1}{j!} \beta^j)  }{(\sum_{i=0}^{C} \frac{1}{i!} \beta^i)^2} < 0.
\end{align*}
The inequality holds since for $0\leq i,j \leq C-1$, $ \sum\limits_{i+j = C+k-1} \frac{1}{i!j!} = \sum_{i=k}^{C-1} \frac{1}{i!(C-1+k-i)!} \geq \frac{C-k}{(C-1)! k!}  $, $ k=0,\ldots,C-1$. Via a well known result for the Erlang loss system, we know that the carried load (expected number of busy servers), $\omega \cdot \frac{\sum_{i=0}^{C-1} \omega^i/i! }{\sum_{i=0}^{C} \omega^i/i!}$, converges to $C$ as $\omega \to \infty$ (\citealp{harel1990convexity}) and the service level, $\frac{\sum_{i=0}^{C-1} \omega^i/i! }{\sum_{i=0}^{C} \omega^i/i!}$, converges to 0 as $\omega \to \infty$
Therefore, we have
\begin{align*}
    \lim_{\alpha \to 0^+} R^\beta(\alpha)   =& \lim_{ \alpha \to 0^+} \left(\frac{C(\frac{1}{\alpha}-1) + \beta}{C} + \frac{C-\beta}{C} \right) \frac{\sum_{i=0}^{C-1} \frac{1}{i!} (C (\frac{1}{\alpha}-1) + \beta)^{i}}{\sum_{i=0}^C \frac{1}{i!} {(C(\frac{1}{\alpha}-1) + \beta)^{i}}} \\
    =& \frac{1}{C} \lim_{\alpha \to 0^+} \left(C(\frac{1}{\alpha}-1) + \beta \right) \frac{\sum_{i=0}^{C-1} \frac{1}{i!} (C(\frac{1}{\alpha}-1) + \beta)^{i}}{\sum_{i=0}^C \frac{1}{i!} {(C(\frac{1}{\alpha}-1) + \beta)^{i}}} + \left( \frac{C-\beta}{C} \right) \lim_{ \alpha \to 0^+}  \frac{\sum_{i=0}^{C-1} \frac{1}{i!} (C(\frac{1}{\alpha}-1) + \beta)^{i}}{\sum_{i=0}^C \frac{1}{i!} {(C(\frac{1}{\alpha}-1) + \beta)^{i}}} \\
    =& \frac{1}{C}\cdot C + 0 = 1
\end{align*}

% Finally, combining $(R^\beta(\alpha))'|_{\alpha
% =1} < 0$, $\lim_{\alpha \to 0^+} R^\beta(\alpha) =1$, $R^\beta(1) =\frac{\sum_{i=0}^{C-1} \beta^i/i! }{\sum_{i=0}^{C} \beta^i/i!} \leq 1 $, and the fact that $R^\beta(\alpha)$ has at most one stationary point in $(0,1)$, we conclude that $R(\alpha,\beta) = R^\beta(\alpha) \geq R^\beta(1) = R(1,\beta) $ for all $ \alpha \in [0,1],\beta \in [0,C-1]$. 
This completes the proof.
\hfill \Halmos \\
\end{proof}

\begin{proof}{\textbf{Proof of Lemma \ref{lem:mono}.} }
 %    Let $q(w) := \frac{\sum_{i=0}^{C-1} w^i/i!  }{\sum_{i=0}^C w^i/i! }$. We next show that the blocking probability $q(w)$ is decreasing in $[0,\infty)$.
 %    \begin{align*}
	% \frac{d q(w)}{d w} =& \frac{\sum_{i=0}^{C-2} \frac{1}{i!}w^{i}\sum_{i=0}^C \frac{1}{i!}w^{i} - [\sum_{i=0}^{C-1} \frac{1}{i!}w^{i}]^2 }{(\sum_{i=0}^C \frac{1}{i!}w^{i})^2 } \\
	% =& \frac{w^{C-1}}{C!}\frac{\sum_{i=0}^{C-1} (i-C)\frac{1}{i!}w^{i}  }{(\sum_{i=0}^C \frac{1}{i!}w^{i})^2} \\
	% \leq & 0.
	% \end{align*}
	% The first two equalities follow from the definition and combining terms. 
    Taking the first partial derivative of $R(\alpha,\beta)$ w.r.t. $\beta $ gives,
	\begin{align*}
	\frac{\partial R(\alpha,\beta)}{\partial \beta} =\frac{1}{\alpha} \frac{(C(\frac{1}{\alpha}-1) + \beta)^{C-1}}{C!}\frac{\sum_{i=0}^{C-1} (i-C)\frac{1}{i!}[C(\frac{1}{\alpha}-1) + \beta]^{i}  }{[\sum_{i=0}^C \frac{1}{i!}(C(\frac{1}{\alpha}-1)+\beta)^{i}]^2} 
	\leq 0.
	\end{align*}
 The inequality holds since $\alpha \leq 1$ and $(i-C) <  0$ for $i=0,\ldots,C-1$. 
 \hfill \Halmos \\
\end{proof}

\begin{proof}{\textbf{Proof of Lemma \ref{lem:tight}}. } 
Consider a 1-class system with $F_1(p) = 1 - \frac{a/\Lambda}{p-b} $, where $p \geq b + \frac{a}{\Lambda} $ for any $a,b>0$. \cor{This distribution is a shifted and scaled version of the limiting triangular distribution studied in \cite{jin2019tight}.} Then, the effective arrival rate is given by $\lambda = \Lambda (1-F_1(p) ) = \frac{a}{p-b} \in [0, \Lambda] $. For $\lambda >0 $, the revenue rate function is $ r(\lambda) = \lambda p(\lambda) = a + b \lambda$. At $\lambda = 0$,  $r(0) = \lim_{\lambda \to 0} r(\lambda) = a  $. We assume the service time distribution is exponential with rate $\mu$. In this case, the optimal fully dynamic policy is equivalent to the optimal inventory-based policy, denoted as $(\lambda_0^*, \ldots, \lambda_{C-1}^*)$. Note that from \eqref{eq:steady}, the revenue rate maximization problem can be simplified as
\begin{align}\label{counter_C}
	\max \limits_{ 0 \leq  \lambda_0,\ldots,\lambda_{C-1}\leq \Lambda} \mathcal{R}(\lambda_0,\ldots,\lambda_{C-1}) =  \frac{ (a+b\lambda_0) +  \sum_{i=1}^{C-1}   (a+b\lambda_i) \frac{1}{i!}  \Pi_{j=1}^i \frac{\lambda_{j-1}}{\mu} }{1 + \sum_{i=1}^C \frac{1}{i!} \Pi_{j=1}^i \frac{\lambda_{j-1}}{\mu}   }.
\end{align}
We clarify that $\Pi_{j=0}^{-1} (\lambda_j/\mu):=1 $.
% For notation simplicity, we denote  $z_{i}=\Pi_{j=0}^i (\lambda_j/\mu) $, $i=0,\ldots,C-1$ and $z_{-1} = 1$.
Taking the first order derivatives of $\mathcal{R}(\lambda_0,\ldots,\lambda_{C-1})$ w.r.t. $\lambda_k$ gives
\begin{align*}
	\frac{\partial \mathcal{R} }{\partial \lambda_{C-1} } =&\frac{1}{\mu C!}
	\frac{ \Pi_{j=0}^{C-2}\frac{\lambda_j}{\mu}  \cdot \sum_{i=0}^{C-1}  [(C-i) b\mu -a ] \frac{1}{i!}\Pi_{j=0}^{i-1}  \frac{\lambda_j}{\mu}   }{(\sum_{i=0}^C \frac{1}{i!} \Pi_{j=0}^{i-1} \frac{\lambda_j}{\mu}  )^2} \\
	\frac{\partial \mathcal{R} }{\partial \lambda_{k} } =&\frac{1}{\lambda_k }
	\frac{ -\frac{a}{C!}  \Pi_{j=0}^{C-1}\frac{\lambda_j}{\mu} \sum_{i=0}^k \frac{1}{i!} \Pi_{j=0}^{i-1} \frac{\lambda_j}{\mu} + b\mu \sum_{i=0}^{k} \sum_{j=k+1}^{C} \frac{1}{i!j!} (j-i)  \Pi_{l=0}^{i-1} \frac{\lambda_l}{\mu}\cdot \Pi_{l=0}^{j-1} \frac{\lambda_l}{\mu}    }{(\sum_{i=0}^C \frac{1}{i!} \Pi_{j=0}^{i-1} \frac{\lambda_j}{\mu}  )^2}, \qquad k= 0,\ldots,C-2. 
\end{align*}
When $\mu$ goes to 0, it is clear that $[(C-i)b\mu -a]< 0 $, implying that $\frac{\partial \mathcal{R} }{\partial \lambda_{C-1} }$ is non-positive. Note that all $\frac{\partial \mathcal{R} }{\partial \lambda_k }$, $k=0,\ldots,C-2$ are non-negative when $\lambda_{C-1} = 0$. Therefore, for any $0  \leq \lambda_i\leq \Lambda$, $i=0,\ldots,C-1$, we have
\begin{align*}
	\mathcal{R}(\lambda_0,\ldots,\lambda_{C-1}) \leq \mathcal{R}(\lambda_0,\ldots,\lambda_{C-2},0) \leq \mathcal{R}(\Lambda,\ldots,\Lambda,0),
\end{align*}
which implies that the optimal solution to problem (\ref{counter_C}) is $\lambda_{C-1}^{*}=0$ and $\lambda_0^{*}=\cdots =\lambda_{C-2}^{*}=\Lambda$. Also, the corresponding optimal prices are $p_{C-1}^* = + \infty$ and $p_0^* = \cdots = p_{C-2}^* = b + a/\Lambda$.
Because $\lambda p(\lambda) = a+b\lambda$ is linear, the Jensen's inequality we used in Lemma~\ref{lem:prof_serv} is actually an equality. Thus, for this instance 
%. Note that $t = 0$ in $R(s,t)$ when $\lambda_{C-1}^*=0$, we then have
\begin{align*}
	\frac{\mathcal{R}^{\tilde{\lambda}}}{\mathcal{R}^{*}} = \frac{1-\mathbb{P}_C(\tilde{\lambda})}{1-\mathbb{P}_C(\bm{\lambda}^{*})} = R(\alpha^{*},\beta^{*}).
\end{align*}
By definition of $\alpha^{*}$ and $\beta^{*}$, we plug in the optimal policy and observe that $\alpha^{*} = 1$ and $\beta^{*} = \frac{\sum_{i=1}^{C-1} (\Lambda/\mu)^{i}/(i-1)! }{1 + \sum_{i=1}^{C-1}(\Lambda/\mu)^{i} /i! }$. Since $\beta^{*} \to C-1 $ as $\mu \to 0$, then
%where $s^* = \frac{\sum_{i=1}^{C-1} (\Lambda/\mu)^{i}/(i-1)! }{1 + \sum_{i=1}^{C-1}(\Lambda/\mu)^{i} /i! } \to C-1$ as $\mu \to 0$. Hence,
\begin{align*}
	\lim_{\mu \to 0}  \frac{ \mathcal{R}^{\tilde{\lambda}} }{ \mathcal{R}^{*} } = \lim_{\mu \to 0}  R(\alpha^{*},\beta^{*}) = \lim_{\mu \to 0} \frac{\sum_{i=0}^{C-1} \frac{1}{i!}(\beta^{*})^{i} }{\sum_{i=0}^C \frac{1}{i!}(\beta^{*})^{i} } = \frac{\sum_{i=0}^{C-1} \frac{1}{i!}(C-1)^{i} }{\sum_{i=0}^C  \frac{1}{i!}(C-1)^{i} } = G(C). 
\hfill \Halmos \\
\end{align*} 
\end{proof}

\begin{proof}{\textbf{Proof of Lemma~\ref{lem:C_incr}.}}
	Since $ G(3) \approx 0.789, G(4)\approx0.794 $, and $ G(5) \approx 0.801 $, it holds that $ G(C+1)>G(C) $ for $ C =3,4 $. When $ C\geq5 $, we directly compute $ G(C+1) -G(C) $, i.e.,
	\begin{align*}
	&	G(C+1) - G(C)  \\
	=& \frac{\frac{1}{C!}(C-1)^C}{ \sum_{i=0}^C \frac{1}{i!} (C-1)^{i}  } -  \frac{\frac{1}{(C+1)!}C^{C+1}}{ \sum_{i=0}^{C+1} \frac{1}{i!} C^{i}  }  \\ 
	%=& \frac{1}{(C+1)!} \frac{(C+1) (C-1)^C \sum_{i=0}^{C+1} \frac{1}{i!} C^{i} - C^{C+1} \sum_{i=0}^{C} \frac{1}{i!} (C-1)^{i} }{[ \sum_{i=0}^C \frac{1}{i!} (C-1)^{i}] [\sum_{i=0}^{C+1} \frac{1}{i!} C^{i}]} \\
	%=&   \frac{1}{(C+1)!} \frac{(C+1) (C-1)^C + (C+1)(C-1)^C\sum_{i=0}^{C} \frac{1}{(i+1)!} C^{i+1} - C^{C+1} \sum_{i=0}^{C} \frac{1}{i!} (C-1)^{i} }{[ \sum_{i=0}^C \frac{1}{i!} (C-1)^{i}] [\sum_{i=0}^{C+1} \frac{1}{i!} C^{i}]}\\
	=&   \frac{1}{(C+1)!} \frac{(C+1) (C-1)^C + \sum_{i=0}^{C} \frac{1}{i!} C^{i+1}(C-1)^C[\frac{(C+1)}{i+1} - (\frac{C}{C-1})^{C-i} ]  }{[ \sum_{i=0}^C \frac{1}{i!} (C-1)^{i}] [\sum_{i=0}^{C+1} \frac{1}{i!} C^{i}]}\\
	=&   \frac{1}{(C+1)!} \frac{(C+1) (C-1)^C + \sum_{i=0}^{C-5} \frac{1}{i!} C^{i+1}(C-1)^C[\frac{(C+1)}{i+1} - (\frac{C}{C-1})^{C-i} ]  + \frac{C^{C-3}(C-1)^{C-4}}{(C-2)!}  (4C^3-12C^2+8C-2)}{[ \sum_{i=0}^C \frac{1}{i!} (C-1)^{i}] [\sum_{i=0}^{C+1} \frac{1}{i!} C^{i}]}.
	\end{align*}
	The last equality follows from combining terms for $i=C,C-1,C-2,C-3$, and $C-4$.
	Note that $ 4C^3 -12C^2 +8C-2 = 4C^2(C-3) + 2(4C-1) > 0$ when $ C\geq 5 $. To prove that $ G(C+1) > G(C) $, it is sufficient to show $ (\frac{C}{C-1})^{C-i} < \frac{C+1}{i+1} $ for $ 0\leq i \leq C-5 $  by backward induction. 
		\begin{itemize}
		\item $ i=C-5 $: $ (C-1)^5(C+1)-C^5(C-4) = 5C^2(C^2-1)+4C-1 >0  $ implies that $ (\frac{C}{C-1})^5 < \frac{C+1}{C-4} $.
		\item Assume $ \left(\frac{C}{C-1} \right)^{C-i} < \frac{C+1}{i+1} $ holds and we want to prove the inequality for $i-1$. We have that
	\end{itemize}
	\begin{align*}
	\left(\frac{C}{C-1}\right)^{C-i+1} -\left(\frac{C}{C-1} \right)^{C-i} =  \left(\frac{C}{C-1}\right)^{C-i} \cdot \frac{1}{C-1} < \frac{C+1}{i+1} \cdot \frac{1}{C-1}
	<\frac{C+1}{i+1} \cdot \frac{1}{i} = \frac{C+1}{i} - \frac{C+1}{i+1}, 
	\end{align*}
	which shows that $ \left(\frac{C}{C-1} \right)^{C-i+1} - \frac{C+1}{i} < \left(\frac{C}{C-1} \right)^{C-i} - \frac{C+1}{i+1} <0  $.
	Thus, $ G(C+1) >G(C) $  for $ C\geq5 $.
    \hfill \Halmos \\
\end{proof}

\section{Proofs in Section~\ref{sec:MHR}}\label{proof:mhr}

\begin{proof}{\textbf{Proof of Lemma \ref{lem:mhr-1-class-positive}}. }
We note that the optimal inventory-based arrival rates are non-increasing in the number customers in the system \citep{Paschalidis2000}, i.e., $ \lambda_0^{inv*} \geq \cdots \geq \lambda_{C-1}^{inv*}  $.  We first have $\lambda_0^{inv*}>0$, since otherwise all arrival rates are zero and the revenue is zero.

Suppose otherwise, and let $k \geq 1$ be the smallest index such that $\lambda_k^{ inv*}=0$. The expression of $\mathcal{R}^{inv*}$ from \eqref{eq:steady} and the conditions of the optimal inventory-based policy are given by
\begin{align*}
\mathcal{R}^{inv*} = \frac{ \sum_{i=0}^{k-1} \frac{1}{i!}\lambda_i^{inv*} p(\lambda_i^{inv*}) \prod_{j=1}^{i}\frac{\lambda^{inv*}_{j-1}}{\mu}}{ 1 + \sum_{i=1}^{k} \frac{1}{i!} \prod_{j=1}^{i}\frac{\lambda_{j-1}}{\mu}   } , \quad
\lambda_0^{  inv*}\geq \cdots\geq\lambda_{k-1}^{  inv*}>0,\quad \lambda_k^{  inv*}=\cdots=\lambda_{C-1}^{  inv*}=0.
\end{align*}
Now, consider the policy obtained by replacing $\lambda_k^{inv*}=0$ with any $\epsilon\in(0,\lambda_{k-1}^{  inv*})$ and leaving all other $\lambda_i^{inv*}$ unchanged. The revenue under this policy is computed as
\begin{align*}
\mathcal{R}(\lambda_0^{inv*},\ldots, \lambda_{k-1}^{inv*} , \epsilon, 0, \ldots, 0 ) = \frac{ \sum_{i=0}^{k-1} \frac{1}{i!}\lambda_i^{inv*} p(\lambda_i^{inv*}) \prod_{j=1}^{i}\frac{\lambda^{inv*}_{j-1}}{\mu} + \frac{\epsilon p(\epsilon)}{k!}\prod_{j=1}^{k} \frac{\lambda_{j-1}^{inv*}}{\mu}  }{ 1 + \sum_{i=1}^{k} \frac{1}{i!} \prod_{j=1}^{i}\frac{\lambda_{j-1}}{\mu} +  \frac{\epsilon}{(k+1)! \mu}\prod_{j=1}^{k} \frac{\lambda_{j-1}^{inv*}}{\mu}  } > \mathcal{R}^{inv*},
\end{align*}
where the inequality uses the fact that  $p(\epsilon)>\frac{\mathcal R^{inv*}}{(k+1)\mu}$. To see this, note that 
\begin{align*}
\mathcal R^{inv*}
=\sum_{i=0}^{k-1}\lambda_i^{inv*}p(\lambda_i^{  inv*}) \mathbb P_i(\bm\lambda^{  inv*}) 
\leq p(\lambda_{k-1}^{  inv*})\sum_{i=0}^{k-1}\lambda_i^{inv*}\mathbb P_i(\bm\lambda^{  inv*}) 
=p(\lambda_{k-1}^{  inv*})\mu\sum_{i=1}^{k}i\mathbb P_i(\bm\lambda^{  inv*})
\leq k\mu p(\lambda_{k-1}^{  inv*}),
\end{align*}
where the first inequality follows from $p(\lambda)$ being non-increasing and $\lambda_i^{inv*}\geq \lambda_{k-1}^{inv*}$ for $i=0,\ldots,k-1$, and the second equality holds by the balance equations $\lambda_i^{  inv*}\mathbb P_i(\bm\lambda^{  inv*})=(i+1)\mu\mathbb P_{i+1}(\bm\lambda^{  inv*})$, for $i=0,\ldots,k-1$. Then, since $\epsilon< \lambda_{k-1}^{inv*}$, it holds that
\begin{align*}
    p(\epsilon)\geq p(\lambda_{k-1}^{  inv*})\geq\frac{\mathcal R^{  inv*}}{k\mu}>\frac{\mathcal R^{  inv*}}{(k+1)\mu}.
\end{align*}
Therefore, $(\lambda_0^{inv*},\ldots, \lambda_{k-1}^{inv*} , \epsilon, 0, \ldots, 0 )$ earns strictly higher revenue, contradicting the optimality of $\bm\lambda^{inv*}$. Hence $\lambda_i^{inv*}>0$ for every $i=0,\ldots,C-1$.  \hfill \Halmos \\
\end{proof}

\begin{proof}{\textbf{Proof of Lemma \ref{lem:0.9755-(2,974)}}.} 
Because the right hand side of \eqref{eq:mhr-d-lower} is a two-layer optimization problem and does not have a simple monotonicity property in $d$, solving it directly is challenging. A naive brute force search over $d$ can approximate the optimal value of $ \min_{0 < d < C }  \frac{ A_C ( e^{\mathcal{I}_C (d) /d}  )}{d} $ but does not provide a provable guarantee. We therefore propose a modified brute-force method that leverages the dual problem of \eqref{min-convex-d} to provide a provable lower bound. To this end, for any $ d \in (0,C) $,  we consider the dual of \eqref{min-convex-d}:
% $x\log\left(\frac{x}{y}\right)=\sup_{s\in\mathbb R}\left\{sx-ye^{s-1}\right\},$
\begin{equation}\label{max-dual}
\begin{aligned}
 \ \max_{s_1,\ldots,s_C,a,b}\quad &a+bd \\
\textrm{s.t.} \quad 
& a \leq - e^{s_1-1}  \\
& a+bi\leq i s_i-e^{s_{i+1}-1}, \qquad i=1,\ldots,C-1, \\
& a+bC\leq Cs_C.
\end{aligned}
\end{equation}
The key observation is that the feasible region of \eqref{max-dual} does not depend on $d$. Therefore, any dual-feasible solution $(s_1,\ldots,s_C,a,b)$ remains feasible for every $d\in(0,C)$, and weak duality gives $\mathcal{I}_C(d)\geq a+bd$ throughout this range. 

Based on this observation, we partition the range of $d$ into finitely many intervals and use one dual solution to obtain the lower bound $\mathcal{I}_C(d)\geq a+bd$ over each interval.
% Based on this observation, we partition the range of $d$ into small intervals. For each interval, we solve \eqref{max-dual} at its midpoint to obtain a dual solution $(a,b)$. Since this solution remains feasible for every $d$ in the interval, it provides a valid line lower bound  on $\mathcal{I}_C(d)$ over the entire interval, rather than only at the midpoint. We can then use this line low bound to obtain a provable lower bound on $\mathcal{I}_C(d)/d$ throughout the interval. 
To reduce the range of $d$ that must be examined, Lemma~\ref{lem:d-range} shows that, for a target guarantee $x\in(0,1)$, it suffices to consider the closed interval $[\underline d_C(x),\bar d_C(x)]$, where $\underline d_C(x):=\sup\left\{d\in[0,C]:\frac{\sum_{i=0}^{C-1}d^i/i!}{\sum_{i=0}^C d^i/i!}\geq x\right\}$ and $\bar d_C(x):=\frac{C^2-C/x}{C-1}$.
 \begin{lemma}\label{lem:d-range}
 Fix $ x \in (0, 1)$. Then $ \frac{ A_C ( e^{\mathcal{I}_C (d) /d}  )}{d} \geq x $ for any $d \in (0, \underline d_C(x)) \cup ( \bar d_C(x), C)  $.
\end{lemma}

\textbf{Algorithm.} Fix $ x \in (0,1) $. By Lemma~\ref{lem:d-range}, it remains to lower bound $ A_C ( e^{\mathcal{I}_C (d) /d}  )/d$ over $d\in[\underline d_C(x),\bar d_C(x)]$. We partition this range into $N$ intervals with endpoints $d_n:=\underline d_C(x)+\frac{n}{N}\left(\bar d_C(x)-\underline d_C(x)\right)$ for $n=0,\ldots,N$. For each $n \in [N] $, we solve \eqref{max-dual} at the midpoint $d=(d_{n-1}+d_n)/2$, yielding a verified dual-feasible solution $(s_{1,n},\ldots, s_{C,n},a_n,b_n)$. 
% For each $n\in[N]$, we use CVXPY to solve \eqref{max-dual} at the midpoint $d=(d_{n-1}+d_n)/2$ and obtain candidate values $(s_{1,n},\ldots,s_{C,n},b_n)$. To ensure dual feasibility despite numerical error, we round $s_{1,n},\ldots,s_{C,n},b_n$ to six decimal places and set $ a_n=\min\left\{-e^{s_{1,n}-1},\ \min_{1\leq i\leq C-1}\left\{is_{i,n}-e^{s_{i+1,n}-1}-ib_n\right\},\ Cs_{C,n}-Cb_n\right\}-10^{-6} $. By construction, $(s_{1,n},\ldots,s_{C,n},a_n,b_n)$ is dual feasible. 
Since the dual feasible region is independent of $d$, weak duality gives $\mathcal{I}_C(d)\geq a_n+b_nd$ for every $d\in[d_{n-1},d_n]$. Consequently, for every $d\in[d_{n-1},d_n]$, we have that
\begin{align*}
\frac{A_C \left(e^{\mathcal{I}_C(d)/d}\right)}{d}
\geq \frac{A_C\left(e^{b_n+a_n/d}\right)}{d}
\geq \frac{A_C\left(\exp\left(\min\left\{b_n+\frac{a_n}{d_{n-1}},b_n+\frac{a_n}{d_n}\right\}\right)\right)}{d_n}: = \mathcal{L}_{C,n},
\end{align*}
where the second inequality follows from the monotonicity of $A_C(\cdot)$. Hence, we conclude that
\begin{align*}
\min_{0< d < C}\frac{A_C\left(e^{\mathcal{I}_C(d)/d}\right)}{d}\geq\min\left\{x ,\min_{1\leq n\leq N} \mathcal{L}_{C,n} \right\}.
\end{align*}
Algorithm~\ref{alg:dual-certificate} below summarizes this procedure.

\begin{algorithm}[htp!]
\caption{Provable lower bound for fixed $x$, $C$, and $N$}\label{alg:dual-certificate}
\begin{algorithmic}[1]
\STATE \textbf{Input:} target $x\in(0,1)$, number of units $C\ge2$, number of intervals $N\ge1$
\STATE Compute $\underline{d}_C(x)$ and $\bar d_C(x)$
\IF{$\underline{d}_C(x)\ge\bar d_C(x)$}
\RETURN $x$
\ENDIF
\STATE Set $  d_n = \underline{d}_C(x)+\frac{n}{N}(\bar d_C(x)- \underline{d}_C(x) ),\ n=0,\ldots,N $
\FOR{$n=1,\ldots,N$}
    \STATE Solve \eqref{max-dual} at $d=(d_{n-1}+d_n)/2$ and obtain the solution $a_n $ and $ b_n$
    \STATE Compute $  \mathcal{L}_{C,n} = \frac{A_C\left(\exp\left(\min\left\{b_n+\frac{a_n}{d_{n-1}},b_n+\frac{a_n}{d_n}\right\}\right)\right)}{d_n} $
\ENDFOR
\RETURN $ \min\left\{x,\min_{1\le n\le N} \mathcal{L}_{C,n}\right\} $
\end{algorithmic}
\end{algorithm}

For each $2\leq C\leq974$, we implement Algorithm~\ref{alg:dual-certificate} with $x=0.9906$. The resulting lower bounds are reported in Appendix~\ref{app:9755 tables}. The smallest value is $0.975501$, attained at $C=22$.  
% For $C\geq975$, Theorem~\ref{Thm: Mregular} and Lemma~\ref{lem:C_incr} immediately imply that the guarantee is at least $G(975) \geq 0.9755$. Together, these results establish the $0.9755$ guarantee for every $C\geq2$.

\subsection{Numerics}\label{app:9755 tables}

For $C = 2, 3,\ldots,100$, we apply Algorithm \ref{alg:dual-certificate} with $x = 0.9906$ and $N = 12000$. The results are presented in Table \ref{table:bounds-2-100}, where all values are truncated to six decimal places. To ensure dual feasibility, for each $n\in[N]$, we use CVXPY to solve \eqref{max-dual} at the midpoint $d=(d_{n-1}+d_n)/2$ and obtain candidate values $(s_{1,n},\ldots,s_{C,n},b_n)$. We then round $s_{1,n},\ldots,s_{C,n},b_n$ to six decimal places and set $ a_n=\min\left\{-e^{s_{1,n}-1},\ \min_{1\leq i\leq C-1}\left\{is_{i,n}-e^{s_{i+1,n}-1}-ib_n\right\},\ Cs_{C,n}-Cb_n\right\}-10^{-6} $. By construction, $(s_{1,n},\ldots,s_{C,n},a_n,b_n)$ is dual feasible. 

\begin{table}[htbp]
\centering
\begin{tabular}{cc@{\quad}|cc@{\quad}|cc@{\quad}|cc@{\quad}cc}
% \toprule
$C$ & Value & $C$ & Value & $C$ & Value & $C$ & Value & $C$ & Value \\
\midrule
2  & 0.990600 & \textbf{22} & \textbf{0.975501} & 42 & 0.976332 & 62 & 0.977430 & 82  & 0.978406 \\
3  & 0.985854 & 23 & 0.975509 & 43 & 0.976388 & 63 & 0.977482 & 83  & 0.978452 \\
4  & 0.982962 & 24 & 0.975524 & 44 & 0.976444 & 64 & 0.977534 & 84  & 0.978496 \\
5  & 0.981034 & 25 & 0.975545 & 45 & 0.976500 & 65 & 0.977585 & 85  & 0.978541 \\
6  & 0.979676 & 26 & 0.975571 & 46 & 0.976556 & 66 & 0.977636 & 86  & 0.978585 \\
7  & 0.978683 & 27 & 0.975602 & 47 & 0.976612 & 67 & 0.977687 & 87  & 0.978629 \\
8  & 0.977937 & 28 & 0.975637 & 48 & 0.976668 & 68 & 0.977738 & 88  & 0.978672 \\
9  & 0.977367 & 29 & 0.975676 & 49 & 0.976724 & 69 & 0.977788 & 89  & 0.978716 \\
10 & 0.976926 & 30 & 0.975717 & 50 & 0.976779 & 70 & 0.977837 & 90  & 0.978758 \\
11 & 0.976581 & 31 & 0.975761 & 51 & 0.976835 & 71 & 0.977887 & 91  & 0.978801 \\
12 & 0.976310 & 32 & 0.975807 & 52 & 0.976890 & 72 & 0.977936 & 92  & 0.978843 \\
13 & 0.976098 & 33 & 0.975855 & 53 & 0.976946 & 73 & 0.977985 & 93  & 0.978885 \\
14 & 0.975931 & 34 & 0.975904 & 54 & 0.977000 & 74 & 0.978033 & 94  & 0.978927 \\
15 & 0.975802 & 35 & 0.975955 & 55 & 0.977055 & 75 & 0.978081 & 95  & 0.978968 \\
16 & 0.975703 & 36 & 0.976007 & 56 & 0.977110 & 76 & 0.978128 & 96  & 0.979009 \\
17 & 0.975628 & 37 & 0.976059 & 57 & 0.977164 & 77 & 0.978176 & 97  & 0.979049 \\
18 & 0.975574 & 38 & 0.976113 & 58 & 0.977218 & 78 & 0.978222 & 98  & 0.979090 \\
19 & 0.975536 & 39 & 0.976167 & 59 & 0.977271 & 79 & 0.978269 & 99  & 0.979130 \\
20 & 0.975513 & 40 & 0.976222 & 60 & 0.977324 & 80 & 0.978315 & 100 & 0.979170 \\
21 & 0.975502 & 41 & 0.976277 & 61 & 0.977377 & 81 & 0.978361 &     &          \\
\end{tabular}
% \bottomrule
\caption{Lower bounds on $\min_{0 < d < C }  \frac{ A_C ( e^{\mathcal{I}_C (d) /d}  )}{d}$ for $C=2,\ldots,100$.}\label{table:bounds-2-100}
\end{table}

For $C=101,102,\ldots, 974$, we apply Algorithm \ref{alg:dual-certificate} with  $x = 0.9906$ and $N = 100$. We only choose $N=100$ to facilitate the computation as the guarantee is already higher than 0.9755. One can set a larger value of $N$ to obtain a better guarantee for each $C \geq 101 $. The results are shown in Figure \ref{Fig:mhr>=100}.

\begin{figure}[!]
\FIGURE{\centering
\includegraphics[width=5.7 in]{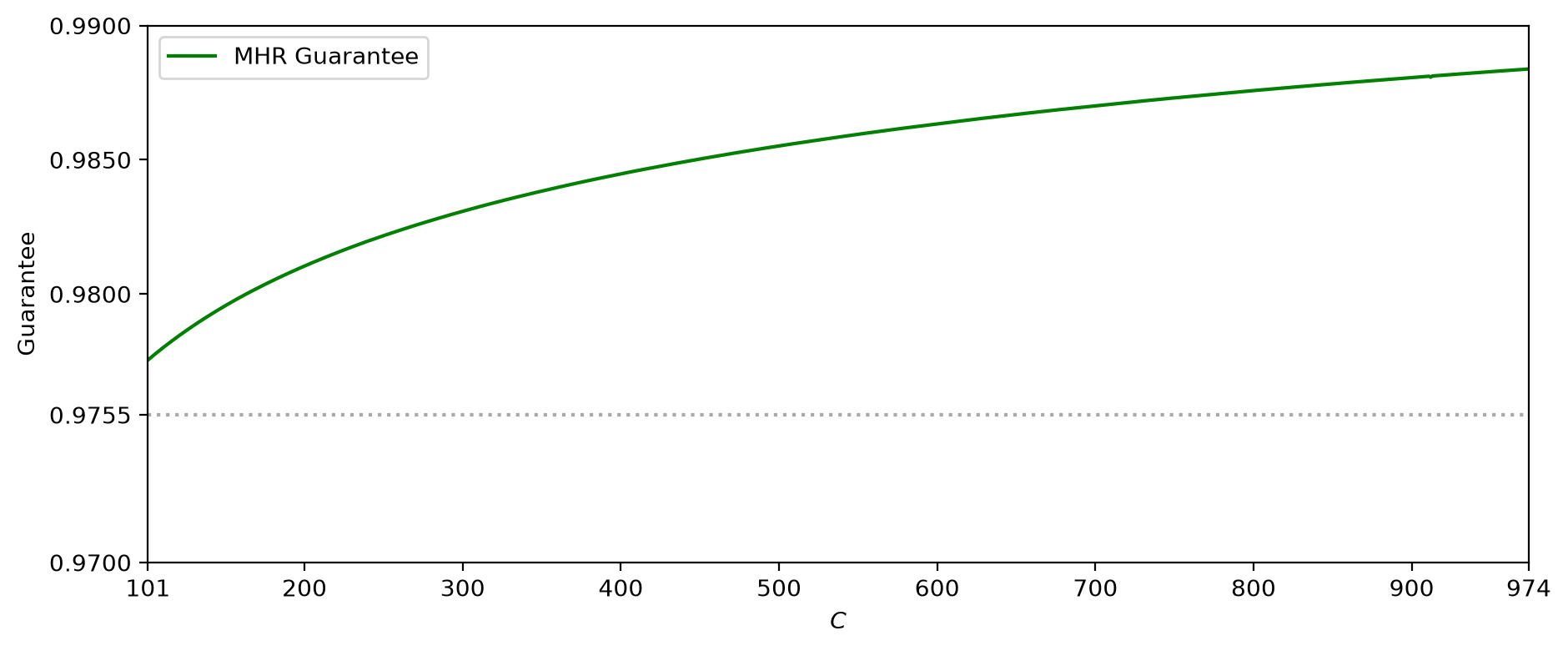}\label{Fig:mhr>=100}
}{Lower bounds on $\min_{0 < d < C }  \frac{ A_C ( e^{\mathcal{I}_C (d) /d}  )}{d}$ for $C=101,\ldots,974$. }{}
\end{figure}

\end{proof}

\begin{proof}{\textbf{Proof of Lemma \ref{lem:d-range}}.} 
For any $x \in (0,1) $, recall that $ \underline d_C(x):=\sup\left\{d\in[0,C]: \frac{\sum_{i=0}^{C-1} d^i/i! }{\sum_{i=0}^{C} d^i/i! }  \geq x \right\} $ and $ \bar d_C(x):=\frac{C^2-C/x}{C-1} $. Assume $d = \sum_{i=1}^C i \mathbb{P}_i  \in (0, \underline{d}_C(x)) $. By Jensen's inequality applied to the convex function $x\log x$, 
\begin{align*}
\log\omega^g
=\frac{\sum_{i=0}^{C-1}\mathbb{P}_i \omega_i\log\omega_i}{d}
\ge \log\left(\frac{\sum_{i=0}^{C-1}\mathbb{P}_i \omega_i}{\sum_{i=0}^{C-1}\mathbb{P}_i} \right)
=\log\left(\frac{d}{1- \mathbb{P}_C} \right).
\end{align*}
Hence, $ \omega^g \geq d/(1-\mathbb{P}_C)$, which holds for any feasible $(\mathbb{P}_0, \ldots, \mathbb{P}_C) $ in \eqref{min-convex-d}. Take the minimum in \eqref{min-convex-d} yields that $ e^{\mathcal{I}_C(d)/d} \geq  d $. 
Since $A_C(\cdot)$ is increasing, we have that 
\begin{align*}
\frac{A_C( e^{\mathcal{I}_C(d)/d} )}{d}\geq \frac{A_C(d)}{d}=\frac{\sum_{i=0}^{C-1} d^i/i!}{\sum_{i=0}^{C} d^i/i! } \geq x , \quad d \in (0, \underline d_C(x) ).  
\end{align*}

For the upper region $d \in (\bar{d}_C(x),C)$, since $ d\leq (C-1)(1-\mathbb P_C)+C\mathbb P_C=C-1+\mathbb P_C$, we have that $1-\mathbb P_C \leq C-d$ and hence $\omega^g \geq d/ (1- \mathbb{P}_c) \geq   d/(C-d)$, which holds for any feasible $(\mathbb{P}_0, \ldots, \mathbb{P}_C) $ in \eqref{min-convex-d}. Take the minimum in \eqref{min-convex-d} yields that $ e^{\mathcal{I}_C(d)/d} \geq  d /(C-d)$. Moreover, it holds that
\begin{align*}
    C-A_C(x)= C - \frac{\sum_{i=0}^{C-1} x x^i/i!}{\sum_{i=0}^{C} x^i/i! }  = \frac{ C \sum_{i=0}^{C} x^i/i! - \sum_{i=1}^C i x^i/i! }{\sum_{i=0}^{C} x^i/i! } \leq C \frac{\sum_{i=0}^{C-1} x^i/i!}{\sum_{i=0}^{C} x^i/i! } =\frac{C \cdot A_C(x)}{x},
\end{align*}
 which implies $A_C(x)\geq Cx/(C+x)$. Therefore, 
 \begin{align*}
     \frac{A_C(e^{\mathcal{I}_C (d) /d})}{d} \geq \frac{A_C\left(d/(C-d)\right)}{d}\geq \frac{C d/(C-d) }{d (C + d/(C-d) } = \frac{C}{C^2-(C-1)d}.
 \end{align*}
The right hand side is at least $x$ whenever $ d \geq  \frac{C^2-C/x}{C-1}  $. \hfill \Halmos \\
\end{proof}

\begin{proof}{\textbf{Proof of Lemma \ref{lem:geo>G(C)}}.}
Recall that $d=\sum_{i=1}^C i\mathbb P_i=\sum_{i=0}^{C-1}\mathbb P_i\omega_i$. Following Section~\ref{sec:proofmulticlass}, we let $ \alpha:=1-\mathbb P_C $ and $
\beta:= \frac{\sum_{i=1}^{C-1}i\mathbb P_i}{1-\mathbb P_C} $. Then, $\alpha\in(0,1)$, $\beta\in[0,C-1]$, and $ \frac{d}{\alpha}
=\frac{\sum_{i=1}^{C-1}i\mathbb P_i+C\mathbb P_C}{1-\mathbb P_C}
=\beta+C\left(\frac{1}{\alpha}-1\right) $. By Jensen's inequality applied to the convex function $x\log x$, we have that
\begin{align*}
\log\omega^g =\frac{\sum_{i=0}^{C-1}\mathbb P_i\omega_i\log\omega_i}{d}
=\frac{\alpha}{d}\sum_{i=0}^{C-1}\frac{\mathbb P_i}{\alpha}\omega_i\log\omega_i
\geq\frac{\alpha}{d}\left(\frac{\sum_{i=0}^{C-1}\mathbb P_i\omega_i}{\alpha}\right)
\log\left(\frac{\sum_{i=0}^{C-1}\mathbb P_i\omega_i}{\alpha}\right)
=\log\left(\frac{d}{\alpha}\right).
\end{align*}
Thus, $\omega^g\geq d/\alpha$. Since $A_C(\cdot)$ is increasing, Lemma~\ref{lem:geo-policy} shows that
\begin{align*}
\frac{\mathcal R^{\lambda^g}}{\mathcal R^{inv*}}
\geq \frac{A_C(\omega^g)}{d}
\geq \frac{A_C(d/\alpha)}{d}
=\frac{1}{\alpha}
\frac{\sum_{i=0}^{C-1}\frac{1}{i!}\left(C\left(\frac{1}{\alpha}-1\right)+\beta\right)^i}
{\sum_{i=0}^{C}\frac{1}{i!}\left(C\left(\frac{1}{\alpha}-1\right)+\beta\right)^i}
=R(\alpha,\beta)
\geq G(C).
\end{align*}
The first equality follows from the definition of $A_C(\cdot)$ and $ d/\alpha = \beta+C\left(\frac{1}{\alpha}-1\right)   $. The second equality follows from Lemmas~\ref{lem:0_ineq} and  \ref{lem:mono}, and the same argument as in \eqref{eq:C_bound}. \hfill \Halmos \\
\end{proof}

\begin{proof}{\textbf{Proof of Theorem \ref{Thm:linear}}.} The 0.9906 guarantee under MHR valuation distributions is shown in Table~\ref{table:bounds-2-100}. 

Next, W.l.o.g., we assume that the uniform valuation distribution is supported on $[0, \bar{v}]$. Since the valuation distribution is uniform, we can write $p(\lambda)=b-a\lambda$ for some $a,b>0$. Recall that  $\omega_i= \lambda_i^{inv*}/\mu$ for $i=0,1$ and $\omega^g = \lambda^g/\mu$. From \eqref{eq:steady}, the revenue expressions are
\begin{align}\label{eq:uniform-inv-revenue}
\frac{\mathcal R^{inv*}}{a\mu^2}
= \frac{\omega_0\left[\frac{b}{a\mu}-\omega_0+\omega_1\left(\frac{b}{a\mu}-\omega_1\right)\right]}{1+\omega_0+\frac{1}{2}\omega_0\omega_1}, \quad
\frac{\mathcal R^{\lambda^g}}{a\mu^2}
= \omega^g\left(\frac{b}{a\mu}-\omega^g\right)\frac{1+\omega^g}{1+\omega^g+\frac{1}{2}(\omega^g)^2}.
\end{align}
The balance equation $\mathbb P_0\omega_0=\mathbb P_1$ and the definition of $\lambda^g$ shows that
\begin{align}\label{eq:uniform-geometric}
\omega^g
= \exp\left(\frac{\log\omega_0+\omega_1\log\omega_1}{1+\omega_1}\right).
\end{align}
The optimal inventory-based policy satisfies $\omega_0\ge\omega_1$ \citep{Paschalidis2000}, and Lemma \ref{lem:mhr-1-class-positive} implies $\omega_0,\omega_1>0$. Differentiating the revenue expression gives
\begin{align*}
\frac{\partial}{\partial\omega_0}\frac{\mathcal{R}(\omega_0, \omega_1)}{a\mu^2}
=\frac{\frac{\mathcal{R}(\omega_0, \omega_1)}{a\mu^2}-\omega_0^2}{\omega_0\left(1+ \omega_0+\frac12\omega_0\omega_1\right)}, \quad
\frac{\partial}{\partial\omega_1}\frac{\mathcal{R}(\omega_0, \omega_1)}{a\mu^2}
=\frac{\omega_0}{1+\omega_0+\frac12\omega_0\omega_1}
\left(\frac{b}{a\mu}- 2\omega_1 -\frac12\frac{ \mathcal{R}(\omega_0, \omega_1)}{a\mu^2}\right).
\end{align*}
Therefore, the KKT conditions imply that
\begin{align*}
\frac{\R^{inv*}}{a\mu^2}\geq& \omega_0^2, \\
\frac{\R^{inv*}}{a\mu^2}=&\frac{2b}{a\mu}-4\omega_1\qquad \text{if }\omega_1<\Lambda/\mu,
\end{align*}
where the first inequality holds with equality if $\omega_0<\Lambda/\mu$. If $\omega_1=\Lambda/\mu$, then $\omega_0=\omega_1$ by $\omega_0\ge\omega_1$ and $\omega_0\le\Lambda/\mu$. Hence, the optimal inventory-based policy is static and $\R^{\lambda^g}/\R^{inv*}=1$. We therefore suppose that $\omega_1<\Lambda/\mu$.
% Recall that $m:=\max\{i\in\{0,1\}:\omega_i=\Lambda/\mu\}$. For simplicity, we further denote $m=-1$ if $\omega_i<\Lambda/\mu$ for all $i=0,1$. We first consider $m<1$. In this case, $\lambda_1^{inv*} < \Lambda$, so the second KKT condition in Lemma~\ref{lem:sim_KKT} holds with equality for $j=1$. 
% Since $\gamma_1= -p'(\lambda) = a$ and $\omega_2= -p'(\lambda) =\gamma_2=0$, we have that
% \begin{align*}
% 2\left(a\omega_1-\frac{b-a\omega_1\mu}{\mu}\right)
% = -\frac{\mathcal R^{inv*}}{\mu^2}
% \Longleftrightarrow
% \frac{\mathcal R^{inv*}}{a\mu^2}
% = \frac{2b}{a\mu} - 4 \omega_1.
% \end{align*}
Combining the above equality and \eqref{eq:uniform-inv-revenue} shows that
\begin{align}\label{eq:uniform-b}
\frac{b}{a\mu}
= \frac{\omega_0\omega_1^2 + 4\omega_0\omega_1 + 4\omega_1-\omega_0^2}{\omega_0+2}.
\end{align}
% Moreover, the first KKT inequality in Lemma~\ref{lem:sim_KKT} and $\gamma_0=a$ imply that $\mathcal R^{inv*}\geq a\mu^2\omega_0^2$, which is equivalent to
Then, the first KKT inequality is equivalent to
\begin{align*}
\frac{2\omega_0(\omega_1^2+2\omega_1-\omega_0)}{\omega_0+2}
\geq \omega_0^2
\Longleftrightarrow \omega_0 \geq
\omega_1 \geq \sqrt{\frac{\omega_0^2}{2}+2\omega_0+1}-1.
\end{align*}
% Lemma~\ref{lemma Structural} also gives $\omega_1\leq\omega_0$. Thus,
% \begin{align}\label{eq:uniform-domain}
% \sqrt{\frac{\omega_0^2}{2}+2\omega_0+1}-1
% \leq \omega_1\leq\omega_0.
% \end{align}

We next show that $\omega_0 < \Lambda/\mu $. Suppose otherwise that $\omega_0=\Lambda/\mu=b/(a\mu)$ and $\omega_1<\Lambda/\mu$. The equality $\mathcal R^{inv*}/(a\mu^2)= 2b/(a\mu)-4\omega_1$, together with $\mathcal R^{inv*}\geq a\mu^2\omega_0^2$ show that $\omega_1\leq\frac{\omega_0}{2}-\frac{\omega_0^2}{4}$. If $\omega_0\geq2$, the right-hand side is non-positive, contradicting the positivity of $\omega_1$ in Lemma~\ref{lem:mhr-1-class-positive}. If $0<\omega_0<2$, then
\begin{align*}
1+2\omega_0+\frac{\omega_0^2}{2}
- \left(1 + \frac{\omega_0}{2}-\frac{\omega_0^2}{4}\right)^2
= \omega_0 + \frac{3}{4}\omega_0^2 + \frac{1}{4}\omega_0^3 -\frac{1}{16}\omega_0^4
>0 \Longrightarrow \sqrt{\frac{\omega_0^2}{2}+2\omega_0+1}-1
> \frac{\omega_0}{2}-\frac{\omega_0^2}{4}.
\end{align*}
This contradicts $ \omega_1 \geq \sqrt{\frac{\omega_0^2}{2}+2\omega_0+1}-1$. 
Therefore, $\omega_0 < \Lambda/\mu $. In this case, the first KKT inequality holds with equality. Hence, we have that
\begin{align}\label{eq:uniform-reduction}
\frac{\mathcal R^{inv*}}{a\mu^2}
=\omega_0^2, \quad
\omega_1=\sqrt{\frac{\omega_0^2}{2}+2\omega_0+1}-1, \quad
\frac{b}{a\mu} = \frac{\omega_0^2}{2}+2\omega_1.
\end{align}
Using \eqref{eq:uniform-inv-revenue} and \eqref{eq:uniform-reduction}, the ratio of revenue rates becomes
\begin{align*}
\frac{\mathcal R^{\lambda^g}}{\mathcal R^{inv*}} =\frac{\omega^g \left(\frac{\omega_0^2}{2} + 2\omega_1-\omega^g\right)(1 + \omega^g)}
{\omega_0^2\left(1 + \omega^g+\frac{1}{2}(\omega^g)^2\right)}, 
\quad \omega_1 = \sqrt{\frac{\omega_0^2}{2} + 2\omega_0+1}-1, \ 
\omega^g = \exp \left(\frac{\log\omega_0 + \omega_1\log\omega_1}{1 + \omega_1}\right).
\end{align*}
Therefore, the ratio is a function of $\omega_0$ alone. A direct differentiation shows that its minimum over $\omega_0>0$ is at least 0.9953725 and is attained at $\omega_0 \approx2.1103, \ \omega_1 \approx1.7290, \ \omega^g \approx1.8600 $.

% Finally, it remains to consider $m=1$, where $\omega_0=\omega_1$. In this case, the optimal inventory-based policy is static, $\omega^g=\omega_0=\omega_1$, and $\mathcal R^{\lambda^g}/\mathcal R^{inv*}=1$. 
\hfill\Halmos
% The minimizing values above are feasible for the uniform instance satisfying $b/(a\mu)\approx5.6847$, and direct maximization of \eqref{eq:uniform-inv-revenue} gives the corresponding optimal inventory-based policy. Hence, the analysis is tight. 
\end{proof}

\section{Proofs in Section~\ref{sec:opt_sta} and \ref{sec: irregular} } \label{Proofs_static}

\begin{proof}{\textbf{Proof of Theorem \ref{thm:optimal-static}.}}
\cor{
Under regular valuation distributions, $ \sum_{j=1}^M \lambda^j p^j(\lambda^j) $ is concave in $(\lambda^1, \ldots,\lambda^M)$. We next show that the reciprocal of service level, defined as $Q_C(\omega):= \frac{\sum_{i=0}^{C} \omega^i/i!}{\sum_{i=0}^{C-1} \omega^i/i!}$, is convex in $\omega$. Then, $Q_C(\sum_{j=1}^M \frac{\lambda^j}{\mu^j} )$ is convex in $(\lambda^1, \ldots,\lambda^M)$. Since
\begin{align*}
\cal R^{sta}\left(\lambda^1,\ldots,\lambda^M \right)  =  \frac{\sum_{j=1}^M \lambda^jp^j(\lambda^j) }{Q_C\left(\sum_{j=1}^M \frac{\lambda^j}{\mu^j} \right)},
\end{align*}
we conclude that \eqref{op_sta} is a concave fractional program. To this end, when $C=1$, $Q_1(\omega)=1+\omega$, which is convex. When $C\geq2$, let $B_{C-1}(\omega):= \frac{\omega^{C-1}/(C-1)!}{\sum_{i=0}^{C-1} \omega^i/i! }$ denote the blocking probability for a system with $C-1$ units. We observe that
\begin{align*}
Q_C(\omega) &=
1+ \frac{\omega^C/C!}{\sum_{i=0}^{C-1}\omega^i/i!} = 1 + \frac{\omega}{C} \cdot B_{C-1}(\omega).
\end{align*}
Because the carried load $\omega(1-B_{C-1}(\omega))$ is concave in $\omega$ (Corollary 1 in \citealt{harel1990convexity}), $\omega B_{C-1}(\omega) = \omega-\omega(1-B_{C-1}(\omega))$ is convex, which implies that $Q_C(\omega)$ is convex. \hfill \Halmos \\
}

\end{proof}

\begin{proof}{\textbf{Proof of Theorem \ref{thm:irregular}: Tightness.}}
\cor{
We show that the ratio $C/(2C-1)$ is tight. We consider the 1-class ($M=1$) system and exponential service times with $\mu=1$. Fix $0<\lambda_H<\Lambda$. We select $v_H=\frac{1}{\lambda_H}$ and $v_L$ such that
\begin{align*}
\Lambda v_L \frac{\sum_{i=0}^{C-1}\Lambda^i/i!}{\sum_{i=0}^{C}\Lambda^i/i!} = \lambda_H v_H \frac{\sum_{i=0}^{C-1}(\lambda_H)^i/i!}{\sum_{i=0}^{C}(\lambda_H)^i/i!} = \frac{\sum_{i=0}^{C-1}(\lambda_H)^i/i!}{\sum_{i=0}^{C}(\lambda_H)^i/i!}.
\end{align*}
For sufficiently large $\Lambda$ and sufficiently small $\lambda_H$, this implies that $v_L<v_H$. For $ 0<\epsilon<
\min\{\Lambda - \lambda_H, \frac{v_H-v_L}{\Lambda - \lambda_H} \} $, we consider the following price function
\begin{align*}
p_\epsilon(\lambda) =
\begin{cases}
v_H+\epsilon(\lambda_H-\lambda), &0\leq \lambda \leq \lambda_H,\\
v_H-\dfrac{v_H-v_L-\epsilon(\Lambda-\lambda_H-\epsilon)}{\epsilon}(\lambda-\lambda_H),
&\lambda_H<\lambda<\lambda_H+\epsilon,\\
v_L+\epsilon(\Lambda-\lambda),
&\lambda_H+\epsilon\leq\lambda\leq\Lambda.
\end{cases}
\end{align*}
This piecewise-linear construction can be approximated arbitrarily closely by price functions induced by irregular valuation distributions satisfying the smoothness assumptions in Section 2, without affecting the limiting ratio below. Let $r_\epsilon(\lambda) = \lambda p_\epsilon(\lambda)$ be the corresponding revenue rate function and let $\bar r_\epsilon$ denote its concave envelope. It is easy to verify that $r_\epsilon(\lambda)$ is not concave in $\lambda$. 
% The revenue rate function is irregular because its derivative jumps upward at $\lambda_H+\epsilon$. Indeed,
% \begin{align*}
% r_\epsilon'((\lambda_H+\epsilon)^+) -r_\epsilon'((\lambda_H + \epsilon)^-) = \frac{\lambda_H+\epsilon}{\epsilon}
% \left( v_H-v_L-\epsilon(\Lambda-\lambda_H) \right) >0.
% \end{align*}
We first compute the optimal static revenue. Recall that the function $ \lambda\frac{\sum_{i=0}^{C-1}\lambda^i/i!}{\sum_{i=0}^{C}\lambda^i/i!}$ is increasing in $\lambda$. We observe that 
\begin{align*}
 r_\epsilon(\lambda) \frac{\sum_{i=0}^{C-1} \lambda^i/i!}{\sum_{i=0}^{C}\lambda^i/i!}
\leq (v_H+\epsilon\lambda_H) \lambda_H
\frac{\sum_{i=0}^{C-1}(\lambda_H)^i/i!}{\sum_{i=0}^{C}(\lambda_H)^i/i!} 
= (1+\epsilon\lambda_H^2) \frac{\sum_{i=0}^{C-1}(\lambda_H)^i/i!}{\sum_{i=0}^{C}(\lambda_H)^i/i!}, \quad 0\leq\lambda\leq\lambda_H,
\end{align*}
\begin{align*}
    r_\epsilon(\lambda)
\frac{\sum_{i=0}^{C-1}\lambda^i/i!}{\sum_{i=0}^{C}\lambda^i/i!}
\leq v_H(\lambda_H+\epsilon)
\frac{\sum_{i=0}^{C-1}(\lambda_H+\epsilon)^i/i!}
{\sum_{i=0}^{C}(\lambda_H+\epsilon)^i/i!}, \quad \lambda_H < \lambda <\lambda_H + \epsilon,
\end{align*}
\begin{align*}
    r_\epsilon(\lambda) \frac{\sum_{i=0}^{C-1}\lambda^i/i!}{\sum_{i=0}^{C}\lambda^i/i!}
\leq (v_L+\epsilon\Lambda)\Lambda \frac{\sum_{i=0}^{C-1}\Lambda^i/i!}{\sum_{i=0}^{C}\Lambda^i/i!}
= \frac{\sum_{i=0}^{C-1}(\lambda_H)^i/i!}{\sum_{i=0}^{C}(\lambda_H)^i/i!} + \epsilon\Lambda^2 \frac{\sum_{i=0}^{C-1}\Lambda^i/i!}{\sum_{i=0}^{C}\Lambda^i/i!}, \quad \lambda_H + \epsilon \leq \lambda\leq \Lambda.
\end{align*}
On the other hand, because $p_\epsilon(\lambda_H)=v_H$, choosing $\lambda=\lambda_H$ yields that $ r_\epsilon(\lambda_H) \frac{\sum_{i=0}^{C-1}(\lambda_H)^i/i!}{\sum_{i=0}^{C}(\lambda_H)^i/i!} = \frac{\sum_{i=0}^{C-1}(\lambda_H)^i/i!}{\sum_{i=0}^{C}(\lambda_H)^i/i!}. $ Therefore, for every fixed $\Lambda$ and $\lambda_H$, the optimal static revenue when $\epsilon \to 0$ is given by
\begin{align*}
\lim_{\epsilon\to0}\mathcal R_{\epsilon}^{sta*}
=\frac{\sum_{i=0}^{C-1}(\lambda_H)^i/i!}{\sum_{i=0}^{C}(\lambda_H)^i/i!}.
\end{align*}
Now consider the inventory-based policy $ \lambda_0=\cdots=\lambda_{C-2}=\Lambda, \lambda_{C-1}=\lambda_H.$ Since $\bar r_\epsilon(\lambda)\geq r_\epsilon(\lambda)$, $p_\epsilon(\Lambda)=v_L$, and $p_\epsilon(\lambda_H)=v_H$, the revenue of this feasible policy satisfies
\begin{align*}
\bar{\mathcal R}_\epsilon(\Lambda, \ldots, \Lambda, \lambda_H) \geq
\frac{ \Lambda v_L\sum_{i=0}^{C-2}  \Lambda^i/i! + \lambda_Hv_H\Lambda^{C-1}/(C-1)! }{ \sum_{i=0}^{C-1}\Lambda^i/i! + \Lambda^{C-1}\lambda_H/C! }
= \frac{ \Lambda v_L\sum_{i=0}^{C-2}\Lambda^i/i! +
\Lambda^{C-1}/(C-1)! }{ \sum_{i=0}^{C-1}\Lambda^i/i! + \Lambda^{C-1}\lambda_H/C! }.
\end{align*}
Since $ \frac{\sum_{i=0}^{C-1}\Lambda^i/i!}{\sum_{i=0}^{C} \Lambda^i/i!} \sim \frac{C}{\Lambda} $
as $\Lambda\to\infty$, the definition of $v_L$ implies that $\Lambda v_L \sim \frac{\Lambda}{C} \frac{\sum_{i=0}^{C-1} (\lambda_H)^i/i!}{\sum_{i=0}^{C}(\lambda_H)^i/i!}.$ Dividing the numerator and denominator of the above lower bound by $\Lambda^{C-1}/(C-1)!$ and taking $\Lambda\to\infty$ gives
\begin{align*}
\liminf_{\Lambda\to\infty}
\bar{\mathcal R}_\epsilon(\Lambda, \ldots,\Lambda, \lambda_H) \geq \frac{ 1 +\frac{C-1}{C} \frac{\sum_{i=0}^{C-1}(\lambda_H)^i/i!}{\sum_{i=0}^{C}(\lambda_H)^i/i!}}{1+\lambda_H/C}.
\end{align*}
Since $\bar{\mathcal R}_{\epsilon}^* \geq \bar{\mathcal R}_\epsilon(\Lambda, \ldots,\Lambda,\lambda_H)$, taking $\lambda_H\to0$ yields that
\begin{align*}
\limsup_{\lambda_H\to0} \limsup_{\Lambda\to\infty}
\limsup_{\epsilon\to 0} \frac{\mathcal R_{\epsilon}^{sta*}}{\bar{\mathcal R}_{\epsilon}^*} \leq \limsup_{\lambda_H\to0} \frac{\sum_{i=0}^{C-1}(\lambda_H)^i/i!} {\sum_{i=0}^{C}(\lambda_H)^i/i!} \cdot \frac{1+\lambda_H/C}{1+\frac{C-1}{C}
\frac{\sum_{i=0}^{C-1}(\lambda_H)^i/i!}
{\sum_{i=0}^{C}(\lambda_H)^i/i!}}
 = \frac{C}{2C-1}.
\end{align*}
% This lower bound does not depend on $\epsilon$. Finally, taking $\lambda_H\to0$ yields
% \begin{align*}
% \lim_{\lambda_H \to 0} \frac{ 1+\frac{C-1}{C} \frac{\sum_{i=0}^{C-1}( \lambda_H)^i/i!}{\sum_{i=0}^{C} (\lambda_H)^i/i!} }{ 1+\lambda_H/C }
% = \frac{2C-1}{C}.
% \end{align*}
% On the other hand, it holds that
% \begin{align*}
% \lim_{\lambda_H\to0} \lim_{\Lambda\to\infty} \lim_{\epsilon\to0} \mathcal R_{\epsilon}^{sta*}
% = \lim_{\lambda_H\to0} \frac{\sum_{i=0}^{C-1}(\lambda_H)^i/i!} {\sum_{i=0}^{C}(\lambda_H)^i/i!} = 1.
% \end{align*}
% Since $\bar{\mathcal R}_{\epsilon}^* \geq \bar{\mathcal R}_\epsilon(\Lambda,\ldots,\Lambda,\lambda_H)$, we obtain that
% \begin{align*}
% \limsup_{\lambda_H\to0} \limsup_{\Lambda\to\infty}
% \limsup_{\epsilon\to 0} \frac{\mathcal R_{\epsilon}^{sta*}}{\bar{\mathcal R}_{\epsilon}^*} \leq \frac{C}{2C-1}.
% \end{align*}
Together with the universal lower bound, this proves that $C/(2C-1)$ is tight. \hfill \Halmos 
}
\end{proof}

\section{Proofs in Section~\ref{sec:network-setting} }\label{sec: network-proof}

\begin{example}\label{ex:constructed-unit-bundle-no-bound}
Consider a unit-bundle network with 3 resources whose capacity vector is $(C_1,C_2,C_3)=(1,2,2)$. There are 3 classes with resource requirements $\bm a_1=(1,1,1), \  \bm a_2=(0,1,0), \ \bm a_3=(0,0,1)$. All service times are exponential with $\mu^1 = \mu^2 = \mu^3 = 1$. Fix $0<\epsilon<1/3$ and $\Lambda\geq 8/\epsilon$. All classes have arrival rate $\Lambda$, and their revenue rate functions are $ r^1(\lambda)=\lambda-\frac{\lambda^2}{\Lambda^3}$ and $ r^2(\lambda) = r^3(\lambda)=\epsilon \lambda - \frac{\lambda^2}{\Lambda^3} $ for $ 0 \leq \lambda\leq\Lambda$. The system state is $\bm x=(x_1,x_2,x_3)$. We first provide a lower bound of the optimal revenue. Consider a dynamic policy that sets $\lambda^j_{\bm x}= \Lambda $ for $  \bm x\in\{0,1\}^3$  and $x_j=0$, $j =1,2,3$,  with all other effective arrival rates equal to 0. Under this policy, the recurrent states are $\{0,1\}^3$, and each class can be independently viewed as a two-state birth-death process with birth rate $\Lambda$ and death rate 1. Therefore, for $\bm x\in\{0,1\}^3$, the steady-state probability is $\mathbb P_{\bm x} = \frac{\Lambda^{x_1+x_2+x_3}}{(1+\Lambda)^3}$, and the service level for each class is $1/(1+ \Lambda)$. The revenue of this dynamic policy is computed as 
\begin{align*}
\frac{r^1(\Lambda)+r^2(\Lambda)+r^3(\Lambda)}{1+\Lambda}
= \frac{(1+2\epsilon)\Lambda-3/\Lambda}{1+\Lambda} \leq \mathcal R^*
\end{align*}

Next, we provide an upper bound. Consider an arbitrary fully dynamic (inventory-based) policy $\bm\lambda$. For simplicity, we further interpret $\lambda^j_{\bm x} = 0$ whenever class $j$ is blocked. Since $r^1(\lambda) \leq \lambda$ and $r^2(\lambda) = r^3(\lambda) \leq\epsilon\lambda$, we have that
\begin{align*}
\mathcal R(\bm\lambda) \leq
\sum_{\bm x} \left( \lambda^1_{\bm x} + \epsilon \lambda^2_{\bm x} + \epsilon \lambda^3_{\bm x} \right) \mathbb P_{\bm x}(\bm\lambda)
= \sum_{\bm x} \left( x_1 + \epsilon x_2 + \epsilon x_3 \right) \mathbb P_{\bm x}(\bm\lambda)
\leq 1 + \epsilon+\epsilon\mathbb P_{111}(\bm\lambda),
\end{align*}
where the equality follows from Little's law applied separately to each class: $\sum_{\bm x}\lambda^j_{\bm x}\mathbb P_{\bm x}(\bm\lambda) = \sum_{\bm x} x_j\mathbb P_{\bm x}(\bm\lambda)$. To see the second inequality, we observe that  $x_1 + \epsilon x_2+\epsilon x_3 \leq 1 + \epsilon+\epsilon\mathbbm{1}\{\bm x=(1,1,1)\}$ for every state $\bm x$. If $x_1=1$ and $\bm x\neq(1,1,1)$, then $x_2+x_3\leq1$, so $x_1+\epsilon x_2+\epsilon x_3\leq1+\epsilon$. If $x_1=0$, then $x_2,x_3\leq2$, so $x_1+\epsilon x_2+\epsilon x_3\leq4\epsilon\leq1+\epsilon$. In state $(1,1,1)$, both sides equal $1+2\epsilon$. Combining the lower and upper bounds, we conclude that
\begin{align*}
1 + \epsilon + \epsilon\mathbb P_{111}(\bm \lambda^*) \geq \frac{(1 + 2\epsilon)\Lambda-3/\Lambda}{1+\Lambda} \Longleftrightarrow 1-\mathbb P_{111}(\bm \lambda^*) \leq \frac{1 + 2\epsilon + 3/\Lambda}{\epsilon(1+\Lambda)}.
\end{align*}
We next lower bound the constructed static arrival rates of classes 2 and 3. Class 2 can be admitted only when $x_1+x_2\leq1$. By Little's law and the definition of the constructed policy, we have that
\begin{align*}
\tilde\lambda^2 = \frac{ \sum_{\bm x:x_1+x_2\leq1} \lambda^{2*}_{\bm x} \mathbb P_{\bm x}(\bm \lambda^*) }{\sum_{\bm x:x_1+x_2\leq1} \mathbb P_{\bm x} (\bm \lambda^*) } = \frac{\sum_{\bm x} x_2\mathbb P_{\bm x}(\bm \lambda^*) }{\sum_{\bm x:x_1 + x_2\leq1}\mathbb P_{\bm x}(\bm \lambda^*)  } \geq
\frac{\mathbb P_{111}(\bm \lambda^*) }{1-\mathbb P_{111}(\bm \lambda^*) }
\geq \frac{\epsilon \Lambda-1-\epsilon-3/\Lambda}
{1+2\epsilon+3/\Lambda} \geq \frac{\epsilon \Lambda/2}{2} = \frac{\epsilon \Lambda}{4}.
\end{align*}
The last inequality holds because $\epsilon \in (0, \frac{1}3)$ and $\Lambda \geq \frac{8}{\epsilon}$ imply that $\epsilon\Lambda-1-\epsilon-\frac{3}{\Lambda}\geq\frac{\epsilon\Lambda}{2} $ and $
1 + 2\epsilon +\frac{3}{\Lambda} \leq 2$. By symmetry, $\tilde\lambda^3 \geq \frac{\epsilon\Lambda}{4}$. Let $\tilde\alpha^j$ denote the service level under the constructed policy. We now upper bound the revenue under the constructed static policy:
\begin{align*}
\mathcal R^{\tilde{\bm\lambda}} = \tilde\alpha^1 r^1(\tilde\lambda^1) + \tilde\alpha^2 r^2(\tilde\lambda^2) + \tilde\alpha^3 r^3(\tilde\lambda^3) \leq \tilde\alpha^1 r^1(\tilde\lambda^1) + 4\epsilon \leq \frac{64}{\epsilon^2\Lambda}  + 4\epsilon.
\end{align*}
The first inequality follows from $\tilde\alpha^2 r^2(\tilde\lambda^2) + \tilde\alpha^3 r^3(\tilde\lambda^3) \leq
\epsilon\tilde\alpha^2 \tilde\lambda^2 + \epsilon\tilde\alpha^3 \tilde\lambda^3 =
\epsilon\sum_{\bm x} x_2\mathbb P_{\bm x}(\tilde{\bm\lambda}) + \epsilon\sum_{\bm x}x_3\mathbb P_{\bm x}(\tilde{\bm\lambda})
\leq 4 \epsilon$ since $x_2,x_3\leq2$. The last inequality is true because
\begin{align*}
 \tilde\alpha^1 r^1(\tilde\lambda^1)  
 =& \frac{ r^1(\tilde\lambda^1) (1+\tilde\lambda^2)(1+\tilde\lambda^3)}{\left(1 + \tilde\lambda^2 + \frac{(\tilde\lambda^2)^2}{2}\right) \left(1 + \tilde\lambda^3 + \frac{(\tilde\lambda^3)^2}{2}\right) +
\tilde\lambda^1(1 + \tilde\lambda^2) (1+\tilde\lambda^3) } \\
\leq &  \frac{ \Lambda(1+\tilde\lambda^2)(1+\tilde\lambda^3) }{\left(1 + \tilde\lambda^2 + \frac{(\tilde\lambda^2)^2}{2}\right) \left(1+\tilde\lambda^3 + \frac{(\tilde\lambda^3)^2}{2}\right)}
\leq \frac{4 \Lambda}{\tilde\lambda^2 \tilde\lambda^3} \leq \frac{64}{\epsilon^2\Lambda},
\end{align*}
where the equality follows from the product-form of steady-state probability, the first inequality holds since $r^1(\lambda)\leq\lambda$ and $\tilde\lambda^1\leq\Lambda$, and the last inequality follows from $\tilde\lambda^2,\tilde\lambda^3\geq\epsilon\Lambda/4$. Combining this upper bound on $\mathcal R^{\tilde{\bm\lambda}} $  with $ \mathcal{R}^* \geq \frac{(1+2\epsilon)\Lambda-3/\Lambda}{1+\Lambda} $ and letting $\Lambda\to\infty$ yields that
\begin{align*}
\limsup_{\Lambda\to\infty}
\frac{\mathcal R^{\tilde{\bm\lambda}}}{\mathcal R^*}
\leq
\frac{4\epsilon}{1+2\epsilon}.
\end{align*}
Since $\epsilon>0$ can be arbitrarily small, the constructed static policy does not admit a positive universal performance guarantee. On the other hand, the static policy that rejects classes 2 and 3 and sets $\lambda^1=\Lambda$ gives revenue $\frac{r^1(\Lambda)}{1+\Lambda} =
\frac{\Lambda-1/\Lambda}{1+\Lambda}$. For fixed $\epsilon$, the ratio of this static revenue to the dynamic lower bound converges to $1/(1+2\epsilon)$ as $\Lambda\to\infty$. Thus, the example shows a limitation of the constructed static policy rather than of optimal static pricing. \hfill \Halmos \\
\end{example}

\begin{proof}{\textbf{Proof of Theorem~\ref{prop: network}. }}
(i). We first prove the fluid upper bound. For the optimal fully dynamic policy $\bm\lambda = \{\lambda^{j*}_{\bm s}\}_{\bm s \in \mathcal{S}, j \in [M]} $, 
% Fix any fully dynamic policy $\bm\lambda = \{\lambda^j_{\bm s}\}_{\bm s \in \mathcal{S}, j \in [M]} $. For a system state $\bm s = (\bm x = (x_1, \ldots, x_M) , \bm y_1, \ldots, \bm y_M)$, let $I^j_{\bm s} = \mathbf{1} \{ \bm{A} \bm x \leq \bm{C} - \bm a_j  \} $ indicate whether adding one class-$j$ customer is allowed.
we let $\hat{\lambda}^j : = \mathbb E_{\bm s}[\lambda^j_{\bm s} \cdot \mathbf{1} \{ \bm{A} \bm x \leq \bm{C} - \bm a_j  \}  ]  $ denote its corresponding unconditional expected effective arrival rate of class $j$. By Little's law, for every resource $k$, it holds that $\sum_{j=1}^M a_{kj} \hat{\lambda}^j/\mu^j\le C_k$. Moreover, since $r^j$ is concave and $r^j(0)\ge0$, Jensen's inequality gives
\begin{align*}
\mathcal R^{*}= \sum_{j=1}^M\mathbb E_{\bm s}[\mathbf{1} \{ \bm{A} \bm x \leq \bm{C} - \bm a_j  \} \cdot r^j(\lambda^{j*}_{\bm s})]
\leq & \sum_{j=1}^M\mathbb E_{\bm s}[r^j( \mathbf{1} \{ \bm{A} \bm x \leq \bm{C} - \bm a_j  \} \cdot  \lambda^{j*}_{\bm s})]  \\
\leq & \sum_{j=1}^M r^j \left(\mathbb E_{\bm s}[\mathbf{1} \{ \bm{A} \bm x \leq \bm{C} - \bm a_j  \} \cdot \lambda^{j*}_{\bm s}] \right) 
=\sum_{j=1}^M r^j(\hat{\lambda}^j) \le \Phi(\bm C).
\end{align*}

Now, let $\mathbb P_{\bm x} (\bm\lambda^F)$ denote the steady-state probability when $\bm{x} = (x_1,\ldots,x_M)$ customers in the system under the fluid policy $\bm\lambda^F$. We also denote $\omega^{j,F} = \frac{\lambda^{j,F}}{\mu^j}$ as the offered load. By the standard product-form distribution for static loss networks, which is insensitive to the service-time distributions beyond their means \citep{kaufman1981blocking}, we have that
\begin{align*}
\mathbb P_{\bm x} (\bm\lambda^F)=\frac{\prod_{j=1}^M(\omega^{j,F})^{x^j}/x^j!}{\mathcal Z(\bm C)}, \qquad \mathcal Z(\bm C):=\sum_{\bm x:\bm A\bm x\le \bm C}\prod_{j=1}^M\frac{(\omega^{j,F})^{x^j}}{x^j!}.
\end{align*}
The goal is to lower bound the service level of each class. The service level of class $j$ under $\bm\lambda^F$ is given by
\begin{align*}
\alpha^{j} (\bm\lambda^F) =\sum_{\bm x:\bm A\bm x\le \bm C-\bm a_j}\mathbb P_{\bm x} (\bm\lambda^F) =\frac{\mathcal Z(\bm C-\bm a_j)}{\mathcal Z(\bm C)} .
\end{align*}
We next lower bound the ratio $\mathcal Z(\bm C-\bm a_j)/\mathcal Z(\bm C)$. Fix any resource $k$. We first lower bound $\mathcal Z(\bm C-\bm e_k)/ \mathcal Z(\bm C).$ To facilitate the analysis, we further define:
\begin{align*}
W_m:=\sum_{\bm x:\bm A \bm x\le\bm C,\ \sum_{l:k \in \mathcal B^l} x_l=m} \ \prod_{j=1}^M \frac{(\omega^{j,F})^{x_j}}{x_j!},\quad m=0,\ldots,C_k.
\end{align*}
Note that $\mathcal Z(\bm C)=\sum_{m=0}^{C_k}W_m$ and $\mathcal Z(\bm C-\bm e_k)=\sum_{m=0}^{C_k-1} W_m$. For $m=0,\ldots,C_k-1$, we observe that
\begin{align*}
(m+1)W_{m+1} &= \sum_{\bm x:\bm A\bm x\le\bm C,\ \sum_{l:k\in\mathcal B^l}x_l=m+1} \ \left(\sum_{l:k\in\mathcal B^l} x_l\right) \prod_{j=1}^M\frac{(\omega^{j,F})^{x_j}}{x_j!} \\
&= \sum_{l:k\in\mathcal B^l} \sum_{\bm x:\bm A\bm x\le\bm C,\ \sum_{h:k\in\mathcal B^h}x_h=m+1} \ x_l\prod_{j=1}^M\frac{(\omega^{j,F})^{x_j}}{x_j!} \\
&= \sum_{l :k\in\mathcal B^l}\omega^{l,F}\sum_{\bm x:\bm A(\bm x+\bm e_l)\le\bm C,\ \sum_{h:k\in\mathcal B^h}x_h=m} \ \prod_{j=1}^M\frac{(\omega^{j,F})^{x_j}}{x_j!} \\
&\le \sum_{l:k\in\mathcal B^l}\omega^{l,F}\sum_{\bm x:\bm A\bm x\le\bm C,\ \sum_{h:k\in\mathcal B^h}x_h=m} \ \prod_{j=1}^M\frac{(\omega^{j,F})^{x_j}}{x_j!} = \left(\sum_{l:k\in\mathcal B^l}\omega^{l,F}\right)W_m.
\end{align*}
The first equality follows from $\sum_{l:k\in\mathcal B^l}x_l=m+1$ in $W_{m+1}$. The second equality expands this total number of customers using resource $k$ class by class. The third equality follows from algebra and $x_l(\omega^{l,F})^{x_l}/x_l!=\omega^{l,F}(\omega^{l,F})^{x_l-1}/(x_l-1)!$. The inequality follows because $\bm A(\bm x+\bm e_l)\le\bm C$ implies $\bm A\bm x\le\bm C$, so the summation set is enlarged. The final equality is by the definition of $W_m$. For every $m=0,\ldots,C_k$, iterating the above inequality gives
\begin{align*}
W_m\ge W_{C_k}\frac{C_k!}{m! \left(\sum_{l:k\in\mathcal B^l}\omega^{l,F}\right)^{C_k-m}} \Longrightarrow \frac{W_{C_k}}{\sum_{m=0}^{C_k}W_m}\le \frac{\left(\sum_{l:k\in\mathcal B^l}\omega^{l,F}\right)^{C_k}/C_k!}{\sum_{m=0}^{C_k}\left(\sum_{l:k\in\mathcal B^l}\omega^{l,F}\right)^m/m!}.
\end{align*}
Since $\mathcal Z(\bm C-\bm e_k)/\mathcal Z(\bm C)=1-W_{C_k}/\sum_{m=0}^{C_k}W_m$, we obtain that 
\begin{align*}
\frac{\mathcal Z(\bm C-\bm e_k)}{\mathcal Z(\bm C)}\ge \frac{\sum_{m=0}^{C_k-1}\left(\sum_{l:k\in\mathcal B^l}\omega^{l,F}\right)^m/m!}{\sum_{m=0}^{C_k}\left(\sum_{l:k\in\mathcal B^l}\omega^{l,F}\right)^m/m!}.
\end{align*}
Now fix class $j$ and express the set of resources used by class $j$ as $\mathcal B^j=\{k_1,\ldots,k_q\}$.
Since the multi-resource network is unit-bundle, i.e., $\bm a_j=\bm e_{k_1}+\cdots+\bm e_{k_q}$, applying the above lower bound yields that
\begin{align*}
\alpha^{j}(\bm{\lambda}^F)=\frac{\mathcal Z(\bm C-\bm a_j)}{\mathcal Z(\bm C)}=\prod_{r=1}^q\frac{\mathcal Z(\bm C-\sum_{t=1}^r\bm e_{k_t})}{\mathcal Z(\bm C-\sum_{t=1}^{r-1}\bm e_{k_t})} \geq \prod_{k\in\mathcal B^j}\frac{\sum_{m=0}^{C_k-1}\left(\sum_{l:k\in\mathcal B^l}\omega^{l,F}\right)^m/m!}{\sum_{m=0}^{C_k}\left(\sum_{l:k\in\mathcal B^l}\omega^{l,F}\right)^m/m!}.
\end{align*}
Recall that the revenue of $\bm\lambda^F$ is
\begin{align*}
\mathcal R^{\bm\lambda^F}=\sum_{j=1}^M r^j(\lambda^{j,F})\alpha^{j}(\bm\lambda^F) \ge \sum_{j=1}^M r^j(\lambda^{j,F})\prod_{k\in\mathcal B^j}\frac{\sum_{m=0}^{C_k-1}\left(\sum_{l:k\in\mathcal B^l}\omega^{l,F}\right)^m/m!}{\sum_{m=0}^{C_k}\left(\sum_{l:k\in\mathcal B^l}\omega^{l,F}\right)^m/m!}.
\end{align*}
Because $\mathcal R^{\mathrm{sta}*}\ge\mathcal R^{\bm\lambda^F}$ and $\mathcal R^*\le\Phi(\bm C)$, we have that
\begin{align*}
\frac{\mathcal R^{\mathrm{sta}*}}{\mathcal R^*}\ge \sum_{j=1}^M\frac{r^j(\lambda^{j,F})}{\Phi(\bm C)}\prod_{k\in\mathcal B^j}\frac{\sum_{m=0}^{C_k-1}\left(\sum_{l:k\in\mathcal B^l}\omega^{l,F}\right)^m/m!}{\sum_{m=0}^{C_k}\left(\sum_{l:k\in\mathcal B^l}\omega^{l,F}\right)^m/m!}.
\end{align*}
It remains to derive the simpler universal bound. The fraction appearing in the product is the service level of a one-resource Erlang loss system with capacity $C_k$ and offered load $\sum_{l:k\in\mathcal B^l} \omega^{l,F}$. This service level is decreasing in the offered load. Since the fluid solution satisfies $\sum_{l:k\in\mathcal B^l}\omega^{l,F}\le C_k$ for every resource $k$, we obtain that
\begin{align*}
\frac{\sum_{i=0}^{C_k-1}\left(\sum_{l:k\in\mathcal B^l}\omega^{l,F}\right)^i/i!}{\sum_{i=0}^{C_k}\left(\sum_{l:k\in\mathcal B^l}\omega^{l,F}\right)^i/i!} \ge \frac{\sum_{i=0}^{C_k-1}C_k^i/i!}{\sum_{i=0}^{C_k}C_k^i/i!}  .
\end{align*}
% Since $\frac{\sum_{i=0}^{C-1}C^i/i!}{\sum_{i=0}^{C}C^i/i!}\ge \frac{3}{5} $ for $C \geq2$ and $\sum_{j=1}^M r^j(\lambda^{j,F})=\Phi(\bm C)$, we conclude that
% \begin{align*}
%     \frac{\mathcal R^{\mathrm{sta}*}}{\mathcal R^*} \ge \sum_{j=1}^M\frac{r^j(\lambda^{j,F})}{\Phi(\bm C)} \prod_{k\in\mathcal B^j}\frac{\sum_{i=0}^{C_k-1}C_k^i/i!}{\sum_{i=0}^{C_k}C_k^i/i!} \geq \left(\frac{3}{5}\right)^{\max_j|\mathcal B^j|}, \forall C_k \geq 2 , \quad  \frac{\mathcal R^{\mathrm{sta}*}}{\mathcal R^*} \ge  \left(\frac{1}{2}\right)^{\max_j|\mathcal B^j|}, \exists \ C_k =1. 
% \end{align*}
Since $\frac{\sum_{i=0}^{C-1}C^i/i!}{\sum_{i=0}^{C}C^i/i!}\ge \frac{1}{2} $ for $C \geq 1$ and $\sum_{j=1}^M r^j(\lambda^{j,F})=\Phi(\bm C)$, we conclude that
\begin{align*}
    \frac{\mathcal R^{\mathrm{sta}*}}{\mathcal R^*} \ge \sum_{j=1}^M\frac{r^j(\lambda^{j,F})}{\Phi(\bm C)} \prod_{k\in\mathcal B^j}\frac{\sum_{i=0}^{C_k-1}C_k^i/i!}{\sum_{i=0}^{C_k}C_k^i/i!}  \geq  \left(\frac{1}{2}\right)^{\max_j|\mathcal B^j|}, 
\end{align*}
Moreover, we show that our analysis is tight by providing a matching upper bound.

\begin{example}\label{ex:fluid-tight}
Fix $C\ge 1 $. Consider a unit-bundle network with $K$ resources, each with $C$ units. There are $K+1$ classes. Class $0$ is the high-value class and requires all resources, i.e., $\mathcal B^0=\{1,\ldots,K\}$.  Class $k$ requires only resource $k$, i.e., $\mathcal B^k=\{k\}$. Let all service rates equal one. For small $\epsilon>0$ and large $H$ such that $H/\epsilon > K $, the revenue rate functions are
\begin{align*}
    r^0(\lambda)=H \cdot \min\{\lambda/\epsilon,1\}, \quad r^k(\lambda)=\lambda,\ k=1,\ldots,K
\end{align*}
The above concave revenue rate functions can be approximated arbitrarily closely by revenue rate functions induced by regular valuation distributions satisfying the assumptions in Section \ref{Model}. The fluid relaxation is $\max_{\bm\lambda\ge0}\{H\min\{\lambda^0/\epsilon,1\}+\sum_{k=1}^{K}\lambda^k:\lambda^0+\lambda^k\le C,\ k=1,\ldots,K\}$. When $H/\epsilon>K$, an optimal fluid solution is $\lambda^{0,F}=\epsilon$ and $\lambda^{k,F}=C-\epsilon$ for $k=1,\ldots,K$, with fluid value $\Phi(\bm C)=H+K(C-\epsilon)$. We now compute the service level of class $0$ under $\bm{\lambda}^F$. Recall that $\bm C=(C,\ldots,C)$ and $\bm a_0=(1,\ldots,1)$. Since class $0$ uses all resources, the service level of class 0 is given by $ \alpha^{0}(\bm\lambda^F)=\frac{\mathcal Z(\bm C-\bm a_0)}{\mathcal Z(\bm C)}.$ Under $\bm\lambda^F$, we compute that $\mathcal Z(\bm C)=\sum_{m=0}^{C}\frac{\epsilon^m}{m!}\left(\sum_{i=0}^{C-m}\frac{(C-\epsilon)^i}{i!}\right)^K$ and $ \mathcal Z(\bm C-\bm a_0)=\sum_{m=0}^{C-1}\frac{\epsilon^m}{m!}\left(\sum_{i=0}^{C-1-m}\frac{(C-\epsilon)^i}{i!}\right)^K$. Hence, the service level of class 0 and its limit when $\epsilon \to 0$ are given by
\begin{align*}
\alpha^{0}(\bm\lambda^F)=\frac{\sum_{m=0}^{C-1}\frac{\epsilon^m}{m!}\left(\sum_{i=0}^{C-1-m}\frac{(C-\epsilon)^i}{i!}\right)^K}{\sum_{m=0}^{C}\frac{\epsilon^m}{m!}\left(\sum_{i=0}^{C-m}\frac{(C-\epsilon)^i}{i!}\right)^K}, \quad \lim_{\epsilon\to 0}\alpha^{0}(\bm\lambda^F)=\left(\frac{\sum_{i=0}^{C-1}C^i/i!}{\sum_{i=0}^{C}C^i/i!}\right)^K
\end{align*}
 Now consider a feasible dynamic policy that offers only class $0$ at rate $\epsilon$ and blocks all other classes. This dynamic policy yields revenue $H \cdot \frac{\sum_{i=0}^{C-1}\epsilon^i/i!}{\sum_{i=0}^{C}\epsilon^i/i!}$, implying that $\mathcal R^*\ge H \cdot \frac{\sum_{i=0}^{C-1}\epsilon^i/i!}{\sum_{i=0}^{C}\epsilon^i/i!}$. On the other hand, $\mathcal R^{\bm\lambda^F}$ is at most $H\alpha^{0}(\bm\lambda^F)+K(C-\epsilon)$. Therefore, we have that
\begin{align*}
\lim_{\epsilon\to0}\limsup_{H\to\infty}\frac{\mathcal{R}^{\bm\lambda^F}}{\mathcal R^*}\le \lim_{\epsilon\to0}\frac{\alpha^{0}(\bm\lambda^F)}{\frac{\sum_{i=0}^{C-1}\epsilon^i/i!}{\sum_{i=0}^{C}\epsilon^i/i!}}=\left(\frac{\sum_{i=0}^{C-1}C^i/i!}{\sum_{i=0}^{C}C^i/i!}\right)^K.
\end{align*}
When $ C=1$, this upper bound is $(1/2)^K$. Hence no universal guarantee for the fluid policy $\bm{\lambda}^F$ can improve upon $(1/2)^K$ as a function of the maximum bundle size $\max_j|\mathcal B^j| $ (note that $K=\max_j|\mathcal B^j|$ in this example). 

Moreover, for $K=1$, the upper bound is $ 1 - \frac{\frac{1}{C!} C^C }{ \sum_{i=0}^{C} \frac{1}{i!} C^{i} } \geq 3/5 $, showing that this guarantee is tight in Proposition \ref{prop:fluid-bound} in Section \ref{sec:fluid-bound}. 

% When $C=1$,  this upper bound is $(1/2)^K$. 
% and when  $C=2$, this upper bound is exactly $(3/5)^K$. Hence no universal guarantee for the fluid policy $\bm{\lambda}^F$ can improve upon $(3/5)^K$ as a function of the maximum bundle size $\max_j|\mathcal B^j| $ (note that $K=\max_j|\mathcal B^j|$ in this example). \hfill \Halmos
\end{example}

(ii). The following example shows that the fluid-based static policy can have an arbitrarily small performance ratio, even with one resource.

\begin{example}
Consider one reusable resource with capacity $C=n^3$($n\ge2$). There are two classes. Class $1$ requires one unit, while class $2$ requires $n^2$ units.  Service times are exponential with $\mu^1=\mu^2=1$. Fix $0< \epsilon <1$ and large $H$ such that $H/\epsilon > n^2$. The revenue rate functions are $r^1(\lambda)=\lambda$ and $r^2(\lambda)=H \cdot \min\{\lambda/\epsilon,1\}$. The above concave revenue rate functions can be approximated arbitrarily closely by revenue rate functions induced by regular valuation distributions satisfying the assumptions in Section \ref{Model}. Then, the fluid relaxation is $\Phi(C)=\max_{\lambda^1,\lambda^2 \ge 0}\left\{\lambda^1 + H \min\left\{\frac{\lambda^2}{\epsilon},1\right\}: \lambda^1+ n^2 \lambda^2 \le C\right\}.$ Since $H/\epsilon>n^2$, an optimal fluid solution is $\lambda^{1,F}=C- n^2\epsilon$ and $\lambda^{2,F}=\epsilon$ with optimal value $\Phi(C)=C-n^2\epsilon + H$. Under the fluid policy, the service level of class $2$ and its limit when $\epsilon \to 0$ are given by
\begin{align*}
    \alpha^2(\bm\lambda^F)= \frac{\sum_{m=0}^{n-1}\frac{\epsilon^m}{m!}\sum_{i=0}^{C-n^2(m+1)}\frac{(C-n^2\epsilon)^i}{i!}}{\sum_{m=0}^{n}\frac{\epsilon^m}{m!}\sum_{i=0}^{C-n^2m}\frac{(C-n^2\epsilon)^i}{i!}}, \quad  \lim_{\epsilon\to0} \alpha^2(\bm\lambda^F)= \frac{\sum_{i=0}^{C-n^2}C^i/i!}{\sum_{i=0}^{C}C^i/i!}.
\end{align*}
Since $r^1(\lambda^{1,F}) \leq C $, we have that $\mathcal R^{\bm\lambda^F}\le C + H\cdot \alpha^2(\bm\lambda^F)$. On the other hand, a feasible dynamic policy offers only class $2$ at rate $\epsilon$ and blocks class $1$, which has revenue $H \cdot \frac{\sum_{i=0}^{n-1}\epsilon^i/i!}{\sum_{i=0}^{n}\epsilon^i/i!}.$ Therefore, we have that
\begin{align*}
\lim_{\epsilon\to0}\limsup_{H\to\infty} \frac{\mathcal R^{\bm\lambda^F}}{\mathcal R^*}
\le  \lim_{\epsilon\to0} \limsup_{H\to\infty} \frac{C + H \alpha^2(\bm{\lambda}^F)}{H\frac{\sum_{i=0}^{n-1}\epsilon^i/i!}{\sum_{i=0}^{n}\epsilon^i/i!}}  
= \frac{\sum_{i=0}^{C-n^2} \frac{C^i}{i!} }{\sum_{i=0}^{C}\frac{C^i}{i!} }  \leq n^3 e^{-\frac{n}{2}+\frac{1}{2n}} ,
\end{align*}
which converges to zero as $ n \to\infty$. The last inequality follows from 
\begin{align*}
\frac{\sum_{i=0}^{C-n^2} \frac{C^i}{i!} }{\sum_{i=0}^{C}\frac{C^i}{i!} } 
\leq \frac{(C-n^2 +1 ) \frac{C^{C-n^2}}{(C-n^2)!} }{ \frac{C^C}{C!} } = (C-n^2+1) \prod_{r=0}^{n^2-1}\left[1-\frac{r}{C}\right] \le Ce^{-\frac{n^2(n^2-1)}{2C}} = n^3 e^{-\frac{n}{2}+\frac{1}{2n}} ,
\end{align*}
where the first inequality holds since $C^i/i!$ is increasing in $i$ for $i\le C$ and the second inequality follows from $1 - x \leq e^{-x}$. Hence the fluid policy does not admit a positive universal constant guarantee in general non-unit-bundle networks. \hfill \Halmos

% In contrast, the buffered fluid bound is nontrivial for this sequence of instances. Here $D_1=n^2$ and $\eta=C_n/D_1=n$. Let $\epsilon=n^{-1/3}$ be the buffer parameter. Since $\eta=n$, we have $\epsilon\in(1/\eta,1)$ for all sufficiently large $n$. Also, $\max_j|\mathcal B^j|=1$ and $r^1(0)=r^2(0)=0$. Therefore, Proposition~\ref{prop: network} gives
% \begin{align*}
% \frac{\mathcal R^{\bm\lambda^{\epsilon}}}{\mathcal R^*} \ge  \left[ 1-e^{-(1-\epsilon)n\psi\left(\frac{\epsilon-1/n}{1-\epsilon} \right)} \right]_+ \cdot (1-\epsilon).
% \end{align*}
% Hence the buffered fluid policy is asymptotically optimal on this sequence, while the exact fluid policy fails. 
\end{example}

\end{proof}

\end{APPENDICES}

\end{document}